\documentclass[preprint,12pt, compress]{elsarticle}

\usepackage{amssymb}
\usepackage{hyperref}
\usepackage{xcolor}
\usepackage{listings}
\usepackage{bm}
\usepackage{relsize}
\usepackage{tabularx}
\usepackage{booktabs}

\usepackage{amsmath}
\usepackage{braket}

\definecolor{codegreen}{HTML}{faac5e}
\definecolor{codegray}{rgb}{0.5,0.5,0.5}
\definecolor{codepurple}{HTML}{45c2c8}
\definecolor{backcolour}{rgb}{0.94,0.94,0.92}
\definecolor{codemagenta}{HTML}{f15e64}

\lstdefinestyle{mystyle}{
    backgroundcolor=\color{backcolour},   
    commentstyle=\color{codegreen},
    keywordstyle=\color{codemagenta},
    numberstyle=\tiny\color{codegray},
    stringstyle=\color{codepurple},
    basicstyle=\ttfamily\footnotesize,
    breakatwhitespace=false,         
    breaklines=true,                 
    captionpos=b,                    
    keepspaces=true,                 
    numbers=left,                    
    numbersep=5pt,                  
    showspaces=false,                
    showstringspaces=false,
    showtabs=false,                  
    tabsize=2
}
\newcommand{\ionion}{_\mathrm{ii}}
\newcommand{\eleion}{_\mathrm{ei}}
\newcommand{\eleele}{_\mathrm{ee}}
\newcommand{\ion}{_\mathrm{i}}
\newcommand{\ele}{_\mathrm{e}}
\newcommand{\br}{\bm{r}}
\newcommand{\bR}{\bm{R}}
\newcommand{\ubr}{\underline{\bm{r}}}
\newcommand{\ubR}{\underline{\bm{R}}}

\newcommand{\bmmatrix}[1]{\bm{\mathrm{#1}}}
\newcommand{\bk}{\bm{k}}
\newcommand{\s}{_\mathrm{{\scriptscriptstyle S}}}

\newcommand{\xc}{_\mathrm{{\scriptscriptstyle XC}}}

\newcommand{\Hartree}{_\mathrm{{\scriptscriptstyle H}}}
\newcommand{\kB}{k_\mathrm{B}}

\DeclareMathOperator{\Tr}{Tr}

\journal{Computer Physics Communications}
\newcolumntype{b}{X}
\newcolumntype{s}{>{\hsize=.4\hsize}X}

\begin{document}

\begin{frontmatter}



\title{Materials Learning Algorithms (MALA): Scalable Machine Learning for Electronic Structure Calculations in Large-Scale Atomistic Simulations}


\author[hzdr,casus]{Attila Cangi\corref{cor1}\fnref{fn1}}

\author[hzdr,casus]{Lenz Fiedler\fnref{fn1}}

\author[hzdr,casus]{Bartosz Brzoza}

\author[hzdr,casus]{Karan Shah}

\author[hzdr,casus]{Timothy J. Callow}

\author[hzdr,casus]{Daniel Kotik}

\author[hzdr]{Steve Schmerler}

\author[snl]{Matthew C. Barry}

\author[snl]{James M. Goff}

\author[snl]{Andrew Rohskopf}

\author[snl]{Dayton J. Vogel}

\author[snl]{Normand Modine}

\author[snl]{Aidan P. Thompson}

\author[snl]{Sivasankaran Rajamanickam}

\affiliation[hzdr]{organization={Helmholtz-Zentrum Dresden-Rossendorf},%Department and Organization
            %addressline={}, 
            city={Dresden},
            postcode={01328}, 
            %state={},
            country={Germany}}

\affiliation[casus]{organization={Center for Advanced Systems Understanding},%Department and Organization
            %addressline={}, 
            city={Görlitz},
            postcode={02826}, 
            %state={},
            country={Germany}}

\affiliation[snl]{organization={Sandia National Laboratories},%Department and Organization
            %addressline={}, 
            city={Albuquerque},
            postcode={87123}, 
            state={NM},
            country={United States}}

\cortext[cor1]{Corresponding author: a.cangi@hzdr.de}
\fntext[fn1]{Both authors contributed equally to this manuscript.}

\begin{abstract}
We present the Materials Learning Algorithms (\texttt{MALA}) package, a scalable machine learning framework designed to accelerate density functional theory (DFT) calculations suitable for large-scale atomistic simulations. Using local descriptors of the atomic environment, \texttt{MALA} models efficiently predict key electronic observables, including local density of states, electronic density, density of states, and total energy. The package integrates data sampling, model training and scalable inference into a unified library, while ensuring compatibility with standard DFT and molecular dynamics codes. We demonstrate \texttt{MALA}'s capabilities with examples including boron clusters, aluminum across its solid-liquid phase boundary, and predicting the electronic structure of a stacking fault in a large beryllium slab. Scaling analyses reveal \texttt{MALA}'s computational efficiency and identify bottlenecks for future optimization. With its ability to model electronic structures at scales far beyond standard DFT, \texttt{MALA} is well suited for modeling complex material systems, making it a versatile tool for advanced materials research.
\end{abstract}




\end{frontmatter}

{\bf PROGRAM SUMMARY/NEW VERSION PROGRAM SUMMARY}
  %Delete as appropriate.

% LF: This is a required section/field for CPiC papers
\begin{small}
\noindent
{\em Program Title: }Materials Learning Algorithms (\texttt{MALA})       \\
{\em CPC Library link to program files:} (to be added by Technical Editor) \\
{\em Developer's repository link:} \url{https://github.com/mala-project/mala} \\
{\em Code Ocean capsule:} (to be added by Technical Editor)\\
{\em Licensing provisions(please choose one):} BSD 3-clause  \\
{\em Programming language:} Python, Fortran                                   \\
{\em Supplementary material:} None                                 \\
{\em Nature of problem(approx. 50-250 words):} Predicting material properties at experimentally relevant scales remains a significant challenge in advanced materials research. Traditional electronic structure methods, particularly density functional theory (DFT), while widely used, are limited by their steep computational demands. DFT calculations typically scale cubically with system size, restricting them to simulations of only a few thousand atoms. Consequently, large-scale simulations often require approximations, classical models, or machine learning methods. Yet, current machine learning approaches fall short in replicating the full predictive accuracy of DFT, thereby constraining their broader applicability in materials science.\\
{\em Solution method(approx. 50-250 words):} The Materials Learning Algorithms (\texttt{MALA}) software package~\cite{1} provides a scalable and efficient machine learning framework designed to accelerate electronic structure calculations. \texttt{MALA} replaces direct DFT calculations with machine learning models, enabling simulations at unprecedented scales. By employing a universal electronic structure representation, \texttt{MALA} models recover key observables from which a broad range of material properties can be derived. The package provides a unified library that integrates data sampling, model training, and scalable inference, while offering seamless compatibility with state-of-the-art simulation tools such as \texttt{Quantum ESPRESSO}~\cite{2} and \texttt{LAMMPS}~\cite{3}. This makes \texttt{MALA} a versatile solution for diverse computational materials research needs.\\
{\em Additional comments including restrictions and unusual features (approx. 50-250 words):} None\\
   \\

\end{small}

\section{Introduction}
\label{s.introduction}

The electronic structure problem, governed by the quantum interactions between ions and electrons in materials, represents one of the most fundamental challenges in computational science. These interactions determine material properties at all scales, from the atomic to the macroscopic, making their accurate prediction crucial for both advancing fundamental science and driving technological innovation. However, solving the full many-body quantum problem for real-world systems remains computationally prohibitive due to the exponential scaling of complexity with system size.

A major breakthrough in addressing this challenge came with the development of density functional theory (DFT)~\cite{HoKo1964, KoSh1965}. By reformulating the electronic structure problem in terms of the electronic density rather than the many-body wavefunction, DFT significantly reduces computational demands, scaling cubically with system size, i.e., the number of atoms. This balance between computational efficiency and reasonable accuracy has made DFT an indispensable tool for predicting a wide range of material properties and chemical phenomena.

Today, DFT is one of the most widely used methods in computational materials science, with applications spanning condensed matter physics, chemistry, biology, and even astrophysics. Tens of thousands of research papers rely on DFT each year~\cite{Bu2012}. However, despite its broad utility, DFT remains limited by its computational scaling, which restricts simulations to systems of only a few thousand atoms. This limitation motivates the search for methods that can extend the reach of electronic structure calculations to larger systems and longer timescales.

Resolving the electronic structure on scales beyond the capability of DFT calculations is critical for advancing several fields: (1) in biomolecular systems, accurate modeling of enzyme active sites is essential for drug discovery~\cite{WeMc24}; (2) at electrochemical interfaces, understanding electron transport~\cite{ToFuDoDe24,Ha23} is key to developing better batteries; (3) in catalysis, capturing electronic state changes is vital for designing efficient catalysts~\cite{Th08,ZhLiZhCh19}; and (4) in organic and flexible electronics based on polymers, charge transport and polymer structure relationships are critical for material optimization~\cite{HuWaLiZh22,ZhWaSuTo24}. These examples underscore the need for methods that go beyond DFT’s computational limits.

Machine learning has emerged as a powerful tool in computational chemistry and materials science, offering new avenues to enhance both accuracy and efficiency. In particular, machine learning techniques have shown promise in accelerating electronic structure calculations by learning to predict key observables directly from atomic configurations. Given that DFT calculations account for a significant portion of the computational workload on high-performance computing platforms, acceleration based on machine learning could dramatically increase throughput for material screening and optimization and extend the accessible scales of simulations.

In this work, we present the Materials Learning Algorithms (\texttt{MALA}) software package~\cite{malagithub}, a scalable machine learning framework designed to accelerate standard DFT calculations. Due to its linear computational scaling with the number of atoms, \texttt{MALA} enables large-scale simulations by generating electronic structure properties directly from atomic snapshots which are the basic input of any DFT calculation. It produces key quantities, including the local density of states, charge density, band energy, and total energy, while bypassing the need for computationally intensive DFT calculations. By integrating data sampling, model training, and inference in a modular framework, \texttt{MALA} extends the reach of electronic structure calculations to systems far beyond the capabilities of standard DFT methods.

\section{Theoretical background}
\label{s.background}

\subsection{Electronic structure problem}
We begin by considering a system of $N\ele$ electrons and $N\ion$ ions with collective coordinates $\ubr={\br_1,\dots,\br_{N\ele}}$ and $\ubR={\bR_1,\dots,\bR_{N\ion}}$, where $\br_j \in \mathbb{R}^3$ refers to the position of the $j$th electron, and $\bR_\alpha \in \mathbb{R}^3$ denotes the position of the $\alpha$th ion of mass $M_\alpha$ and charge $Z_\alpha$. 
Working within the scope of non-relativistic quantum mechanics and adopting the Born-Oppenheimer approximation~\cite{born_zur_1927}, the time-independent Schrödinger equation is given by 
\begin{align} 
\label{eq:se} 
\hat{H}(\ubr; \ubR)\, \Psi(\ubr; \ubR) &= E\, \Psi(\ubr; \ubR)\;, 
\end{align} 
where $\Psi(\ubr; \ubR)$ denotes the many-body wavefunction and the Hamiltonian
\begin{align} 
\label{eq:hamiltonian} 
\hat{H}(\ubr; \ubR) &= \hat{T}\ele(\ubr) + \hat{V}\eleele(\ubr) + \hat{V}\eleion(\ubr; \ubR) + E\ionion(\ubR)\,,
\end{align} 
where 
\begin{align} 
\hat{T}\ele(\ubr) &= -\frac{1}{2} \sum_{j}^{N\ele} \nabla_j^2 
\end{align} denotes the kinetic energy of the electrons, 
\begin{align} 
\hat{V}\eleele(\ubr) 
&= \frac{1}{2}\sum_{j}^{N\ele} \sum_{k\neq j}^{N\ele} \frac{1}{|\br_j -\br_k|}
\end{align} 
the electron-electron interaction, 
\begin{align} 
\hat{V}\eleion(\ubr; \ubR) &= -\sum_{j}^{N\ele} \sum_{\alpha}^{N\ion} \frac{Z_\alpha}{|\br_j-\bR_\alpha|}
\end{align}
the electron-ion interaction, and 
\begin{align} 
E\ionion(\ubR) &= \frac{1}{2}\sum_{\alpha}^{N\ion} \sum_{\beta \neq \alpha}^{N\ion} \frac{Z_\alpha Z_\beta}{|\bR_\alpha -\bR_\beta|}
\end{align} 
the ion-ion interaction, which amounts to a constant shift in energy within the Born-Oppenheimer approximation. Additionally, the Hamiltonian and the many-body wavefunction depend only parametrically on the coordinates of the ions indicated by the semicolon. In the following we will drop this parametric dependence to simplify our notation.  
Also note that we adopt atomic units here and in the following, where $\hbar = m\ele = e^2 = 1$, such that energies are expressed in Hartrees and lengths in Bohr radii.

\subsection{Density functional theory}
The electron-electron interaction term complicates finding solutions for $\Psi$ in Eq.~\eqref{eq:se}. Instead, a computationally feasible approach is provided by the Kohn-Sham scheme~\cite{KoSh1965} of DFT~\cite{HoKo1964}. 
This approach maps the interacting electron problem onto a fictitious system of non-interacting electrons governed by the Kohn-Sham equations
\begin{equation} 
\label{eq:dft} 
\left[-\frac{1}{2}\nabla^2 + v\s(\br)\right] \psi_j(\br) = \epsilon_j \psi_j(\br) 
\end{equation} 
which form a set of single-particle Schrödinger equations, where the index $j=1,\dots,N\ele$ labels non-interacting electrons whose number equals that of the electrons in the original many-electron problem as defined in Eq.~\eqref{eq:se}. Note also a subtle difference in the meaning of the spatial coordinate $\br$. While $\ubr$ denotes the position of the electrons, $\br$ is a variable that denotes a spatial grid.  
Solving the Kohn-Sham equations yields the Kohn-Sham orbitals $\psi_j$ and corresponding eigenvalues $\epsilon_j$. 
The Kohn-Sham scheme is designed so that the single-particle electronic density
\begin{equation} 
\label{eq:density_dft} 
n(\br) = \sum_j^{N\ele} |\psi_j(\br)|^2
\end{equation} 
matches the electronic density $n(\br)=\int dr_2\dots \int dr_{N\ele} |\Psi(\br,\br_2,\dots,\br_{N\ele})|^2$ from solving the many-electron Schrödinger equation. 
This correspondence is achieved by the Kohn-Sham potential 
\begin{align} 
v\s(\br) &= \frac{\delta E\Hartree[n]}{\delta n(\br)} + \frac{\delta E\xc[n]}{\delta n(\br)} + v\eleion(\br)\; ,
\end{align} 
an effective single-particle potential that expresses electron-electron interactions at a mean-field level. The components in this expression are the Hartree energy $E\Hartree[n]$, the exchange-correlation (XC) energy $E\xc[n]$, and the electron-ion interaction $v\eleion(\br) = -\sum_\alpha Z_\alpha/|\br-\bR_\alpha|$. Although the Kohn-Sham scheme is formally exact, practical applications use well-established approximations for $E\xc$ whose exact form is unknown.
In the Kohn-Sham scheme, the total energy is given by 
\begin{align} 
\label{eq:energy.ks} 
\begin{split} 
E[n] &= T\s[n] + E\Hartree[n] + E\xc[n] + \int d\br\ n(\br) v\eleion(\br) + E\ionion\,, 
\end{split} 
\end{align} 
where $T\s$ denotes the Kohn-Sham kinetic energy and $E\ionion$ the the ion-ion interaction which only depends on the positions of the ions $\ubR$. \texttt{MALA} uses the generalization of Kohn-Sham DFT to finite temperature. The necessary details are provided in ~\ref{sec:appendix.thermal.electronic.ensembles}, while some relevant concepts, such as the Fermi function and the free energy, are introduced below as needed.

\subsection{Reformulating density functional theory suitable for machine learning}
Although the density $n(\br)$ is the primary quantity in DFT, it is not the most convenient quantity for building a machine learning model. A model of the density alone is insufficient for total energy evaluation because an expression for the Kohn-Sham kinetic energy as an explicit functional of $n$ is not known. 
Consequently, an additional machine learning model would be required to map densities to kinetic energies, introducing further complexity and possible accuracy loss~\cite{SRH2012,SRMB2015,BrVo2017}.

As an alternative, we consider the local density of states (LDOS), defined as 
\begin{equation} 
\label{eq:ldos} 
d(\br, \epsilon) = \sum_j |\psi_{j} (\br)|^2\, \delta(\epsilon-\epsilon_j)\;, 
\end{equation} 
where $\epsilon$ is a continuous energy variable and $\delta(\epsilon)$ is the Dirac delta function. 
The LDOS has the property that integrating it over energy yields the electronic density 
\begin{equation} 
\label{eq:density.from.ldos} 
n(\br) = \int d\epsilon\ f_\beta(\epsilon)\, d(\br, \epsilon)\,,
\end{equation} 
where $f_\beta(\epsilon)$ is an occupation function (typically the Fermi function) appropriate to inverse electronic temperature $\beta = 1/(k_{B}\tau)$ with the Boltzmann constant $k_{B}$ defined in Eq.~\eqref{eq:fermidistribution}. Likewise, integrating the LDOS over space yields the density of states (DOS)
\begin{equation} 
\label{eq:DOS.from.ldos} 
D(\epsilon) = \int d\br\ d(\br, \epsilon)\ . 
\end{equation}

By virtue of the first Hohenberg-Kohn theorem~\cite{HoKo1964}, we can treat the LDOS as the primary quantity, allowing us to rewrite the total energy as 
\begin{align} 
\label{eq:energy.ldos} 
E & = E_\mathrm{b} - E\Hartree + E\xc - V\xc + E\ionion\,,
\end{align} 
where $V\xc = \int d\br\ n(\br)\, v\xc(\br)$ denotes the potential energy component of the XC energy, $v\xc(\br) = \delta E\xc/\delta n(\br)$, and $E_\mathrm{b}$ the band energy defined below~\cite{DrGr90}.
The expressions for the total energy in Eq.~\eqref{eq:energy.ks} and Eq.~\eqref{eq:energy.ldos} are mathematically equivalent.

The primary advantage of this reformulation is that, through Eq.~\eqref{eq:energy.ldos}, the total energy is expressed entirely in terms of the LDOS, making it suitable for machine learning. This allows us to build a machine learning model for the LDOS, which can then be used to obtain the total energy.

The Hartree energy $E\Hartree$, XC energy $E\xc$, and potential XC energy $V\xc$ are all implicitly determined by the LDOS via their dependence on the electronic density. 
The band energy 
\begin{equation} 
\label{eq:band.energy} 
E_\mathrm{b} = \int d\epsilon\ f_\beta(\epsilon)\, \epsilon\, D(\epsilon) 
\end{equation} 
is similarly determined by the LDOS through its dependence on the DOS.

This provides the theoretical basis for the \texttt{MALA} model which is based on learning the LDOS.
For simplicity, the description of the theoretical background is limited to many-electron systems in their ground state.
The generalization to thermal ensembles has been implemented in the MALA model. The corresponding background material is provided in ~\ref{sec:appendix.thermal.electronic.ensembles}.

\subsection{Machine learning with neural networks}

The \texttt{MALA} software framework uses supervised machine learning, where the general task is to find a model $M$ to approximate the ground-truth (data generating) function $f$
\begin{equation}
    \bm{y} = f(\bm{x}) \; ,
\end{equation}
which maps inputs $\bm{x}$ to outputs $\bm{y}$. Both input and output data can either be vectorial or scalar data, while the function $f$ may be analytically representable yet challenging to compute, or it may be unknown. In the context of this work, $f$ represents the electronic structure problem itself. Common input data includes ionic positions or their representations, while typical outputs are observables of interest, such as energies or ionic forces. Since the outputs are continuous variables, the learning task is a regression problem.

The model $M$ seeks to produce an accurate prediction $\bm{\tilde{y}}$ of the actual outputs $\bm{y}$ based on the inputs
\begin{equation}
    \bm{\tilde{y}} = M(\bm{x}) \; , \label{eq:principalMLmodel}
\end{equation}
where labeled data sets are used, with $\bm{y}_j \in  \bm{\underline{y}}$ and $\bm{x}_j \in \bm{\underline{x}}$, where $j$ indexes individual data points within the sets $\bm{\underline{y}}$ and $\bm{\underline{x}}$, each containing $N_j$ data points.

\begin{figure}[ht!]
\centering
\includegraphics[width=0.95\textwidth]{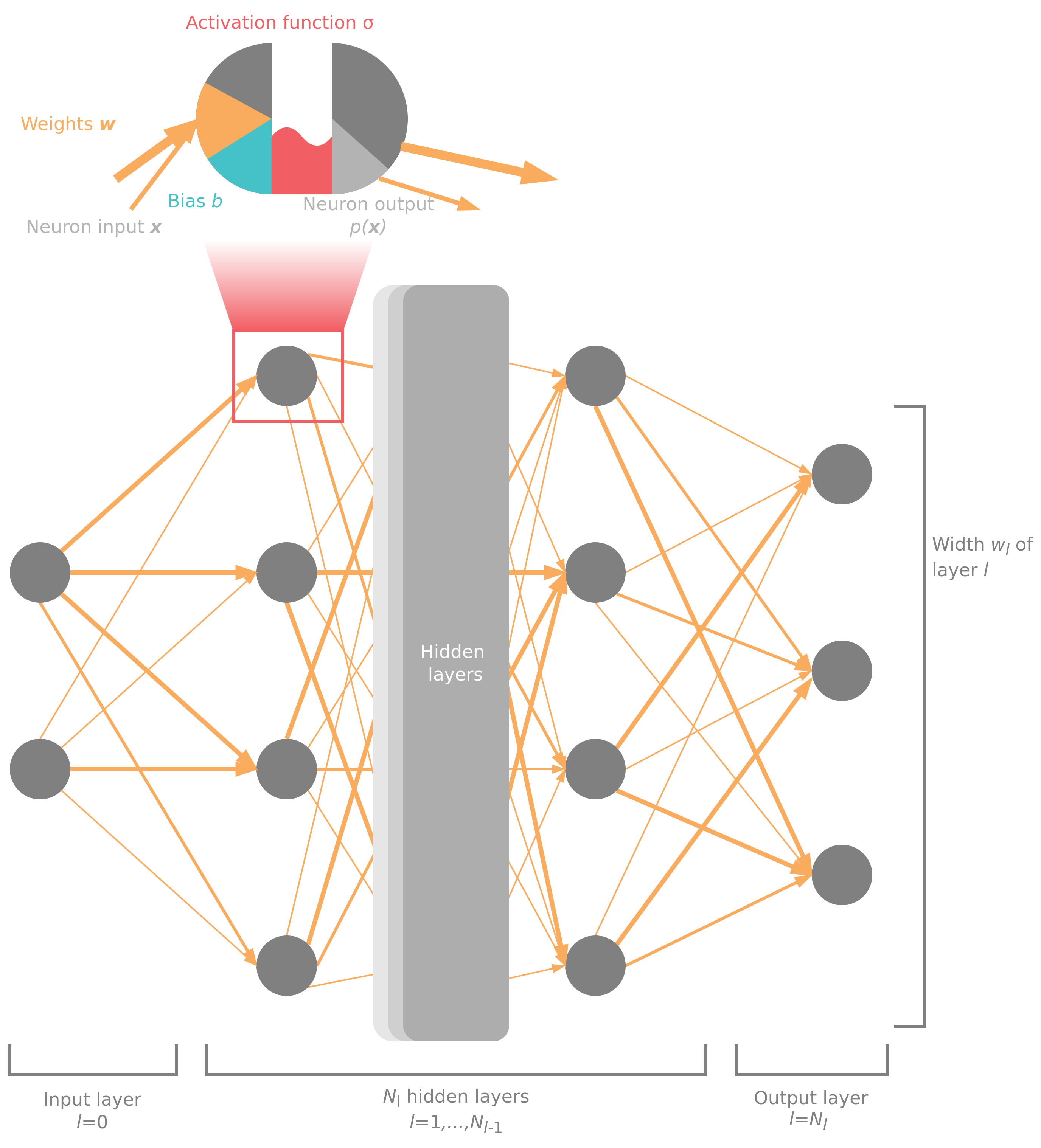}
\caption{Overview of the general workflow of a feed-forward neural network used in the \texttt{MALA} package. A network consists of $N_l$ layers each comprising $w_l$ neurons. Neurons are interconnected by weights $\bm{w}$. In the network, input data $\bm{x}$ is propagated to individual neurons, where it is multiplied by weights, summed, and added to a bias $b$. This value is then passed through a non-linear activation function $\sigma$, producing the neuron’s output. The output is subsequently passed to the next layer via different weights, and this process repeats until the final layer is reached, yielding the neural network’s output.}
\label{fig:nn_basic}
\end{figure}

We construct a machine learning model in terms of neural networks~\cite{DeepLearning1,DeepLearning2} as shown in Fig.~\ref{fig:nn_basic}. The basic building block of the model is an artificial neuron, which combines a linear transformation with a non-linear activation function, such that the output of a neuron is given by 
\begin{equation}
	p(\bm{x}) = \sigma(\bm{w}\cdot\bm{x}+b) \; , \label{eq:perceptron}
\end{equation}
where $\bm{x}$ is the input data, $\bm{w}$ and $b$ are the weights and biases, and $\sigma$ is a non-linear activation function. In a neural network, neurons are organized in layers, with the output of preceding layers serving as input to subsequent layers. A neural network can thus be defined as the transformation
\begin{equation}
	\bm{x}_{l+1} = \sigma(\bmmatrix{W}_l\bm{x}_l + \bm{b}_l) \; , \label{eq:ffnn}
\end{equation}
where $\bm{x}_l$ and $\bm{x}_{l+1}$ represent outputs of the $l$-th and $(l+1)$-th layer, while $\bmmatrix{W}_l$ and $\bm{b}_l$ are the weight matrix and bias vector of the $l$-th layer. This expression defines the simplest form of a neural network, known as feed-forward or fully connected neural network. 

A feed-forward neural network is characterized by its number of layers $N_l$, the width of each layer $w_0,...,w_l$, where $w_0$ and $w_l$ correspond to the input and output dimensions, and the choice of activation function $\sigma$. Layers between input and output, $l=[1,N_l-1]$ are referred to as hidden layers. As illustrated in Fig.~\ref{fig:nn_basic}, the activation function $\sigma$ introduces non-linear transformations, enabling neural networks to serve as powerful function approximators through interconnected transformations via weight matrices and biases.

To achieve practical accuracy, a neural network must be fit to relevant training data, adjusting $\underline{\bmmatrix{W}}$ and $\underline{\bm{b}}$ to minimize the difference between $\bm{y}$ and $\bm{\tilde{y}}$. This minimization is mathematically defined as
\begin{equation}
	\min_{\underline{\bmmatrix{W}},\underline{\bm{b}}} L(\bm{\underline{x}}, \bm{\underline{y}}, \underline{\bmmatrix{W}}, \underline{\bm{b}}) \; , \label{eq:training}
\end{equation}
where $L$ denotes the loss function, often taken as the mean squared error between the ground truth and predictions
\begin{equation}
	L(\bm{\underline{x}}, \bm{\underline{y}}, \underline{\bmmatrix{W}}, \underline{\bm{b}})  = \frac{1}{N_j} \sum_{j=1}^{N_j} {\left( \bm{\tilde{y}}_j - \bm{y}_j\right)}^2 =  \frac{1}{N_j} \sum_{j=1}^{N_j} {\left[ M(\bm{x}_j, \underline{\bmmatrix{W}}, \underline{\bm{b}}) - \bm{y}_j\right]}^2 \; . \label{eq:mseloss}
\end{equation}

During the minimization of $L$ (neural network training), weights and biases are updated iteratively using backpropagation~\cite{RuHiWi86} which is equal to reverse-mode automatic differentiation~\cite{baydin_2018_AutomaticDifferentiationMachine, mml_book_2020}. The latter enables the calculation of the gradient of $L$
with respect to network parameters $\underline{\bmmatrix{W}}$ and $\underline{\bm{b}}$ by propagating the prediction error backwards through the neural network. 

In practice, gradient calculations involve first computing the network's predicted output (forward pass), determining the output error, and then backpropagating this error (backward pass). This process is computationally efficient, especially for neural networks containing many neurons over multiple layers, and is highly optimized on graphical processing units (GPUs)~\cite{NNs_on_GPU}, which are standard for neural network training.

During training, the entire set of $N_j$ data points is processed in portions of $N_m$, with each epoch consisting of $N_j/N_m$ gradient updates. The optimization continues for a specified number of epochs $N_\mathrm{epoch}$ until a stopping criterion, such as a target prediction accuracy, is achieved.

Fitting a neural network model and selecting the optimal architecture for a given data set is complex. Numerous parameters, such as layer width, activation function, and learning rate, must be fine-tuned for optimal performance. These parameters are known as hyperparameters, distinguishing them from model parameters, which are adjusted during training. A detailed discussion of hyperparameter optimization in the context of this work can be found in a previous publication~\cite{hyperparameterpaper} and some details will be discussed in the following section on the numerical implementation of the \texttt{MALA} package.

\section{Numerical implementation}
\label{s.implementation}

\subsection{Code structure}
\label{ss.code.structure}

Developing a broadly applicable machine learning workflow for DFT calculations is challenging, as it requires coordinating a variety of tasks, including data sampling, data preparation, and model training. Furthermore, the resulting code must accommodate a diverse range of user requirements. For example, materials scientists may primarily focus on applying pre-trained models, while researchers who train these models need fine-grained control over the training processes.

\begin{figure}[htp]
    \centering
    \includegraphics[width=1.2\textwidth]{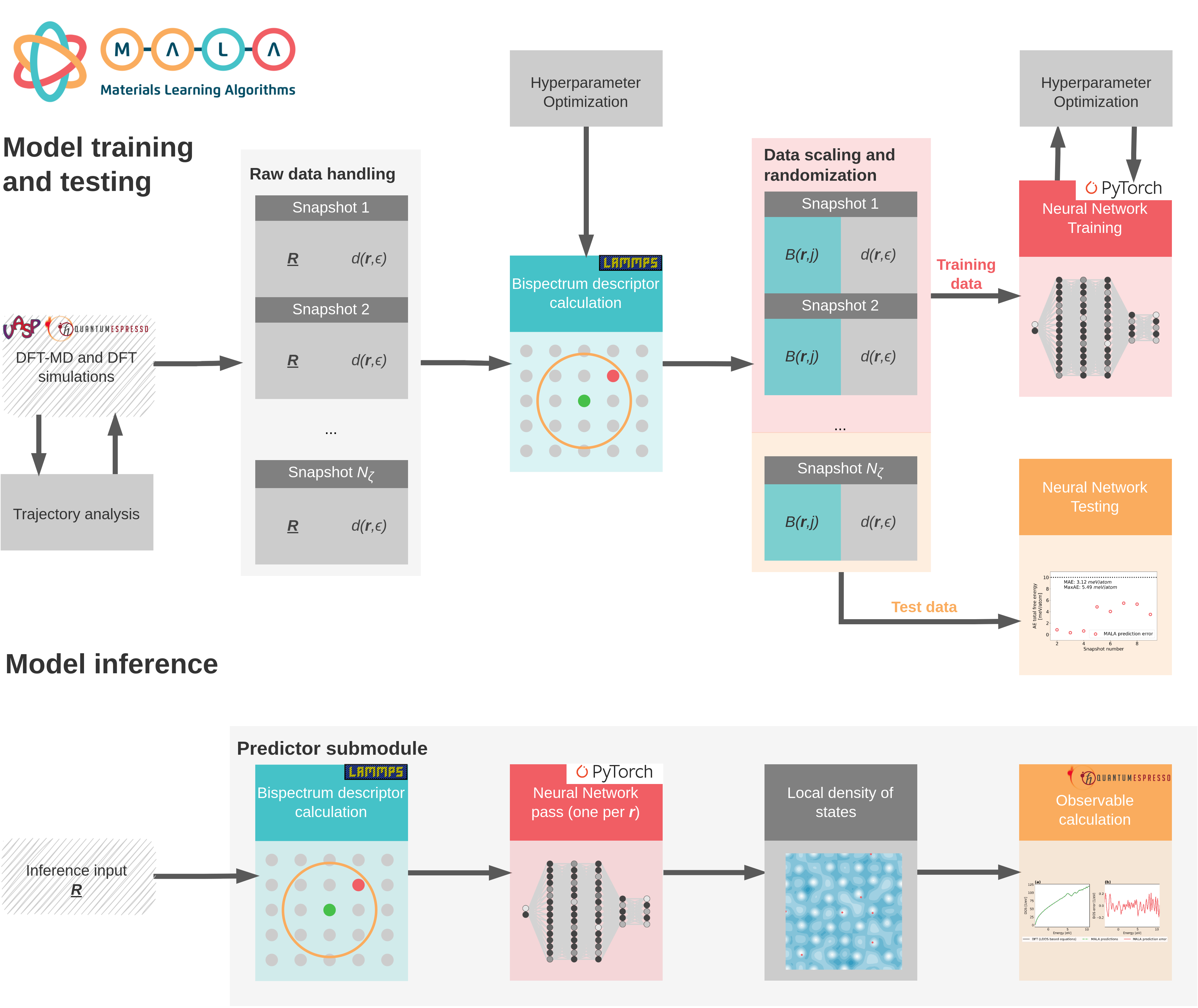}
    \caption{Overview of \texttt{MALA} workflow. Both model training procedure and inference are shown. For the latter, relevant routines are bundled in a separate class within the \texttt{MALA} package to facilitate access. Gray boxes denote pure \texttt{Python}-based routines, while blue, red and orange boxes denote that the majority of the computational workload is offloaded to the external libraries \texttt{LAMMPS}~\cite{LAMMPS,LAMMPS2}, \texttt{PyTorch}~\cite{pytorch} and \texttt{Quantum ESPRESSO}~\cite{giannozzi_quantum_2009,giannozzi_advanced_2017,giannozzi_q_2020}, respectively. Data generation is typically performed with the \texttt{VASP}~\cite{kresse_ab_1993,kresse_efficiency_1996,kresse_efficient_1996} and \texttt{Quantum ESPRESSO} simulation codes, in conjunction with a \texttt{MALA} specific DFT-MD trajectory analysis. \texttt{MALA} internal routines are implemented with the aid of NumPy~\cite{Numpy} and SciPy~\cite{Scipy}.}
    \label{fig.general.code.structure}
\end{figure}

To address these needs, the \texttt{MALA} package has been designed with a modular architecture. Using a central collection of control parameters, users can assemble custom workflows tailored to their specific goals. These workflows are composed of interchangeable building blocks that delegate core computational tasks to external or internal \texttt{Python}-based libraries. Fig.~\ref{fig.general.code.structure} illustrates the range of tasks a workflow might include, as well as the integration of internal and external libraries.

As shown in Fig.~\ref{fig.general.code.structure}, the \texttt{MALA} code covers a range of computational tasks implemented either directly in \texttt{Python} or using interfaces to the \texttt{LAMMPS}~\cite{LAMMPS,LAMMPS2}, \texttt{PyTorch}~\cite{pytorch} and \texttt{Quantum ESPRESSO}~\cite{giannozzi_quantum_2009,giannozzi_advanced_2017,giannozzi_q_2020} libraries. The following sections detail these components of the machine learning workflow based on the LDOS: Sec.~\ref{ss.data.sampling} describes the acquisition of suitable training data, Sec.~\ref{ss.hyperparameters} focuses on model training, and Sec.~\ref{ss.observables} explains the computation of observables from the model predictions. Technical aspects of the workflow, including parallelization and hardware acceleration, are discussed in Sec.~\ref{ss.parallelization}, while Sec.~\ref{ss.data.management} covers data management for volumetric datasets.

\subsection{Data sampling}
\label{ss.data.sampling}

Sampling training data is a crucial task for any machine learning model, as the model's accuracy is inherently constrained by the quality of its training data. Within the \texttt{MALA} workflow, curating datasets involves three main challenges. First, representative ionic configurations must be identified for electronic structure simulations. Second, these configurations’ ionic structures must be accurately represented on a suitable numerical grid. Finally, training data must be correctly extracted from DFT simulations.

\subsubsection{Sampling of ionic configurations} 
\label{ss.data.sampling.ionic}

The first task is addressed by coupling DFT to molecular dynamics simulations (DFT-MD simulations) within the Born-Oppenheimer approximation. DFT-MD simulations are used to model ionic dynamics by decoupling the ionic and electronic problems. DFT calculations are used to predict the electronic structure of an ionic configuration. This prediction involves calculating the forces acting on the ions by solving the classical equations of motion to propagate the ions over a small time interval. This process is then repeated for the resulting new set of ionic coordinates. DFT-MD simulations allow the simulation of a wide range of thermodynamic observables and are thus the primary tool for computational materials scientists to gain insight into materials under relevant conditions~\cite{MDReviewPaper}. 

In the context of \texttt{MALA} models, DFT-MD simulations are employed to generate sets of ionic configuration on which DFT and LDOS calculations are carried out. To select suitable configurations, \texttt{MALA} firstly performs an equilibration analysis as described in Ref.~\citenum{ofdftpaper,Lenz_Dissertation}. For this purpose, \texttt{MALA} computes the radial distribution functions (RDF)~\cite{rdf1,rdf2,rdf3} of an ionic configuration as
\begin{equation}
g(r)=\frac{\Xi(r)}{\rho\ion N\ion V_\mathrm{shell}} \; , \label{eq:RDF1}
\end{equation}
with the ionic density $\rho\ion$ and average number of ions $\Xi(r)$ in a shell of volume $V_\mathrm{shell}$ and radius $r+dr$. From the equilibrated portion at the end of a DFT-MD trajectory, a reasonable equilibration threshold can be determined.

This approach allows for the exclusion of unequilibrated parts of the DFT-MD trajectory, yielding a substantial set of ionic configurations. These configurations are then further refined using real-space distance metrics, as outlined in Ref.~\citenum{Lenz_Dissertation}. Specifically, the minimal Euclidean distance between any two ions in pairs of ionic configurations within a trajectory is used, given as
\begin{align}
D_{c,c'}^\mathrm{min-euc} = \min \big[D^\mathrm{euc}(\bR_{c,1},\bR_{c',1}),\, &D^\mathrm{euc}(\bR_{c,2},\bR_{c',2}), ..., \nonumber \\ & D^\mathrm{euc}(\bR_{c,{N\ion}},\bR_{c',{N\ion}})  \big] \label{eq:minimumeucdistance}
\end{align}
where the indices $c$ and $c'$ denote different time steps within the sampled trajectory and $D^\mathrm{euc}$ the Euclidean distance. By requiring this minimal Euclidean distance to fall outside a 99\% confidence interval for two uncorrelated snapshots, a threshold for snapshot sampling can be established. Based on this threshold and starting from the first equilibrated configuration, the trajectory is sampled until a sufficiently different configuration has been identified. This configuration is then selected for further processing, set as new reference, and the process is repeated. This technique is described in detail in Ref.~\citenum{Lenz_Dissertation} and has been implemented in the \texttt{MALA} package.

This set of ionic configurations forms the basis of a training dataset. For each ionic configuration, volumetric data representing both ionic and electronic structure (particularly the LDOS) must be computed. In the following, the complete set of metadata, along with the volumetric ionic and electronic structure data for each ionic configuration, will be referred to as a snapshot.

To represent the ionic structure of a snapshot on a computational grid, various approaches can be applied, typically following considerations for computing descriptors of entire ionic configurations. Two prominent examples in this regard are the Spectral Neighborhood Analysis Potential (SNAP)~\cite{thompson_spectral_2015,wood_extending_2018,wood_data-driven_2019,cusentino_explicit_2020} and the Atomic Cluster Expansion (ACE)~\cite{lysogorskiy_performant_2021,drautz_atomic_2019} descriptors, both of which were adapted to operate on a computational real-space grid. A general overview of descriptors is provided in Ref.~\citenum{deepdive}.

\subsubsection{Bispectrum descriptors} 
\begin{figure}[htp]
    \centering
    \includegraphics[width=0.95\columnwidth]{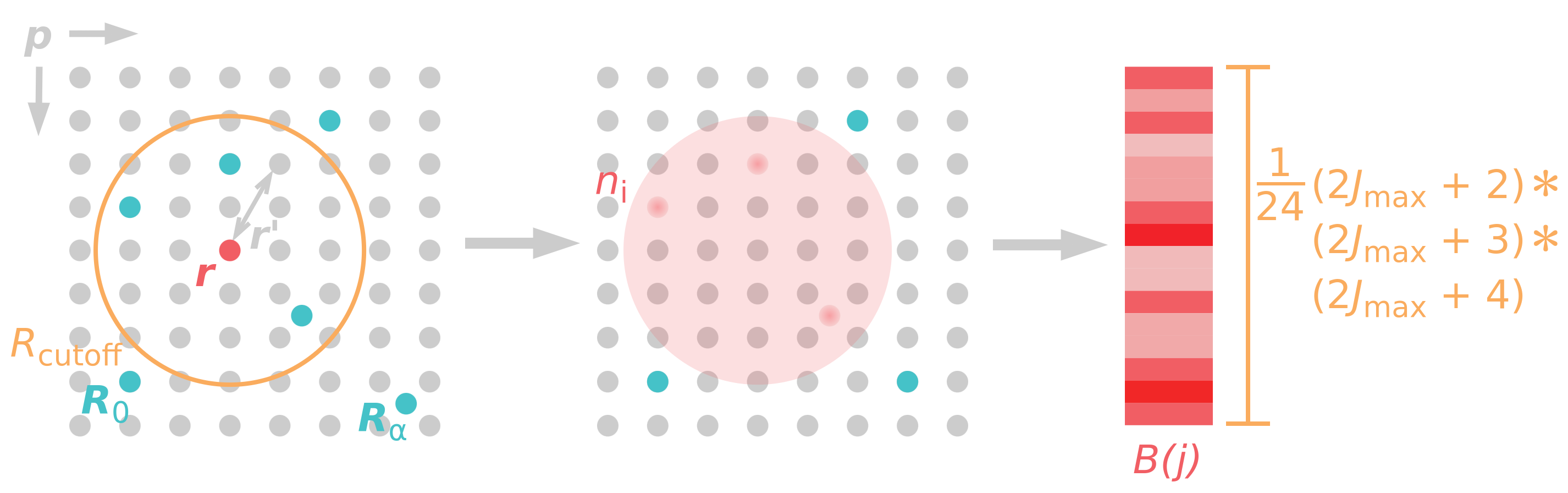}
    \caption{Calculation of bispectrum descriptors, representing the local ionic configuration within the neighborhood surrounding a grid point in real space. Figure adapted from Ref.~\citenum{Lenz_Dissertation}.}
    \label{fig:bisepctrum.descriptors}
\end{figure}

In its original form, the \texttt{MALA} package was developed using bispectrum descriptors based on the SNAP formalism~\cite{malapaper}. This approach begins by computing an ionic density $n\ion$ around each point in space $\br$,
\begin{equation}
	n_{\mathrm{i}}(\br') = \delta(\bm{0})+\mathlarger{\mathlarger{\sum}}_{\alpha=1}^{|\bR_{\alpha}-\br'|<R^{\eta_\alpha}_\mathrm{cutoff}} \Big[ f_\mathrm{c}\left(|\bR_{\alpha}-\br'|, R^{\eta_\alpha}_\mathrm{cutoff}\right) w_{\eta_\alpha}\delta(\bR_{\alpha}-\br')  \Big]\; ,\label{eq:ionicdensity}
\end{equation}
where $\br'=\bm{p}-\br$ and $\bm{p}$ denotes a second, auxiliary numerical grid for each point in space $\br$. This auxiliary grid is necessary to compute the spatially resolved ionic density $n_{\mathrm{i}}$, as shown in Fig.~\ref{fig:bisepctrum.descriptors}. $ R^{\eta_\alpha}_\mathrm{cutoff}$ is a cutoff radius around $\br$, while $w_{\eta_\alpha}$ is a dimensionless element weight, and the cutoff function $f_\mathrm{c}\left(R, R_\mathrm{cutoff}\right)$ ensures that the descriptor values go smoothly to zero at $R_\mathrm{cutoff}$. The chemical species index $\eta$ will be discarded in the following for the sake of brevity. 

The specific method for calculating $n\ion(\br')$ varies across different descriptors~\cite{Descriptor_Characterization}. For instance, a Gaussian function may be used instead of $\delta$-distributions. To construct descriptors from $n\ion(\br')$, further processing is applied to ensure both computational feasibility and conservation of relevant invariances. In the case of bispectrum descriptors, vectors $B(\br, j)$ are computed for each $\br$ by expressing the ionic density
\begin{equation}
	n_{\mathrm{i}}(\br') = \sum_{j=0, \frac{1}{2}, ...}^{\infty} \bmmatrix{u}_j \cdot \bmmatrix{U}_j (\theta_0, \theta, \phi) \label{eq:SNAPExpansion}
\end{equation}
as a sum of hyperspherical harmonic functions $\bmmatrix{U}_j$ with degree $j$ and expansion coefficients $\bmmatrix{u}_j$.  This expression implicitly defines the expansion coefficients $\bmmatrix{u}_j$, which are computed in practice using the orthogonality properties of $\bmmatrix{U}_j$.

The elements of bispectrum vectors $B(\br, j)$ are then computed using scalar triple products of the expansion coefficients from Eq.~\eqref{eq:SNAPExpansion}. The resulting bispectrum descriptors are real-valued, rotationally invariant, and have the dimensionality
\begin{equation}
	B_\mathrm{max} = \frac{1}{24}(2J_\mathrm{max}+2)(2J_\mathrm{max}+3)(2J_\mathrm{max}+4) \; ,
    \label{eq:snap_dim}
\end{equation}
where $J_\mathrm{max}$ denotes the order of the basis set expansion. This process is further illustrated in Fig.~\ref{fig:bisepctrum.descriptors}. The fidelity of this encoding depends on selecting appropriate values for $J_\mathrm{max}$ and $R_\mathrm{cutoff}$. These serve as hyperparameters of the \texttt{MALA} model and can be optimized using hyperparameter optimization methods or, more efficiently, by leveraging the physical principle underlying their derivation~\cite{hyperparameterpaper}.
%\TODO{If we want to show ACE somewhere in more detail, it would have to be here.}

\subsubsection{Atomic cluster expansion descriptors}
Though initial implementations of \texttt{MALA} relied on the bispectrum components following the SNAP formalism, \texttt{MALA} is not restricted to grid-based SNAP descriptors. The Atomic Cluster Expansion (ACE) descriptors~\cite{drautz_atomic_2019} are a more general class of descriptors. Grid-centered ACE descriptors are actively being developed in \texttt{MALA} for more complete descriptions of ion density about grid centers. The grid-centered ACE descriptors are currently being implemented within the \texttt{MALA} framework following a similar formalism used for SNAP bispectrum components. This includes an efficient parallel implementation in \texttt{LAMMPS}.

For the ACE descriptors, we begin with an unweighted ion density.
\begin{equation}
	n\ion^\circ(\br') = \delta(\bm{0})+\mathlarger{\mathlarger{\sum}}_{\alpha=1}^{|\bR_{\alpha}-\br'|<R^{\eta_\alpha}_\mathrm{cutoff}} \delta(\bR_{\alpha}-\br')\; . 
    \label{eq:unweighted_ionicdensity}
\end{equation}
The unweighted ionic density in Eq.~\eqref{eq:unweighted_ionicdensity} removes the cutoff function and element weights from the ionic density in Eq.~\eqref{eq:ionicdensity}. While this removes the element weights that scale the contributions of different ion types to the ionic density, it allows one to treat the element types explicitly with a grid-atom basis.
The grid-atom basis is comprised of a product of bases for the chemical, radial, and angular degrees of freedom as
\begin{equation}
    \phi _{\mu_\circ \mu nlm }  (\boldsymbol{R}_{\alpha}-\br') = e(\mu_\circ,\mu ) \mathcal{R}_{n}(|\boldsymbol{R}_{\alpha}-\br'|) Y_l^m(\boldsymbol{R}_{\alpha}-\br'),
    \label{eq:gridatom_basis}
\end{equation}
where $e(\mu_\circ,\mu )$ is a delta function chemical basis, similar to that in Ref.~\citenum{drautz_atomic_2019}, $\mathcal{R}_{n}$ is a Chebyshev radial basis with index $n$ that only depends on the scalar distance, and the angular basis is comprised of spherical harmonic functions, $Y_l^m$, where the indices $l$ and $m$ obey $l\ge0$ and $-l\le m \le l$. The spherical harmonics depend only on the direction of the interaction, $(\boldsymbol{R}_{\alpha}-\br')/|\boldsymbol{R}_{\alpha}-\br'|$.
The second chemical function index, $\mu$, may take on any value between one and the number of chemical elements. In \texttt{MALA} the chemical basis is used to restrict the models to grid-atom interactions. This is achieved by constraining the first chemical function index to that assigned for grid points, $e(\mu_\circ,\mu )=e(\bm{p},\mu )$.

The grid-atom basis in Eq.~\eqref{eq:gridatom_basis} is not yet permutation- or rotation-invariant and describes only one-body character. 
A complete set of $N$-body functions could be obtained by taking tensor-products of Eq.~\eqref{eq:gridatom_basis}, but this would result in exponential scaling of the descriptors with body order. Instead, the permutation invariance of this grid-atom basis may be enforced while ensuring computational scalability. 
This is done by projecting the unweighted ionic density onto the basis in Eq.~\eqref{eq:gridatom_basis} to yield the grid basis
\begin{equation}
    A_{\mu nlm }(\boldsymbol{R}_{\alpha}-\br') = \langle n\ion^\circ | \phi_{\mu nlm} \rangle\; ,
    \label{eq:grid_base}
\end{equation}
which is invariant with respect to permutations of ions of the same chemical type. 
Products of grid functions in Eq.~\eqref{eq:grid_base},
\begin{equation}
    A_{\boldsymbol{\mu nlm}} (\boldsymbol{R}_{\alpha}-\br') = \prod_\kappa^N A_{\mu_\kappa n_\kappa l_\kappa m_\kappa}  (\boldsymbol{R}_{\alpha}-\br') \; ,
    \label{eq:product_grid_basis}
\end{equation}
are used to obtain permutation-invariant $N$-body functions that scale linearly with the number of ions. For brevity, index multisets of length $n$ are defined for the chemical indices $\boldsymbol{\mu}=\{ \mu_1 \mu_2 \ldots \mu_N \}$, the radial function indices $\boldsymbol{n}=\{ n_1 n_2 \ldots n_N\}$, the angular momentum indices $\boldsymbol{l}=\{ l_1 l_2 \ldots l_N\}$, and the projection indices $\boldsymbol{m}=\{ m_1 m_2 \ldots m_N\}$.

Rotational invariance, or equivariance, of the $N$-body product functions in Eq.~\eqref{eq:product_grid_basis} is achieved by contracting with generalized Clebsch-Gordan coefficients, $C_{\boldsymbol{m}}^{\boldsymbol{l}}(\boldsymbol{L})$, as
\begin{equation}
    B_{\boldsymbol{\mu nlL}} = \sum_{\boldsymbol{m}}C_{\boldsymbol{m}}^{\boldsymbol{l}}(\boldsymbol{L})A_{\boldsymbol{\mu nlm}} (\boldsymbol{R}_{\alpha}-\br'),
    \label{eq:ace_grid}
\end{equation}
where the sum in Eq.~\eqref{eq:ace_grid} runs over all the possible combinations of $\boldsymbol{m}$ to form the rotation and permutation invariant ACE descriptors, $B_{\boldsymbol{\mu nlL}}$, and the additional index multiset, $\boldsymbol{L} = \{L_1 L_2 \ldots L_{N-1}\}$ specifies the intermediate couplings in the generalized Clebsch-Gordan coefficients. 
In general, multiple $\boldsymbol{L}$ multisets are allowed by angular momentum coupling conditions, but not all of them result in independent ACE descriptors~\cite{dusson2022atomic}.
Relationships derived from quantum angular momentum raising and lowering operations are used to ensure that ACE descriptors with $N\ge$4 are not redundant~\cite{goff_permutation-adapted_2024}.

The dimensionality of ACE descriptor sets depends on the body-order of the product functions and other parameters. More information on ACE descriptor counts and enumeration can be found elsewhere~\cite{drautz_atomic_2019,dusson2022atomic,goff_permutation-adapted_2024}. In \texttt{MALA}, the set of bispectrum or ACE descriptors are controlled by one or more hyperparameters. With the bispectrum component descriptors, the maximum radial and angular character is defined solely by $2 J_{\rm{max}}$ and it is the same for all chemical elements. For ACE, the maximum radial index and the maximum angular index are defined per body-order $N$ and these can be different for different chemical elements. 
ACE offers a richer set of descriptors with tuneable resolution that is better suited to chemically complex systems.

\subsubsection{Calculation of the LDOS}

With the descriptors established, the next step is to sample the LDOS. Unlike Eq.~\eqref{eq:ldos}, which is formally exact, the practical calculation of the LDOS within a plane-wave DFT code takes the form 
\begin{equation}
d(\br, \epsilon) = \sum_{k=1}^{N_k} \sum_{j=1}^{N\ele'} w_k {|\psi_{j,\bk_k}(\br)|}^2 \tilde{\delta}(\epsilon-\epsilon_j) \label{eq:LDOS_practical} \; ,
\end{equation}
where a summation over a set of $\bk$-points in Fourier space is performed with weights $w_k$. For systems without symmetry such as the ones considered here, these weights are often set to $w_k=\frac{1}{N_{k}}$, as has been done throughout this work. The $\delta$-distribution is approximated via a Gaussian function as
\begin{equation}
    	\tilde{\delta}(x) = \frac{e^{-\frac{x^2}{w_\mathrm{d}^2}}}{\sqrt{\pi w_\mathrm{d}^2}} \; ,
\end{equation}
with width $w_\mathrm{d}$. 

\begin{figure}[htp]
    \centering
    \includegraphics[width=0.95\columnwidth]{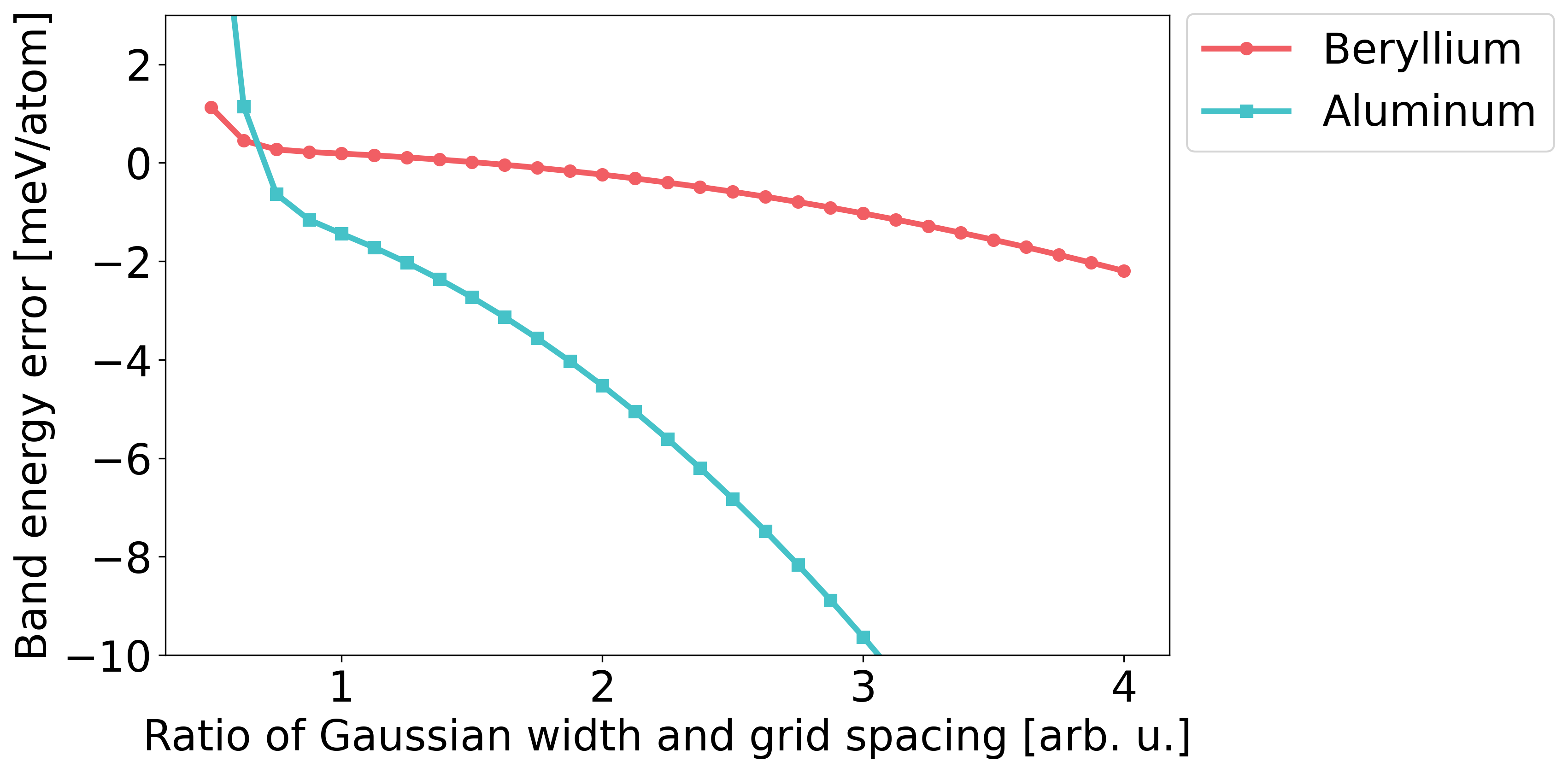}
    \caption{Dependence of band energy error on Gaussian width (expressed in units of energy grid spacing $\delta\epsilon$) for aluminum (256 atoms) and beryllium (128 atoms) at room temperature. Band energy errors are averaged over three ionic configurations per element. A range of optimal width values is evident, beyond which errors increase noticeably. Previous publications~\cite{malapaper,hyperparameterpaper,temperaturepaper,sizetransferpaper,Lenz_Dissertation} have identified $w_\mathrm{d}=2\delta\epsilon$ as a balanced choice for both aluminum and beryllium.}
    \label{fig:ldoswidth}
\end{figure}

Computing the LDOS in this way introduces two additional hyperparameters: the number $N_k$ of $\bk$-points used for summation in Fourier space and the width of the Gaussian $w_\mathrm{d}$, usually expressed in terms of the energy grid spacing on which the LDOS is generated. Control of unphysical oscillations in the LDOS is achieved by adjusting $N_k$ as discussed in Refs.~\citenum{malapaper}, \citenum{Lenz_Dissertation}, and \citenum{hyperparameterpaper}. Generally, $\bk$-grids required to sample the LDOS with sufficient accuracy are larger than those needed for typical DFT calculations. This difference arises because within DFT convergence, total free energy is usually employed as reasonable surrogate for targeted observables to monitor convergence progress. The total free energy benefits from error corrections, whereas direct LDOS sampling requires a finer $\bk$-grid for accurate numerical integration. Since both $N_k$ and $w_\mathrm{d}$ can be determined through relatively simple energy convergence routines~\cite{malapaper,Lenz_Dissertation,hyperparameterpaper}, their optimization is considered trivial in the context of hyperparameter tuning. An example is given in Fig.~\ref{fig:ldoswidth}. It should be noted, however, that due to the larger $N_k$ values, DFT data generation for \texttt{MALA} models can be more computationally intensive than standard DFT calculations. 

\subsection{Hyperparameter optimization and model training}
\label{ss.hyperparameters}

At the core of the \texttt{MALA} package are fully connected feed-forward neural networks that map ionic to electronic information.
Feed-forward neural networks represent a relatively simple subclass of neural networks compared to more complex architectures such as convolutional or graph-based networks. They are well-suited to this task, given the \texttt{MALA} package's point-by-point mapping of bispectrum descriptors $B(\br, j)$ to the LDOS $d(\br, \epsilon)$. The LDOS is numerically treated as a vector at each grid point, with its dimensionality determined by the discretization of the energy $\epsilon$.

The feed-forward neural networks are implemented using the \texttt{PyTorch} library~\cite{pytorch}.
All relevant neural network hyperparameters are controlled through the \texttt{MALA} package’s central parameter collection. For setting up a \texttt{MALA} model, users define a neural network architecture (i.e., width, depth, and activation functions) along with essential training hyperparameters, including learning rate (which controls the step size for gradient updates), minibatch size (the portion of training data processed per step), and the weight optimization algorithm (e.g., the ADAM optimizer~\cite{kingma_adam_2017}). \texttt{MALA} provides sensible defaults for most hyperparameters, facilitating efficient model training when used with published models~\cite{malapaper,hyperparameterpaper}.

Determining an optimal set of hyperparameters can be a challenging and computationally demanding task~\cite{HyperOptCitation}. Since the \texttt{MALA} package is aimed at researchers in materials science, quantum materials, and quantum chemistry, efficient and broadly applicable routines for hyperparameter optimization are essential. In addition to neural network hyperparameters, other parameters to be optimized include those for bispectrum descriptor calculation and LDOS sampling.

For LDOS sampling, as discussed in Sec.~\ref{ss.data.sampling}, a straightforward energy error analysis during data generation suffices, and \texttt{MALA} provides the necessary tools for this. For bispectrum descriptor calculation, hyperparameters may be treated similarly to traditional neural network hyperparameters, with optimization carried out as described below. However, because bispectrum descriptors represent the local ionic environment, their relationship to local electronic structure can be exploited to improve hyperparameter evaluation efficiency.

To this end, Ref.~\citenum{hyperparameterpaper} introduces a method that compares similarity metrics between bispectrum descriptors at different points with similarity metrics of LDOS vectors at corresponding points in space. This comparison yields a distribution of similarity metrics that can be evaluated against an idealized distribution, quantified by a scalar measure called the average cosine similarity distance. Using this approach, bispectrum hyperparameters for simple test systems have been identified with significantly reduced computational cost~\cite{hyperparameterpaper}.

As an alternative we also investigate an approach based on information theory, leveraging the mutual information score to select the optimal descriptor type and its hyperparameters. Mutual information is defined as
\begin{equation}
    I(X, Y) = \int_Ydy \int_Xdx\text{ } p_{XY}(x, y) \log_2\left(\frac{p_{XY}(x, y)}{p_X(x)p_Y(y)}\right)
\end{equation}
and reflects the amount of information (in the unit of bits) present in random variable $X$ useful for the prediction of random variable $Y$. Taking $X$ and $Y$ to be the descriptor and LDOS vectors respectively, $I(X,Y)$ yields a theoretically grounded measure of a descriptor's usefulness for the task of predicting the LDOS. An efficient method of estimating mutual information for the case of high-dimensional continuous random variables has been implemented and will be further discussed in upcoming work.

This leaves the optimization of neural network hyperparameters, also addressed in Ref.~\citenum{hyperparameterpaper}. In machine learning, this process often involves evaluating multiple hyperparameter sets (called trials) and training a neural network model for each trial. The associated prediction errors are then analyzed to identify the optimal hyperparameter sets. The methods used to select and evaluate trials vary across hyperparameter optimization approaches.

Ref.~\citenum{hyperparameterpaper} investigates a range of traditional and advanced hyperparameter optimization techniques, including direct search~\cite{DirectSearch1,DirectSearch2}, the orthogonal array tuning method~\cite{OATM} (rooted in experimental design), the tree-structured Parzen estimator~\cite{TPE} (as implemented in the \texttt{Optuna} library~\cite{optuna}), and a training-free method known as NASWOT (neural architecture search without training)~\cite{NASWOT}. 

\begin{figure}[htp]
\centering
\includegraphics[width=0.95\columnwidth]{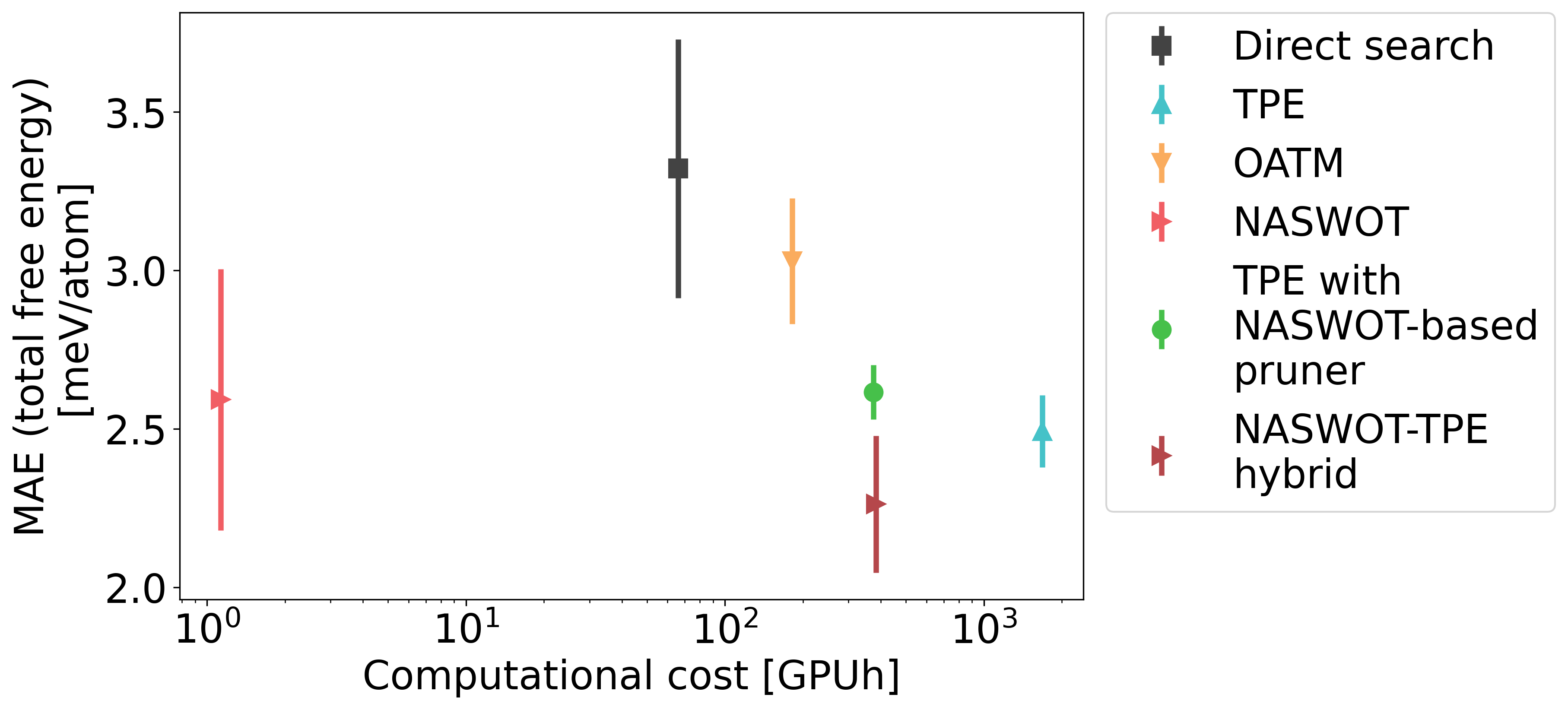}
\caption{Comparison of different hyperparameter optimization methods for tuning \texttt{MALA} models. The MAE of total free energy prediction error across the test set is shown relative to the computational cost of determining model hyperparameters. An aluminum system at room temperature, containing 10 configurations with 256 atoms each, was modeled. For each set of hyperparameters, five neural networks were trained. Error bars denote the standard deviation across these five network initializations, while solid symbols indicate the average MAE. Results are presented for direct search, the tree-structured Parzen estimator (TPE), the orthogonal array tuning method (OATM), and neural architecture search without training (NASWOT). Also shown are two combinations of TPE and NASWOT: using NASWOT to prune unpromising trials during TPE optimization (NASWOT-based pruner) and using TPE to optimize training hyperparameters for which NASWOT provides no insights. Details on hyperparameter optimization methods are available in Ref.~\citenum{hyperparameterpaper}.}
\label{fig:hyperopt}
\end{figure}

The main findings of this analysis are presented in Fig.~\ref{fig:hyperopt}. Prediction errors for models trained using hyperparameters obtained through these optimization schemes are shown. Generally, the accuracy of predictions improves with the computational cost invested in the optimization process. While this trend may seem intuitive, it serves as valuable guidance for researchers selecting a hyperparameter optimization technique for their specific problem. Automated and parallelized routines for the tree-structured Parzen estimator, orthogonal array tuning method, and NASWOT are implemented within \texttt{MALA}.

The performance of the NASWOT technique is particularly noteworthy. This method operates on the assumption that optimal neural networks can be identified without requiring full training. By computing a surrogate metric that estimates model performance after training, NASWOT enables the selection of promising network architectures without the costly need to train a candidate neural network model for each hyperparameter combination. As shown in Fig.~\ref{fig:hyperopt}, this approach sacrifices some accuracy but significantly reduces computational costs. By offering a range of training-free and training-based hyperparameter optimization methods, the \texttt{MALA} package allows users to develop suitable machine learning models for new materials of interest.

\subsection{Observable calculation}
\label{ss.observables}

The principal output of the \texttt{MALA} package is the LDOS. While the equations in Sec.~\ref{s.background} define how observables of interest can, in principle, be evaluated from the LDOS, their numerical implementation requires additional considerations. The observables accessible through \texttt{MALA} include the system’s total energy, total free energy at a given temperature of the electronic system, electronic density, and electronic DOS. The overall processing workflow starts with the LDOS and one first calculates the DOS and density, then the band energy and entropy contribution from the DOS, and finally derives the remaining components of the total free energy from the electronic density.

Computing the DOS from the LDOS is straightforward from a numerical perspective, as the real-space integration can be efficiently handled through appropriate summation. However, further processing of the DOS, as well as calculating the density from the LDOS, requires integration over the energy domain. Specifically, this involves solving integrals of the form
\begin{equation}
    I = \int d\epsilon \; i(\epsilon) \mathcal{D} \; , \label{eq:ldosintegralgeneral}
\end{equation}
where $\mathcal{D}$ can either be the LDOS $d(\br, \epsilon)$, yielding the density in Eq.~\eqref{eq:density.from.ldos} or the DOS $D(\epsilon)$, yielding the band energy in Eq.~\eqref{eq:DOS.from.ldos}, and the entropy contribution in Eq.~\eqref{eq:electronicentropyDOS}, given the respective weights $i(\epsilon)$. Since the weights $i(\epsilon)$ are functions of the Fermi-Dirac distribution $f_\beta(\epsilon)$, which changes rapidly close to the Fermi energy, the numerical integration in Eq.~\eqref{eq:ldosintegralgeneral} is challenging. 

\texttt{MALA} therefore implements an analytical evaluation of these integrals by re-expressing Eq.~\eqref{eq:ldosintegralgeneral} as
\begin{equation}
    I=\sum_{\xi=1}^{N_\epsilon} w_\xi \mathcal{D}(\epsilon_\xi) \; . \label{eq:ldosintegralsummation}
\end{equation}
By expressing the Fermi-Dirac distribution in terms of the polylogarithm $\mathrm{Li}_n$ as
\begin{equation}
	f_\beta(\epsilon-\epsilon_\mathrm{F}) = 1 + \mathrm{Li}_0(-e^{(\epsilon-\epsilon_\mathrm{F})}) \; , \label{eq:fermidiracspecial}
\end{equation} 
appropriate weights $w_\xi$ can be identified. The resulting framework enables numerically accurate and efficient evaluation of all quantities expressed as integrals via Eq.~\eqref{eq:ldosintegralgeneral}. Naturally, evaluating Eq.~\eqref{eq:fermidiracspecial} also requires knowledge of the Fermi energy $\epsilon_\mathrm{F}$. 

To this end, an automatic determination of $\epsilon_\mathrm{F}$ is implemented within \texttt{MALA}. This is possible since the number of electrons $N\ele$ is obtained by integrating the DOS following Eq.~\eqref{eq:ldosintegralgeneral} as 
\begin{equation}
	N\ele = \int d \epsilon \; f_\beta(\epsilon) D(\epsilon) \; . \label{eq:NumberElectrons_from_DOS}
\end{equation}
If this equation is evaluated using a machine learning prediction of the DOS, $\tilde{D}$, obtained through real-space integration of the predicted LDOS, then $f_\beta$ can be determined as the root of 
\begin{equation}
	f_\mathrm{o}(\epsilon_\mathrm{F}) =  N\ele  - \int d \epsilon \; \frac{\tilde{D}(\epsilon)}{1+ \exp{\left[\beta(\epsilon-\epsilon_\mathrm{F})\right]}}   \; .
\end{equation} 
Numerically, this root is found with a number of root-finding algorithms. Within \texttt{MALA}, the \texttt{TOMS748}~\cite{TOMS748} algorithm is employed. Once the Fermi energy is calculated, band energy $E_\mathrm{b}$, entropy contribution $S\s$ and electronic density $n$ are computed from DOS and LDOS. 

When evaluating the total energy as defined in Eq.~\eqref{eq:energy.ldos}, the calculation of the Hartree energy $E\Hartree$, the XC energy $E\xc$, the potential energy contribution to the XC energy $V\xc$, and the ion-ion interaction energy $E\ionion$ needs to be implemented. In principle, all of these quantities can be calculated through an interface with a standard DFT code, as they are components of standard DFT calculations. Note that at finite electronic temperature the total free energy expression in Eq.~\eqref{eq:KSfreeenergy} is evaluated instead.

For $E\Hartree$, this calculation is straightforward and has been implemented in the \texttt{MALA} package via an interface to the \texttt{Quantum ESPRESSO} code. 

For the ion-ion interaction $E\ionion$, the XC energy $E\xc$, and $V\xc$, straightforward evaluation is also possible but not advisable. Standard DFT codes are not designed to handle very large systems, and the quadratic scaling of standard implementations for both $E\ionion$ and the non-linear core corrections in the XC energy would negatively impact \texttt{MALA}'s scalability as the number of atoms grows. To address this bottleneck, \texttt{MALA} employs routines based on a Gaussian representation of $\ubR$, as introduced in Ref.~\citenum{sizetransferpaper}. 

Specifically, both the non-linear core corrections and the long-range interaction component of $E\ionion$~\cite{EwaldHimself,EwaldSummation} can be expressed as functions of the form 
\begin{equation}
	f_\mathrm{S}(\br)=\sum_{\alpha}^{N\ion}f(\br-\bR_\alpha) \; . \label{eq:structfact_function}
\end{equation}
The corresponding Fourier transformation yields
\begin{equation}
	\tilde{f}_\mathrm{S}(\bm{G})=\tilde{S}(\bm{G})\tilde{f}(\bm{G}) \; ,
\end{equation}
with the Fourier transform of the structure factor 
\begin{equation}
	\tilde{S}(\bm{G}) = \sum_{\alpha=1}^{N\ion} \exp{(i\bm{G}\cdot\bR_\alpha)} \; , \label{eq:structurefactor}
\end{equation}
and the Fourier transform $\tilde{f}_\mathrm{S}(\bm{G})$ of $f_\mathrm{S}(\br)$. Efficient calculation of $f_\mathrm{S}(\br)$ is thus achievable through the efficient calculation of $\tilde{S}(\bm{G})$, which is implemented in \texttt{MALA} as 
\begin{equation}
	\tilde{S}(\bm{G})  = \frac{\tilde{G}(\bm{G})}{\tilde{G}_0(\bm{G})} \label{eq:structurefactor_gaussian} \; , 
\end{equation}
where $\tilde{G}$ is the Fourier transform of a Gaussian representation of $\ubR$, i.e., a sum of Gaussians centered at $\ubR$, and $\tilde{G}_0$ is the Fourier transform of a Gaussian function of a single reference atom centered at the origin of the simulation cell. The computation of this Gaussian representation, referred to as Gaussian descriptors hereafter, is offloaded to \texttt{LAMMPS}. This approach enables straightforward parallelization and hardware acceleration, as described in Sec.~\ref{ss.parallelization}. Fourier and inverse Fourier transformations are performed via \texttt{Quantum ESPRESSO}.

By moving from direct evaluation of $E\ionion$ and the XC contributions to this Gaussian representation scheme, linear scaling across relevant length scales is achieved, as shown in Sec.~\ref{ss.scaling}. Note that this introduces an additional hyperparameter into the workflow, namely the width of the Gaussians $w_\mathrm{G}$. As detailed in Ref.~\citenum{Lenz_Dissertation}, this hyperparameter can be determined automatically by setting a reference value for a known numerical grid and adjusting it based on the grid spacing of the inference calculation. This method minimizes aliasing errors and ensures overall accuracy. 

Finally, it is worth noting that the computation of the total (free) energy also enables the calculation of atomic forces. Specifically, the force acting on the $\alpha$-th ion is given by 
\begin{equation}
	\bm{f}_\alpha = -\frac{\partial E(\ubR)}{\partial \bR_\alpha} \; , \label{eq:forces}
\end{equation}
where we have explicitly denoted the parametric dependence of the total energy on the positions of the ions $\ubR$.
In the context of the \texttt{MALA} workflow, one employs the chain rule to express this relation as
\begin{equation}
\bm{f}_\alpha = -\frac{\partial E(\ubR)}{\partial d(\br, \epsilon; B)}\frac{\partial d(\br, \epsilon; B)}{\partial B(\br,j;\ubR)}\frac{\partial B(\br,j;\ubR)}{\partial \bR_\alpha} \; . \label{eq:forcesmala}
\end{equation}
It would be computationally inefficient to calculate the two terms on the right, which would have the form of a matrix at each grid point, and then multiply the terms afterward.  Instead, one can first evaluate the first term $\partial E(\ubR)/\partial d(\br, \epsilon; B)$, which is a vector at each grid point, and then calculate the result of applying the second term $\partial d(\br, \epsilon; B)/\partial B(\br,j;\ubR)$ to this vector followed by the result of applying the third term $\partial B(\br,j;\ubR)/\partial \bR_\alpha$ to the result.  The application of $\partial d(\br, \epsilon; B)/\partial B(\br,j;\ubR)$ to $\partial E(\ubR)/\partial d(\br, \epsilon; B)$ is straightforward to compute, as it involves only neural network backpropagation. The numerical implementation of $\partial E(\ubR)/\partial d(\br, \epsilon; B)$ involves standard quantities calculated in DFT with the exception of a term involving the derivative of the XC potential with respect to the electronic density, i.e., the XC kernel.  Similar quantities appear in density functional perturbation theory, and algorithms to evaluate this type of term efficiently are known and have been implemented in \texttt{Quantum ESPRESSO}. Custom routines for accessing the required term in \texttt{Quantum ESPRESSO} are currently developed. Likewise, we have implemented routines in \texttt{LAMMPS} to apply $\partial B(\br,j;\ubR)/\partial \bR_\alpha$ to an array passed to it from \texttt{MALA}. An early version of this forces interface is currently developed and tested in \texttt{MALA} and will be available in the next \texttt{MALA} release. 

\subsection{Parallelization and hardware acceleration}
\label{ss.parallelization}

Due to the variety of computational tasks involved in the \texttt{MALA} package, a combination of multiple acceleration techniques is necessary to ensure optimal performance. Broadly, two main acceleration strategies can be adopted. First, tasks can be parallelized across multiple computational nodes using a spatial decomposition that communicate via network interfaces, such as through the MPI library~\cite{MPI}. Second, hardware accelerators like graphical processing units (GPUs) attached to computational nodes can be utilized to boost performance, achievable through programming libraries such as CUDA~\cite{CUDA}.

Traditionally, electronic structure theory codes have been accelerated through parallelization across multiple central processing units (CPUs). However, the rising computational power of GPUs has sparked growing interest in GPU-accelerated electronic structure codes, leading to GPU implementations in software such as \texttt{VASP}~\cite{VASPGPU1, VASPGPU2} and \texttt{Quantum ESPRESSO}~\cite{QEGPU}. Likewise, since neural networks are highly efficient on GPUs~\cite{NNs_on_GPU}, GPU-based acceleration has become the standard for accelerating neural network training and inference.

The core strategy for optimizing \texttt{MALA}’s performance, therefore, is to offload as many computations as possible onto GPUs, followed by parallelization across multiple GPU-equipped computational nodes. In this setup, even workflow components that do not directly benefit from hardware acceleration can still leverage CPU parallelization, as each node is necessarily equipped with a CPU.

To illustrate this, Tab.~\ref{tab:gpu_cpu_parallelization} provides an overview of typical compute tasks encountered in the \texttt{MALA} workflow, along with their suitability for different parallelization strategies. Additionally, it presents an estimate of computational cost for a typical target system, based on the results in Sec.~\ref{ss.scaling}, to further justify the chosen implementation strategies.

\renewcommand{\arraystretch}{1.5}

\begin{table}
\centering  
\caption{Overview of parallelization strategies implemented for typical tasks within the \texttt{MALA} package. Computational cost estimates are based on results published in Refs.~\citenum{malapaper}, \citenum{hyperparameterpaper}, \citenum{temperaturepaper}, \citenum{sizetransferpaper}, and \citenum{Lenz_Dissertation}. Numerical benchmarks are discussed in detail in Sec.~\ref{ss.scaling}. Program libraries used for acceleration and/or parallelization are indicated in parentheses.}\label{tab:gpu_cpu_parallelization}		
	% \begin{tabularx}{\textwidth}{XXXX}%cccccccc}
    	\begin{tabularx}{\textwidth}{bsss}%cccccccc}

	\toprule
	\textbf{Task} &  \textbf{GPU acceleration} & \textbf{Parallel-ization} & \textbf{Computa-tional cost}	\ \\\midrule
	Data preprocessing (scaling, randomizing, ...) &  No & Yes (MPI) & low \\
	LDOS conversion &  No & Yes (MPI) & low \\
	Bispectrum descriptor calculation &  Yes (Kokkos~\cite{Kokkos1,Kokkos2,Kokkos3}) & Yes (MPI) & high \\
	Gaussian descriptor calculation &  Yes (Kokkos) & Yes (MPI) & moderate \\
	neural network forward/backward pass &  Yes (\texttt{PyTorch}) & Yes (\texttt{PyTorch DDP})~\cite{PytorchDDPTest1,PytorchDDPTest2} & high (for an entire snapshot) \\
	LDOS integration &  No & Yes (MPI) & low \\
	DOS integration &  No & Yes (MPI) & low \\
	Density integration (direct) &  No & Yes (MPI) & high (for large systems)\\
	Density integration (with Eq.~\eqref{eq:structurefactor} and ~\eqref{eq:structurefactor_gaussian} ) &  No & Partially (MPI) & moderate \\
	Hyperparameter optimization &  Yes (\texttt{PyTorch}) & Yes (MPI, SQL~\cite{optuna}) & high \ \\\bottomrule
\end{tabularx}
\end{table}

The distribution of computational costs, as shown in Tab.~\ref{tab:gpu_cpu_parallelization}, is as expected. The main computational requirements of a \texttt{MALA} model arise from its three core steps: computing a representation of the ionic structure, propagating these descriptors (and backpropagating prediction errors), and integrating the LDOS to relevant observables. 

Among these tasks, parallelizing and accelerating neural network-related operations is the most straightforward. The \texttt{PyTorch} library, extensively used within the \texttt{MALA} package, includes native GPU acceleration via the CUDA library and supports distributed (i.e., parallelized) training through \texttt{PyTorch}’s DDP formalism~\cite{PytorchDDPTest1,PytorchDDPTest2}.

Similarly, because nearly all computations for bispectrum descriptor calculations are handled by the \texttt{LAMMPS} code, its built-in parallelization and acceleration capabilities can be leveraged. \texttt{LAMMPS} uses the Kokkos~\cite{Kokkos1,Kokkos2,Kokkos3} library, enabling GPU usage across an arbitrary number of nodes connected via the MPI library, allowing for large-scale calculations.

The remaining step is the calculation of the observables. Here the bulk of the computational cost is associated with processing the electronic density to energetic contribution. The calculation of density itself from the LDOS is straightforward, as is the computation of the DOS from the LDOS, and DOS-related observables from the DOS. Processing of the density to derive observables involves a custom interface to the \texttt{Quantum ESPRESSO} code and includes the calculation of the Ewald sum (ion-ion interaction) and the exchange correlation energy. As discussed in Sec.~\ref{ss.observables}, direct evaluation of these quantities scales poorly with particle number, creating a computational bottleneck for large systems. While the computation can be parallelized, GPU acceleration is not readily available for this step.

Using Eq.~\eqref{eq:structurefactor} and Eq.~\eqref{eq:structurefactor_gaussian} in place of direct evaluation for ion-ion and XC energies alleviates these issues. With this approach, the primary computational cost of this task shifts to calculating the Gaussian descriptors, which are processed through \texttt{LAMMPS} and can be accelerated using the previously mentioned strategy, as shown in Tab.~\ref{tab:gpu_cpu_parallelization}. 

Thus, all major computational tasks involved in network training and inference can be performed in a multi-GPU setting, as detailed in Tab.~\ref{tab:gpu_cpu_parallelization}. Smaller tasks related to data processing are executed on CPUs and are parallelized for efficiency. Additionally, as shown in Tab.~\ref{tab:gpu_cpu_parallelization}, hyperparameter optimization is also parallelized within the \texttt{MALA} package. Depending on the algorithm used, parallelization is supported either through MPI (e.g., for NASWOT) or a database-driven approach, with individual ranks collecting trials in either an SQL database (for TPE, via \texttt{Optuna}) or a simple file (for OATM).

\subsection{Data management}
\label{ss.data.management}

While the neural networks in the \texttt{MALA} package perform vector-to-vector mappings, the training data itself consists of volumetric data, which entails managing large data volumes. \texttt{MALA} datasets typically include a relatively small number of individual atomic snapshots $N_j$, usually on the order of $10^0$ to $10^1$. However, each snapshot contains a large number of grid points, with both LDOS and bispectrum descriptor information associated with each point. This results in a substantial number of training data points.

Below, we provide a brief overview of the challenges and solutions for managing such extensive, volumetric datasets, particularly in the context of neural network training. In the inference case, no special considerations are required apart from having sufficient memory to store the bispectrum descriptor vectors and LDOS before calculating relevant observables. Since inference is typically performed in parallel, adequate memory can usually be pooled across a suitable number of computational nodes.

\begin{figure}[htp]
\centering
\includegraphics[width=0.95\columnwidth]{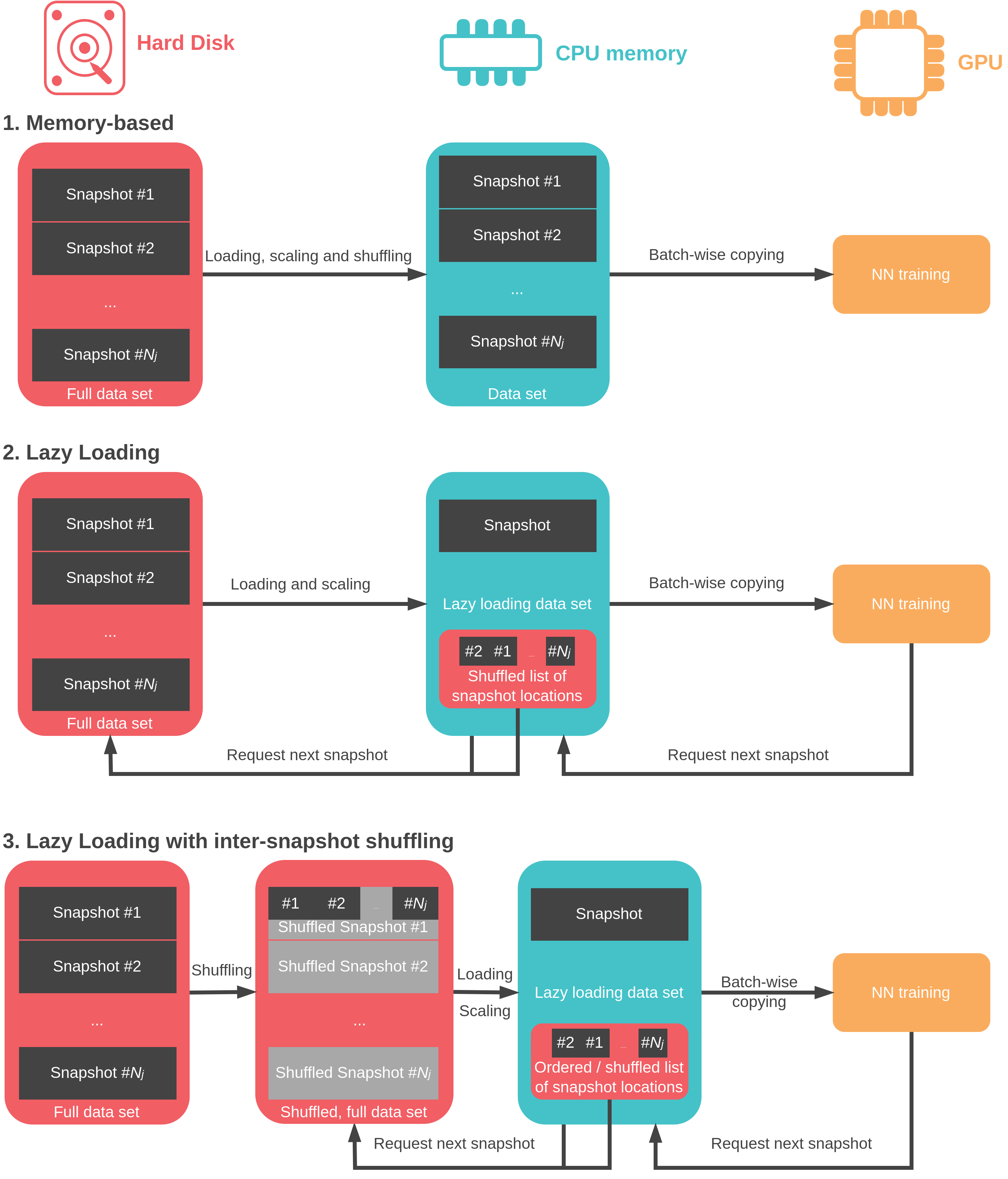}
\caption{Visualization of training data management in \texttt{MALA}, with storage locations indicated by color. From top to bottom, memory-based data management, lazy loading and lazy loading with data shuffling are shown. Figure taken from Ref.~\citenum{Lenz_Dissertation}.}
\label{fig:datamanagement}
\end{figure}

Data management for neural network training comprises three main tasks. First, data must be loaded into CPU memory, from where it is sent to the GPU in batches specified by the training routine. Next, data may need to be scaled (min-max scaling or standardization), depending on the dataset characteristics. Lastly, data must be shuffled to randomize the access order of data points, as training on ordered data can reduce neural network accuracy~\cite{ShuffleReference}. Depending on the size and complexity of the dataset, different techniques are required to ensure these tasks are handled efficiently, as illustrated in Fig.~\ref{fig:datamanagement}.

If sufficient CPU memory is available to accommodate all training snapshots, these tasks can be performed entirely in memory after loading all relevant snapshots. This approach, referred to as ``memory-based'', is shown in the top row of Fig.~\ref{fig:datamanagement}. It is straightforward to implement and aligns well with standard machine learning practices.

However, the data volume required for \texttt{MALA} models makes memory-based data management impractical. To manage larger datasets, \texttt{MALA} employs a lazy loading approach. In this setup, snapshots are loaded into CPU memory and scaled one by one. Each snapshot is processed in memory, offloaded to the GPU for neural network training, and then replaced by the next snapshot. This process is illustrated in the second row of Fig.~\ref{fig:datamanagement}.

One limitation of the lazy loading approach is that it does not support full randomization of the training data. While it is simple, available, and recommended to randomize the order in which snapshots are accessed, this does not achieve true randomization at the scale of the entire dataset. Since only one snapshot is loaded into memory at a time, full dataset shuffling cannot be accomplished in memory. To address this issue, \texttt{MALA} includes a file-based shuffling method. In a separate preprocessing step, all data points from the $N_j$ snapshots are shuffled into an arbitrary number of ``snapshot-like'' objects, which are then stored on disk. The lazy loading algorithm is then applied to these fully shuffled, snapshot-like objects, as shown in the third row of Fig.~\ref{fig:datamanagement}.

Naturally, storing shuffled snapshots requires adequate hard disk space, but this is generally not difficult to achieve. Additionally, reshuffling the data before neural network training is feasible, as it typically takes only a few minutes on a single CPU, as described in Ref.~\citenum{Lenz_Dissertation}. This preprocessing cost is negligible compared to the overall training time (detailed in Sec.~\ref{ss.scaling}). Lazy loading itself introduces a minor computational overhead due to additional file operations, but this impact on training time is minimal, given the structure of the dataset. Specifically, a small number of large (shuffled) data files minimizes this overhead.

Therefore, lazy loading with shuffling is preferable to standard lazy loading when the enhanced accuracy from full dataset randomization is needed. For many applications, including the temperature and revised phase boundary results presented in Ref.~\citenum{temperaturepaper} and Ref.~\citenum{Lenz_Dissertation}, full randomization is beneficial (see Sec.~\ref{ss.accuracy} for details). Future developments in the \texttt{MALA} package aim to replace this file-based approach with streamed hard disk access, for example via the OpenPMD~\cite{OpenPMD} library. Additionally, ongoing work is exploring redundancy removal algorithms for volumetric data files, which could further reduce training dataset size.

\section{Computational workflow example}
\label{s.workflow}

As a computational tool is best understood through practical application, the following is an example of the \texttt{MALA} workflow in action. Past \texttt{MALA} applications have demonstrated results for simple metals, specifically aluminum~\cite{malapaper,hyperparameterpaper,temperaturepaper,Lenz_Dissertation} and beryllium~\cite{hyperparameterpaper,sizetransferpaper,Lenz_Dissertation}. Current research projects are expanding this range to include materials such as carbon structures, silicon oxide, and hydrogen to evaluate the broader applicability of the \texttt{MALA} workflow.

In this publication, we present boron as an example. While boron’s electronic structure itself is not currently a major research focus, studies that use boron as a dopant or impurity often analyze its electronic properties~\cite{Boron1,Boron2,Boron3}. Boron was selected here primarily due to its relatively low number of valence electrons, which simplifies LDOS computations, as well as its ability to form a crystalline structure at room temperature. Boron has several stable allotropes under ambient conditions; aside from amorphous boron, it exists in multiple crystalline forms, most commonly as $\alpha$-rhombohedral, $\beta$-rhombohedral, and $\beta$-tetragonal~\cite{BoronStructures}. Throughout this section, we focus on crystalline $\alpha$-rhombohedral boron.

\subsection{Data generation}
\label{ss.data.generation}

The first step in developing a \texttt{MALA} model is to gather the necessary DFT and LDOS data. For the most part, this follows a standard computational materials science workflow, with no specific adjustments required for the \texttt{MALA} workflow. A work-in-progress collection of data generation functions for \texttt{MALA} is available for early use~\cite{MALADA}, though it has not yet been officially released. Room temperature is assumed throughout this section, and all convergence calculations are conducted with an accuracy threshold of $1\,\mathrm{meV/atom}$, unless otherwise specified.

\subsubsection{Sampling of ionic configurations}
To begin the modeling workflow, a suitable crystal structure is obtained, for example, from the Materials Project database~\cite{MaterialsProject}. For $\alpha$-rhombohedral boron this structure corresponds to \lstinline|mp-160| in the database. In this configuration, boron atoms form interconnected $\mathrm{B}_{12}$ icosahedra. For \texttt{MALA} model data generation, it is advisable to use simulation cells containing a few hundred atoms. Very small cells often require large $\bk$-grids for accurate LDOS sampling, while larger cells increase computational overhead. 

In this example, a boron cell with 144 atoms is used by extending the ideal crystal structure in all three spatial dimensions. To obtain thermalized atomic configurations, DFT-MD simulations are performed using the \texttt{VASP} code. A cutoff energy of 310 eV is applied, based on convergence calculations, along with a PBE~\cite{perdew_generalized_1996} PAW pseudopotential~\cite{VASP_PP2} and a Nosé-Hoover thermostat~\cite{nose,hoover,evans1985nose} with a Nos\'e mass of $Q=0.001$. The resulting trajectory is analyzed using the method outlined in Sec.~\ref{ss.data.sampling.ionic} and described in detail in Refs.~\cite{ofdftpaper, Lenz_Dissertation}. This analysis can be performed with just a few lines of code in \texttt{MALA} with:

\begin{lstlisting}[language=Python]
import mala

# Use default parameters for the analysis.
parameters = mala.Parameters()

# Load the trajectory and create the analyzer.
trajectory_analyzer = (
    mala.TrajectoryAnalyzer(...)

# Perform the analysis.
(trajectory_analyzer.
 get_uncorrelated_snapshots("VASP_output_file"))
\end{lstlisting}

This code example uses default parameters for trajectory analysis, which have been found to yield satisfactory results for boron. These parameters can, of course, be adjusted for different systems, with a comprehensive overview of relevant settings provided in Ref.~\citenum{Lenz_Dissertation}. The output of this trajectory analysis is a set of 833 uncorrelated atomic configurations. Based on the anticipated complexity of the system, DFT simulations are then performed on an appropriately sized subset of these configurations.

\subsubsection{Calculation of the LDOS}
\begin{figure}[htp]
\centering
\includegraphics[width=0.95\columnwidth]{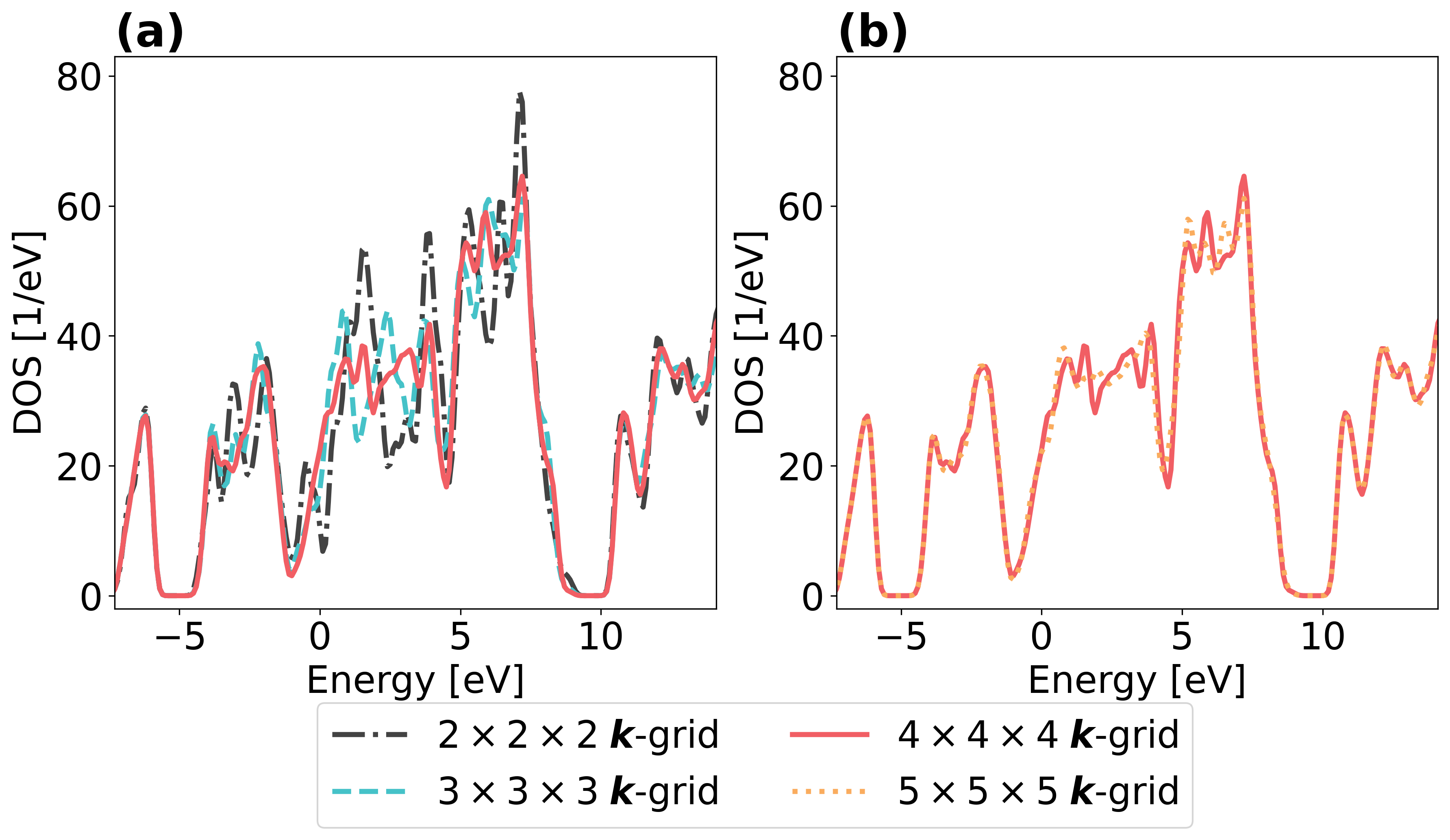}
\caption{DOS of a thermalized Boron configuration containing 144 atoms at varying $\bk$-grids. \textbf{(a)}: As the number of $\bk$-points increases, unphysical oscillations are reduced and eventually disappear. \textbf{(b)}: At $4\times4\times4$ $\bk$-points, these oscillations are sufficiently converged, as an increase of the $\bk$-grid to $5\times5\times5$ $\bk$-points changes the DOS only marginally.}
\label{fig:dos_b}
\end{figure}

For $\alpha$-rhombohedral boron, 15 snapshots were selected for DFT simulation and LDOS calculation. These calculations were carried out with \texttt{Quantum ESPRESSO} with an optimized norm-conserving Vanderbilt pseudopotential~\cite{QEPSP}. The energy cutoff for the simulations was determined through convergence calculations, yielding an optimal energy cutoff of approximately 1089 eV (i.e., 80 Rydberg). For the $\bk$-grid, a higher number of $\bk$-points is necessary for accurate (L)DOS sampling, as discussed in Sec.~\ref{ss.data.sampling}. A Monkhorst-Pack $\bk$-grid~\cite{monkhorst_special_1976} is employed, with the grid resolution determined by examining derived quantities (such as the band energy) or by directly inspecting the (L)DOS computed at various $\bk$-points. A visual representation of this process is provided in Fig.~\ref{fig:dos_b}. 

The analysis shows that a that a $\bk$-grid of $4\times4\times4$ $\bk$-points is required to adequately suppress unphysical oscillations. This grid is significantly larger than the $2\times2\times2$ $\bk$-grid typically sufficient for standard-accuracy DFT calculations, yet it remains manageable even for larger datasets.

A similar analysis can also be conducted by calculating the total free energy from the LDOS and examining relevant trends. Additionally, by accessing the band energy, one can further investigate the Gaussian width and energy spacing used to compute the LDOS, as outlined in Sec.~\ref{ss.data.sampling}. Both quantities are easily accessible in \texttt{MALA} and can be conveniently compared with DFT values through:

\begin{lstlisting}[language=Python]
import mala

# LDOS parameters depend on material
parameters = mala.Parameters()
parameters.targets.ldos_gridsize = 241
parameters.targets.ldos_gridspacing_ev = 0.1
parameters.targets.ldos_gridoffset_ev = -8
parameters.targets.pseudopotential_path = ...

# Other file formats can be used as well.
# QuantumESPRESSO writes a list of cube 
# files, the * is replaced by the running
# counter.
ldos_calculator = (
    mala.LDOS.from_cube_file(parameters,
                              "ldos*.cube"))
(ldos_calculator.
 read_additional_calculation_data(
    "QuantumESPRESSO_output.out"))

# Energy errors are usually reported in meV/atom
print((ldos_calculator.total_energy -
       ldos_calculator.total_energy_dft_calculation) *
      (1000/len(ldos_calculator.atoms)))
print((ldos_calculator.band_energy - 
       ldos_calculator.band_energy_dft_calculation) * 
      (1000/len(ldos_calculator.atoms)))
\end{lstlisting}

\begin{figure}[htp]
\centering
\includegraphics[width=0.95\columnwidth]{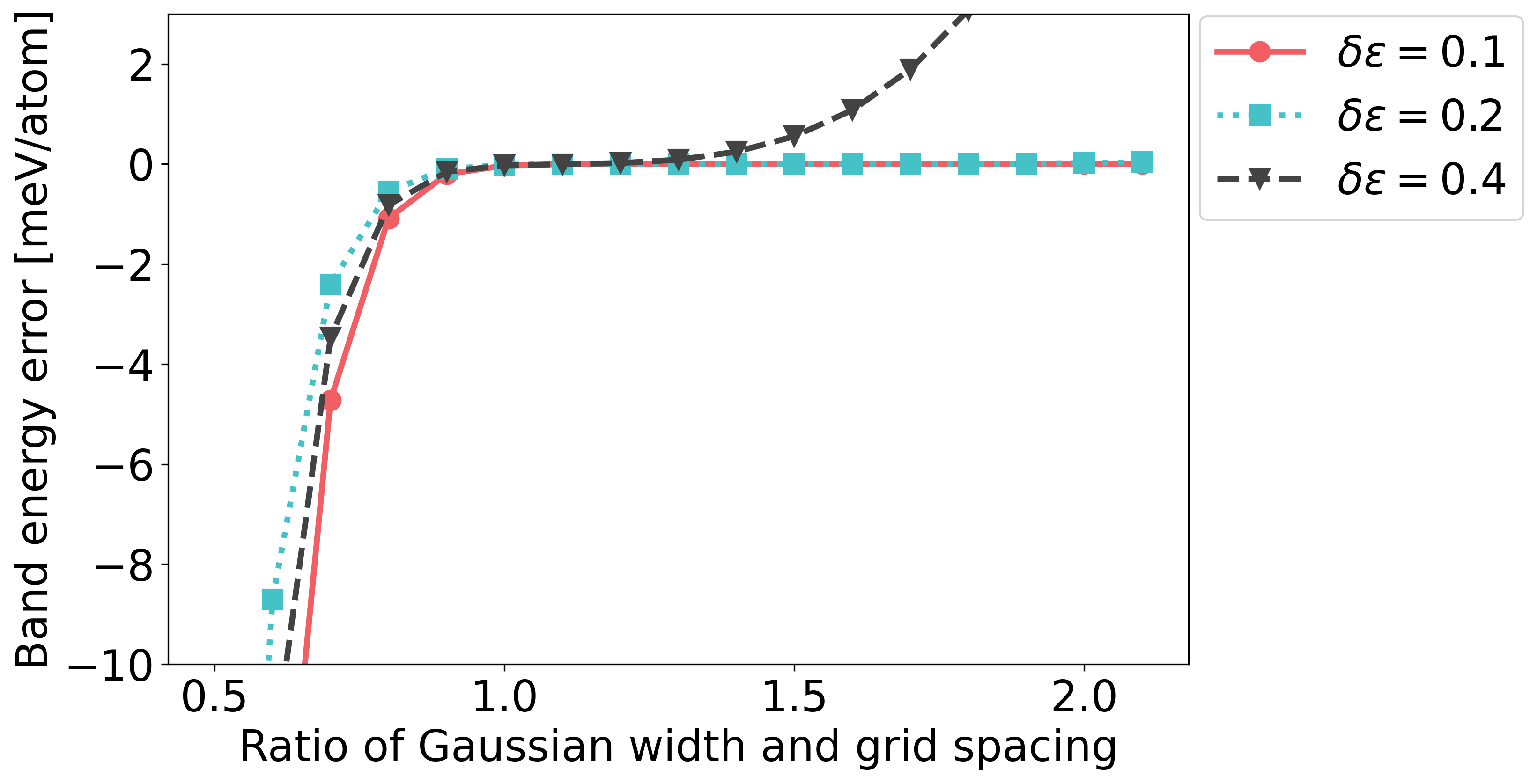}
\caption{Dependence of band energy error on Gaussian width (expressed in units of energy grid spacing) for 144 boron atoms at room temperature, computed at varying values of $\delta\epsilon$. Band energy errors are shown for a single ionic configurations. It is observed that reasonable combinations of $w_\mathrm{d}$ and $\delta\epsilon$ can be identified at different $\delta\epsilon$ levels. To maintain consistency with the values used for aluminum and beryllium, a Gaussian width of $w_\mathrm{d}/\delta\epsilon=2$ and an energy spacing of $\delta\epsilon=0.1\,\mathrm{eV}$ have been selected.}
\label{fig:width_analysis_b}
\end{figure}

The resulting analysis of Gaussian width and energy spacing is shown in Fig.~\ref{fig:width_analysis_b}. Here, a ratio of Gaussian width to energy spacing of $w_\mathrm{d}/\delta\epsilon=2$ is found to be suitable, consistent with the values used for aluminum and beryllium. The energy spacing for (L)DOS sampling can similarly be maintained at $\delta\epsilon=0.1\,\mathrm{eV}$.

Thus, the LDOS is sampled in steps of $\delta\epsilon=0.1\,\mathrm{eV}$ over the range from $-8\,\mathrm{eV}$ to $16\,\mathrm{eV}$, using a $4\times4\times4$ $\bk$-grid on a real-space grid with $108\times108\times135\approx1.5\times10^{6}$ grid points. After completing all 15 DFT and LDOS simulations, errors in computing the band energy and total energy via the LDOS are evaluated by comparing these to the DFT results. For this dataset, the average total free energy error is $1.23\,\mathrm{meV/atom}$ with a standard deviation of $3.30\,\mathrm{meV/atom} $. The average band energy error is $8.26\times10^{-3} \,\mathrm{meV/atom}$, with a standard deviation of $8.33\times10^{-4} \,\mathrm{meV/atom}$. To contextualize these results, note that typical convergence accuracy for DFT calculations, e.g., w.r.t.~$\bk$-grid or cutoff energy, is $1\,\mathrm{meV/atom}$, while \texttt{MALA} models often aim for a prediction accuracy of below $10\,\mathrm{meV/atom}$. Thus, these results indicate that the overall discretization error in transitioning from the Kohn-Sham framework to the LDOS-based framework is minimal, confirming that the resulting LDOS data can reliably be used to predict the electronic structure.

\subsubsection{Preprocessing and descriptor calculation}
From this set of calculations, the bispectrum descriptors are computed, and the LDOS is converted into a more convenient format. By default, \texttt{Quantum ESPRESSO} outputs LDOS data across multiple text-based files, which can be inefficient. These tasks are handled within a single \texttt{MALA} class, which ensures consistent nomenclature and adherence to the required file format. Additionally, if parallelization is applied for the computation of bispectrum descriptors or LDOS parsing, this class automatically initiates the appropriate routines.

To compute the bispectrum descriptors, the relevant bispectrum hyperparameters ($J_\mathrm{max}$ and $R_\mathrm{cutoff}$) must be specified. For boron, the established hyperparameters used in Ref.~\citenum{malapaper} and applied to beryllium~\cite{sizetransferpaper} are employed. This example demonstrates \texttt{MALA}’s prediction capabilities using these previously validated defaults. For more complex systems, optimizing these hyperparameters using the methods outlined in Ref.~\citenum{hyperparameterpaper} is recommended. The code for performing this data conversion is as follows:

\begin{lstlisting}[language=Python]
import mala

parameters = mala.Parameters()

# Parallelize data processing.
parameters.use_mpi = True
parameters.targets.target_type = "LDOS"

# As determined by error analyis.
parameters.targets.ldos_gridsize = 241
parameters.targets.ldos_gridspacing_ev = 0.1
parameters.targets.ldos_gridoffset_ev = -8

# As determined by surrogate metric.
parameters.descriptors.descriptor_type = "Bispectrum"
parameters.descriptors.twojmax = 10
parameters.descriptors.rcutfac = 4.67637

data_converter = mala.DataConverter(parameters)
for snapshot in range(0, 15):
    data_converter.add_snapshot(...)

(data_converter.
 convert_snapshots(descriptor_save_path="bispectrum_folder/",
                   target_save_path="ldos_folder/",
                   additional_info_save_path="additional_info_folder/",
                   naming_scheme="B_snapshot*.h5"))
\end{lstlisting}

This process produces a ready-to-use dataset encoded with the OpenPMD API~\cite{OpenPMD}. Using this library ensures accurate storage of metadata, facilitating the reproducibility of machine learning results. With this dataset, model training can now proceed.

\subsection{Model training}
\label{ss.model.training}

Model training often focuses solely on tuning model parameters, but rigorously, it also requires selecting appropriate hyperparameters. In this example, we use previously established hyperparameters, so hyperparameter optimization will be omitted here. A detailed discussion of this process can be found in Ref.~\citenum{hyperparameterpaper}. Since model architectures identified in Ref.~\citenum{malapaper} through direct search have shown strong performance on elements beyond those for which they were originally developed (likely due to their high information capacity~\cite{hyperparameterpaper}), we anticipate reasonable prediction accuracy for boron as well. Consequently, we employ the model architecture used for room temperature aluminum in Ref.~\citenum{malapaper}.

As outlined in Sec.~\ref{ss.data.management}, full randomization of data access order is recommended. For \texttt{MALA} models trained in lazy loading fashion, this requires creating data files that aggregate data points across multiple ionic configurations. \texttt{MALA} provides an interface for this process, which can be executed directly before neural network training, using the following code:
\begin{lstlisting}[language=Python]
import mala

params = mala.Parameters()

# Setting a seed is useful to make shuffling
# of snapshots reproducible.
params.data.shuffling_seed = ...
data_shuffler = mala.DataShuffler(params)

# Adding snapshots to be shuffled
for i in range(0, 6):
    data_shuffler.add_snapshot(....)


data_shuffler.shuffle_snapshots(
    descriptor_save_path="shuffled_bispectrum_folder/",
    target_save_path="shuffled_ldos_folder",
    save_name="shuffled_snapshot*.h5",
)
\end{lstlisting}

Here, we assume that six atomic snapshots are sufficient for training a boron model. In addition to the training dataset, the validation dataset (the portion of the dataset used to monitor accuracy during training) can also be shuffled. However, shuffling the validation set prevents calculation of certain observables, such as the band energy, during training, as the LDOS must be organized on the numerical grid rather than in a shuffled order. For this example, a single atomic snapshot is used for validation, which makes shuffling unnecessary. Thus, six snapshots are reserved for training, one for validation, and the remaining eight snapshots for testing to confirm that the trained model can generalize to unseen snapshots. Training a neural network model in \texttt{MALA} is straightforward and can be achieved with the following code:

\begin{lstlisting}[language=Python]
import mala

parameters = mala.Parameters()
# Add all relevant parameters for the model.
# Loading could also be done from file.
...

# Add training data
data_handler = mala.DataHandler(parameters)
data_handler.clear_data()

# Add data.
for i in range(0, 6):
    data_handler.add_snapshot(...)

data_handler.prepare_data()

# Specify NN size.
parameters.network.layer_sizes = [
    data_handler.input_dimension,
    ...,
    data_handler.output_dimension,
]

# Train and save model.
network = mala.Network(parameters)
trainer = mala.Trainer(parameters, network, data_handler)
trainer.train_network()
trainer.save_run("boron_model_01")
\end{lstlisting}

An arbitrary number of GPUs can be used for training by enabling \texttt{PyTorch}'s Distributed Data Parallel (DDP) interface with \lstinline|parameters.use\_ddp = True|. Note that in parallel training mode, some training metrics are unavailable and, e.g., the band energy error cannot be reported during training.

Since the prediction accuracy of neural networks can be sensitive to network initialization (although the extent of this dependence varies with the robustness of the chosen hyperparameters), it is recommended to train several models. In this case, five models were trained with different random weight initializations. Accuracy evaluation is facilitated by a specialized \texttt{MALA} class using the following code:

\begin{lstlisting}[language=Python]
import mala

# Load trained network.
(parameters, network, inference_data_handler, tester) = mala.Tester.load_run(
...
)

# Add all snapshots to be tested.
for i in range(0, 8):
    inference_data_handler.add_snapshot(...)
inference_data_handler.prepare_data()
    
# Specify which observables to test.
# The result of test_all_snapshots()
# will be a dict with a list for
# each key which was selected for
# testing.
tester.observables_to_test = ["total_energy", "band_energy"]
results = tester.test_all_snapshots()
\end{lstlisting}

\begin{figure}[ht]
\centering
\includegraphics[width=0.95\textwidth]{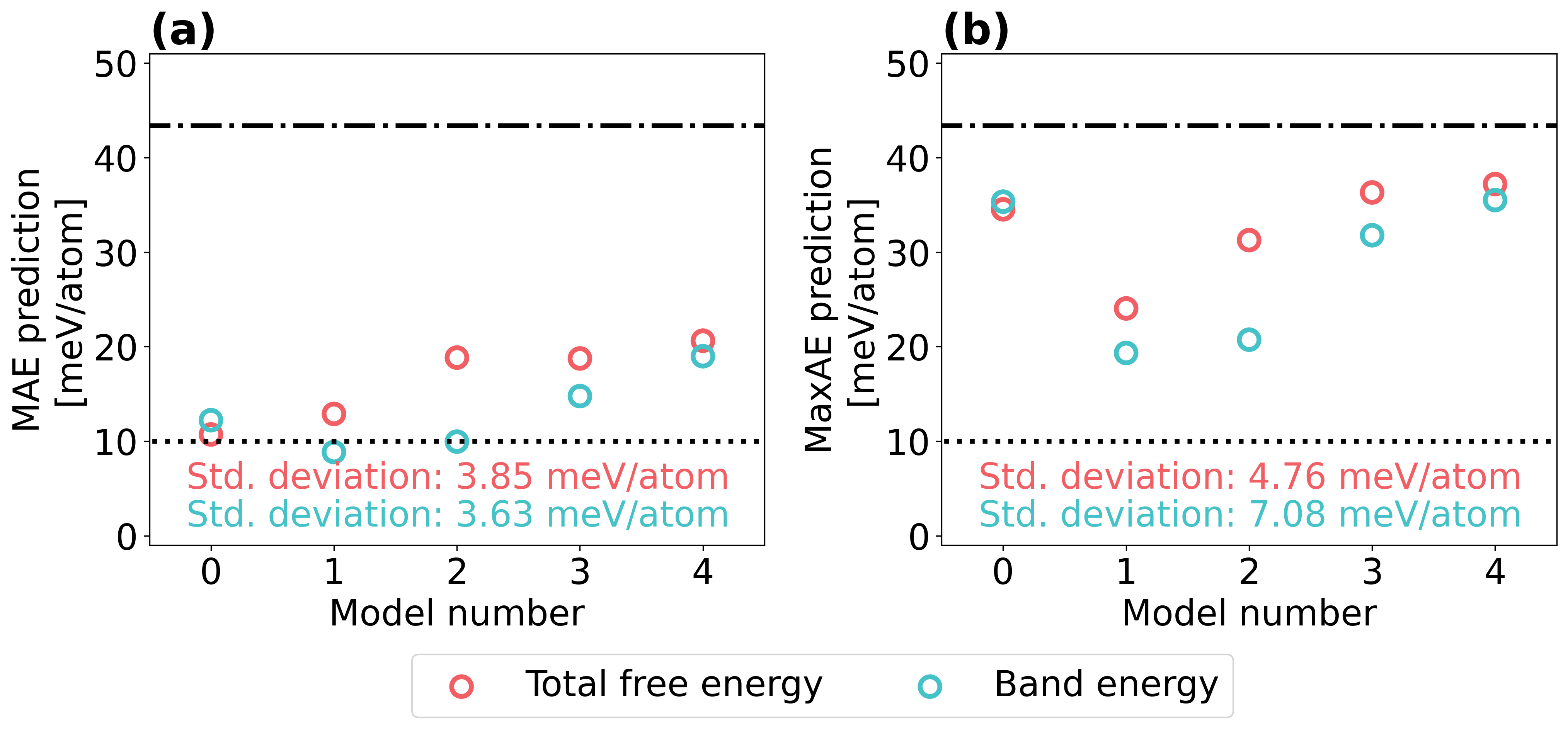}
\caption{Mean and maximum energy inference errors for all five room temperature boron models across the test set of eight atomic snapshots. Panel \textbf{(a)} shows the $\mathrm{MAE}$, while panel \textbf{(b)} displays the $\mathrm{MaxAE}$. The dotted line at $10 \,\mathrm{meV/atom}$ represents a commonly employed accuracy threshold, while the dashed-dotted line at $43.4 \,\mathrm{meV/atom}$ indicates chemical accuracy. The standard deviations given in both panels are calculated across model initializations for each respective energy metric. Note that models have not been ordered by accuracy; the apparent pattern observed in panel \textbf{(a)} is coincidental.}	\label{fig:B_multi_energy_inference}
\end{figure}

The results of these testing operations provide prediction accuracies for both band energies and total energies. These accuracies allow for comparison of individual neural network initializations, as shown in Fig.~\ref{fig:B_multi_energy_inference}. In this figure, both the mean absolute error (MAE, panel \textbf{(a)}) and maximum absolute error (MaxAE, panel \textbf{(b)}) are displayed for inferences across the test set of eight atomic snapshots. Notably, prediction accuracy varies between different neural network initializations.

In general, band energy predictions are found to be more accurate than total free energy predictions. This is unsurprising as total free energy predictions require a highly accurate spatially-resolved electronic density and are therefore more susceptible to inaccuracies related to ML prediction. Band energy predictions are derived from the DOS and may benefit from error cancellations in the spatial domain. A model is selected from Fig.~\ref{fig:B_multi_energy_inference} by seeking both a low MAE and a low MaxAE, the latter being important to ensure robustness against outliers. For boron, model \#1 is chosen and further analyzed in the following section.

\subsection{Inference and application}
\label{ss.post-processing}

\begin{figure}[ht]
\centering
\includegraphics[width=0.95\textwidth]{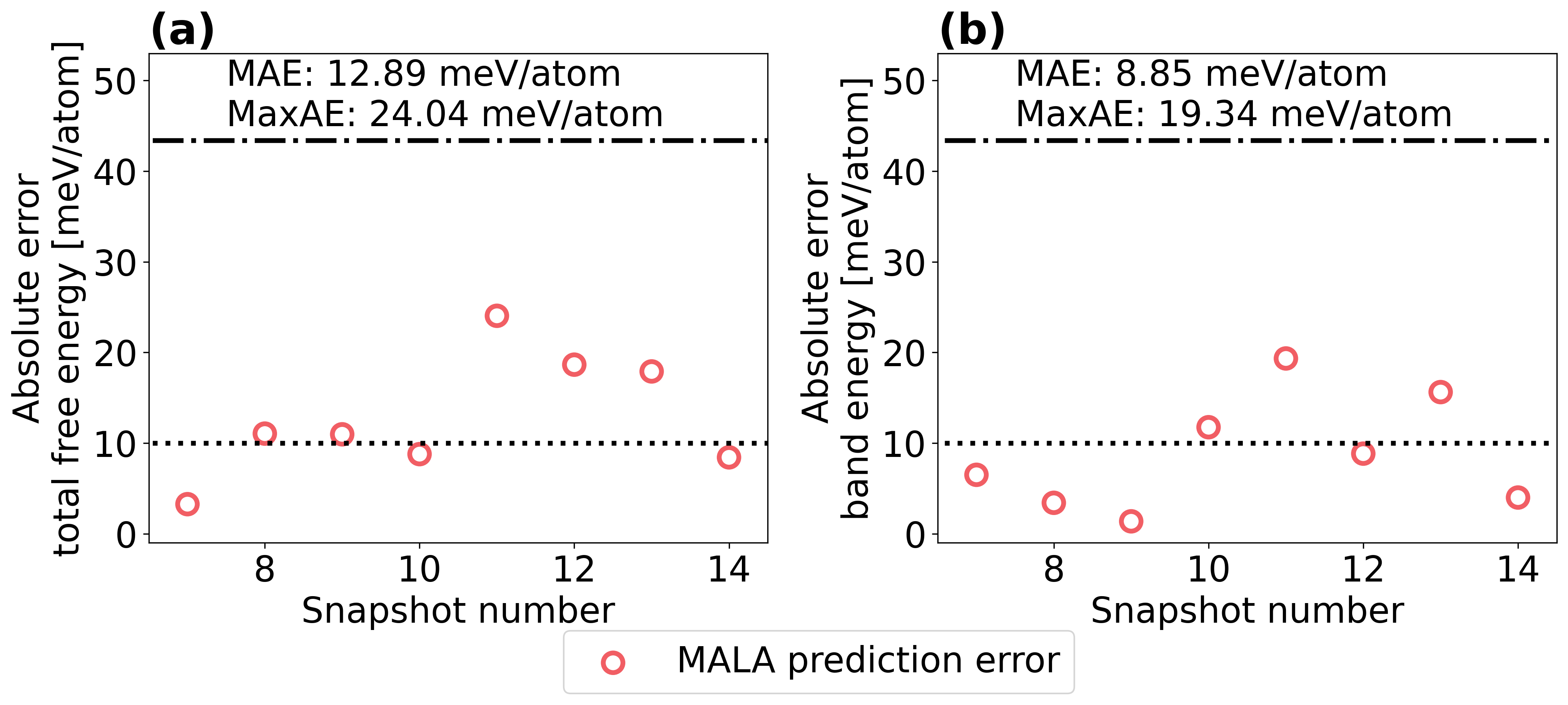}
\caption{Energy inference errors for selected boron model. All results are presented for the testing portion of the boron dataset, labeled snapshots \#7 to \#14 according to the internal structure of the dataset. Panel \textbf{(a)} shows the total free energy error, while panel \textbf{(b)} shows band energy errors. The dotted line at $10 \,\mathrm{meV/atom}$ represents a commonly employed accuracy threshold, while the dashed-dotted line at $43.4 \,\mathrm{meV/atom}$ indicates chemical accuracy.}
\label{fig:B298_energyMAE}
\end{figure}

\begin{figure}[ht]
\centering
\includegraphics[width=0.95\textwidth]{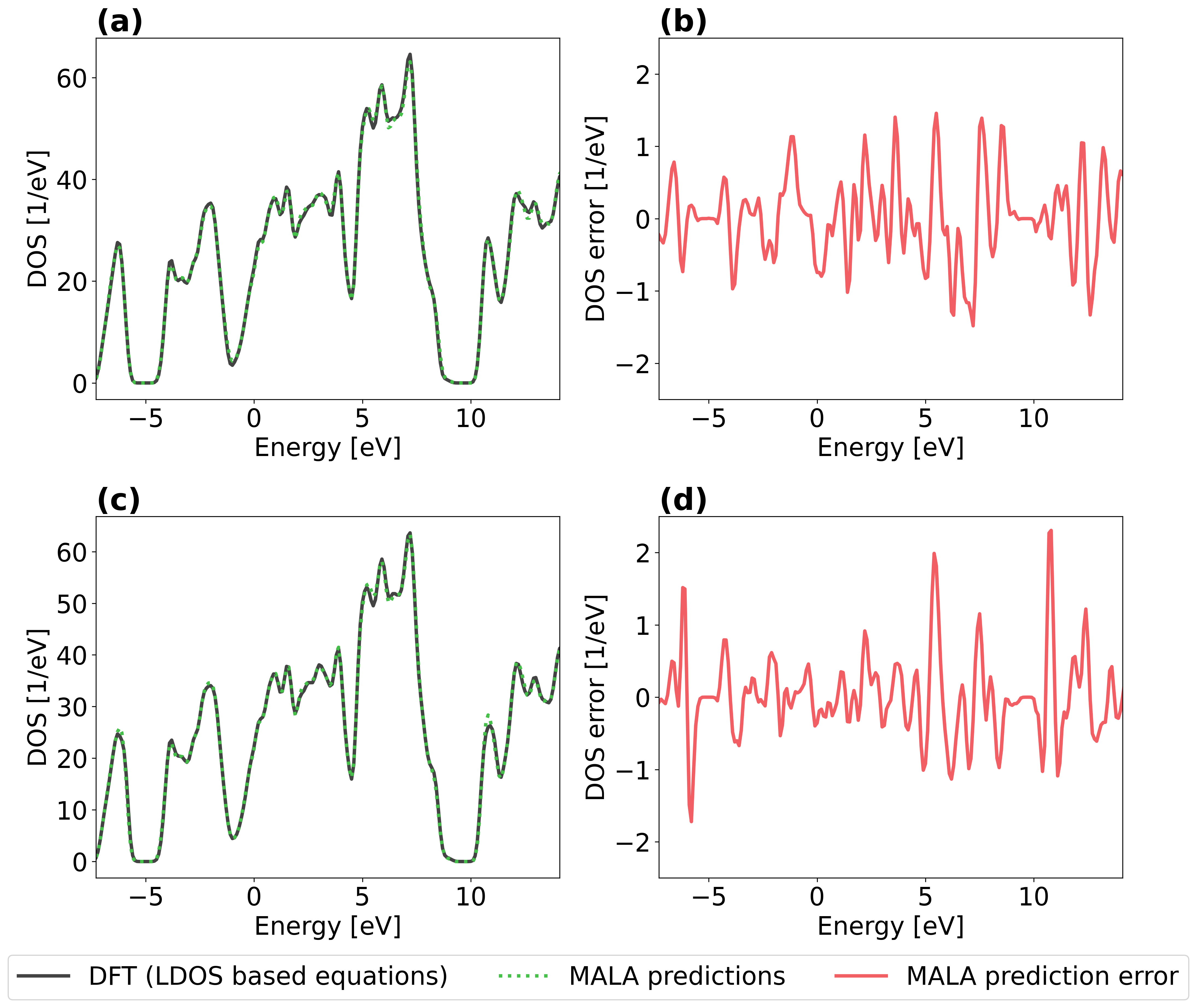}
\caption{DOS prediction and prediction error for two boron configurations (labeled snapshot \#7 and snapshot \#11) drawn from the test dataset. Predictions were made using the same model as in Fig.~\ref{fig:B298_energyMAE}. Panels \textbf{(a)} and \textbf{(c)} show the actual and predicted DOS for snapshot \#7 and snapshot \#11, respectively, while panels \textbf{(b)} and \textbf{(d)} display the DOS prediction error snapshot \#7 and snapshot \#11, respectively. The DFT-reported Fermi energy is $9.68\,\mathrm{eV}$ for snapshot \#7 and $9.71\,\mathrm{eV}$ for snapshot \#11.}
\label{fig:B298_DOS}
\end{figure}

\begin{figure}[ht]
\centering
\includegraphics[width=0.95\textwidth]{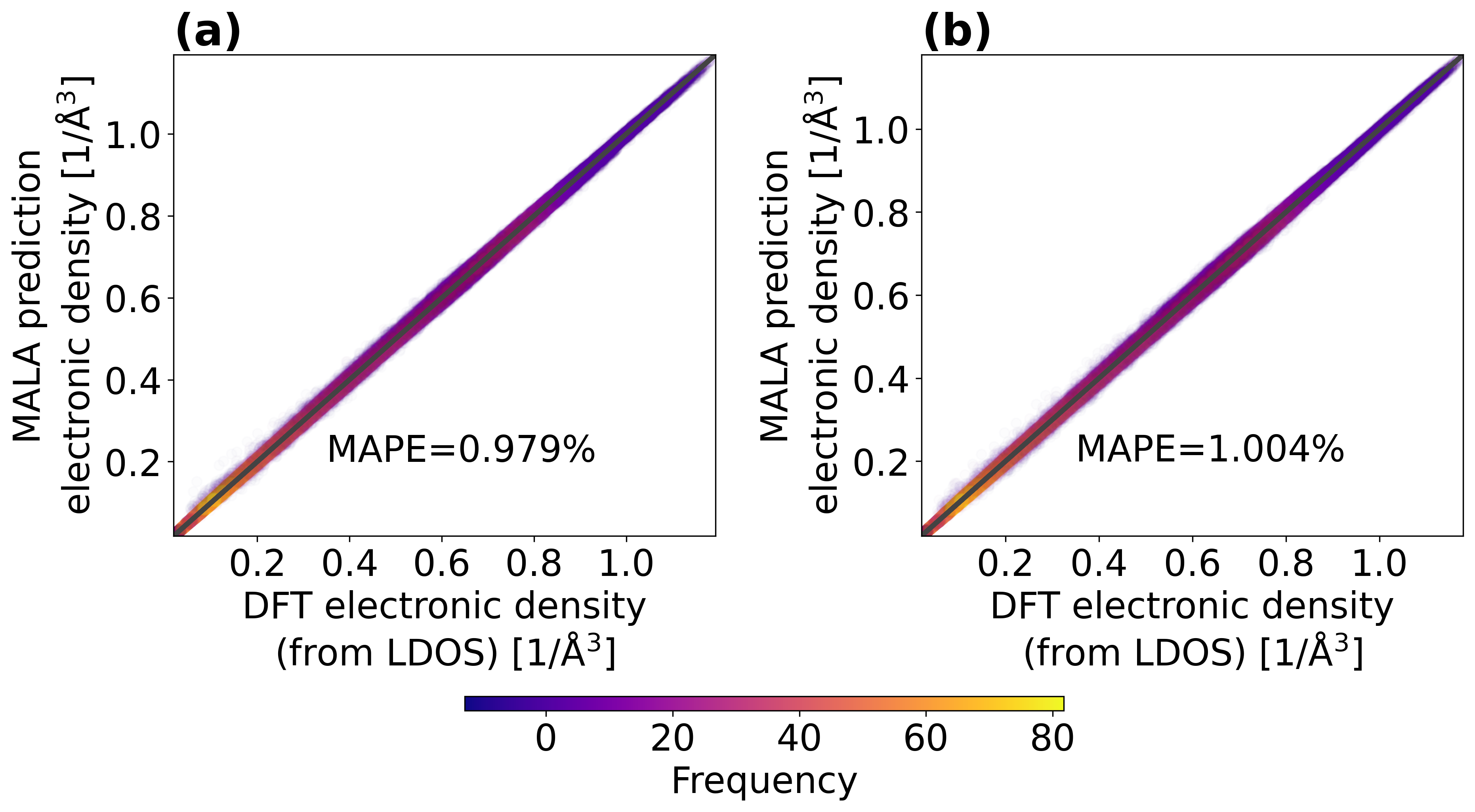}
\caption{Visualization of density prediction with respective MAPE for two boron configurations (labeled snapshot \#7 and snapshot \#11) the test dataset. Predictions were made with the same model as in Fig.~\ref{fig:B298_energyMAE}. Panel \textbf{(a)} shows results for snapshot \#7, while panel \textbf{(b)} shows results for snapshot \#11. Each panel illustrates the correlation between the \texttt{MALA} prediction and the actual electronic density, represented as a colored point cloud, with the color bar indicating the frequency of individual values. An ideal prediction is depicted by a dark gray line.}
\label{fig:B298_Density}
\end{figure}

After selecting a model from the set of trained models, the testing interface described above is used to gain detailed insights into its predictive capabilities. In this analysis, Fig.~\ref{fig:B298_energyMAE} presents the energy inference errors, Fig.~\ref{fig:B298_DOS} shows DOS predictions and errors, and Fig.~\ref{fig:B298_Density} displays electronic density prediction errors.

In the context of \texttt{MALA} models, total free energy prediction errors are often considered a principal quality criterion, given the significance of the total free energy and its derivatives. Two typical thresholds for ``accurate'' predictions are chemical accuracy defined as $1\,\mathrm{kcal/mol} \triangleq 43.4\,\mathrm{meV/atom}$ and a commonly used threshold of $10 \,\mathrm{meV/atom}$. Of course, the accuracy requirements for a model are ultimately governed by its intended application.

As shown in Fig.~\ref{fig:B298_energyMAE}, the strict threshold of $10\,\mathrm{meV/atom}$ is not fully met for all total free energy predictions across atomic configurations and is slightly exceeded for the MAE. However, prediction accuracy remains well within chemical accuracy and is close to this threshold. For the band energy, this threshold is even met on average. Overall, the model demonstrates sufficient accuracy for modeling $\alpha$-rhombohedral boron, especially when considering that previously determined hyperparameters have been employed for much of the workflow.

Since DOS and electronic density are readily accessible with \texttt{MALA} models, their prediction accuracy is also of interest. To relate energy prediction accuracy to DOS and density predictions, Fig.~\ref{fig:B298_DOS} and Fig.~\ref{fig:B298_Density} present results for the atomic snapshots with the lowest (snapshot \#7) and highest (snapshot \#11) energy prediction errors, respectively.

\clearpage

In Fig.~\ref{fig:B298_DOS}, it is evident that the DOS is well reproduced by \texttt{MALA} models, even for snapshot \#11, which exhibits the highest prediction error among the test set. Errors in the DOS reproduction appear as slight over- or underestimation of certain peaks, but the magnitude of these errors is relatively small. This is further quantified by the prediction error curves shown in Fig.~\ref{fig:B298_DOS}\textbf{(b)} and Fig.~\ref{fig:B298_DOS}\textbf{(d)}. Although the errors shown are approximately an order of magnitude higher than those reported for aluminum~\cite{malapaper,Lenz_Dissertation}, they remain small and oscillate around zero, indicating that \texttt{MALA} can effectively reproduce the DOS of $\alpha$-rhombohedral boron, which presents a more complex profile than that of simple metals.

To assess the prediction accuracy of the electronic density, Fig.~\ref{fig:B298_Density} plots the actual electronic density at a given grid point on the $x$-axis against the predicted electronic density at the same grid point on the $y$-axis. For an ideal model, the resulting points would lie on the line $y=x$, though slight deviations are expected in practice. Additionally, the MAPE is calculated as a scalar metric for density prediction accuracy. As shown in Fig.~\ref{fig:B298_Density}, both atomic snapshots exhibit only minor deviations from the ideal prediction, resulting in an MAPE of roughly 1\% in each case. While the MAPE values indicate a slightly lower accuracy for snapshot \#11, this difference is not visually evident in the direct comparison of predicted and actual electronic density.

\clearpage

\begin{figure}[ht]
\centering
\includegraphics[width=0.95\textwidth]{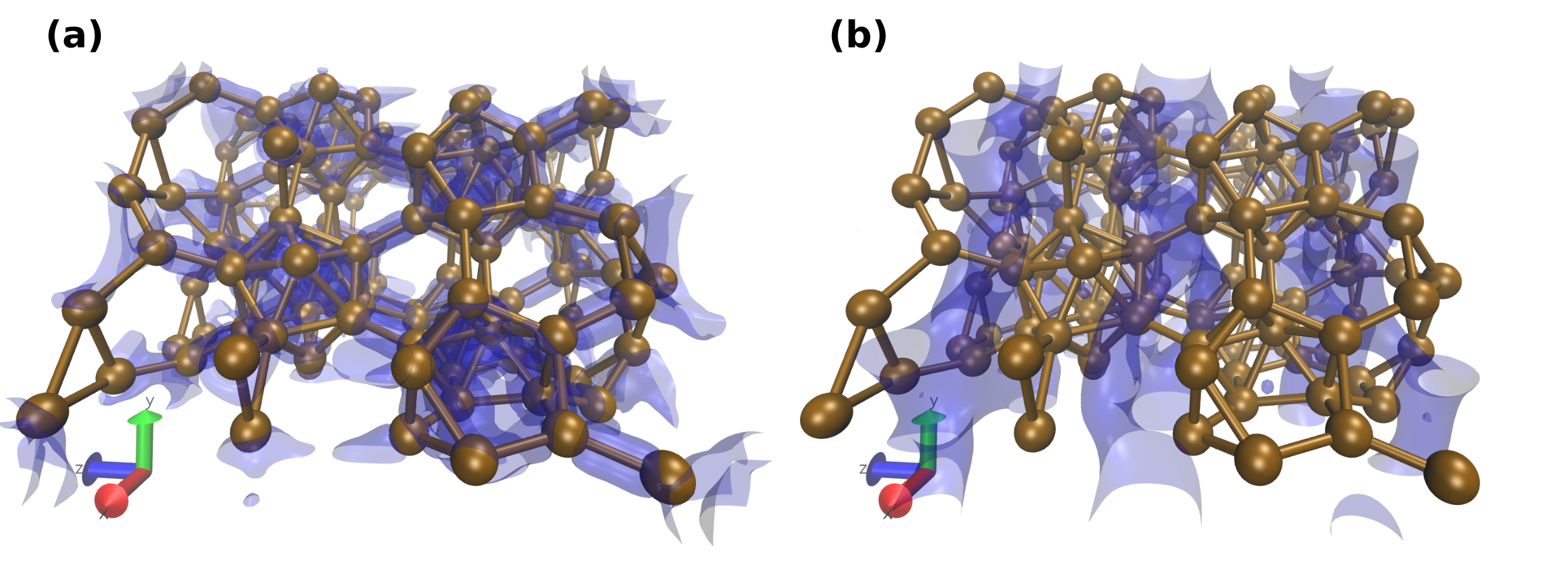}	\caption{Visualization of regions with high \textbf{(a)} and low \textbf{(b)} electronic density in a boron simulation cell containing 144 atoms. The electronic density is represented by blue isosurfaces, while boron atoms are shown in brown. Bonds are added between boron atoms to highlight the ionic structure, which consists of interconnected  $B_{12}$ clusters.}
\label{fig:boron_density_viz}
\end{figure}

Another important insight from Fig.~\ref{fig:B298_Density} concerns the spatial structure of the electronic density. Compared to similar visualizations for aluminum~\cite{malapaper,Lenz_Dissertation}, where most grid points correspond to density values of intermediate magnitude, the boron case shows that the majority of grid points correspond to relatively low electronic density values, with a broader range of density values overall. This difference arises from the $\alpha$-rhombohedral boron structure, where atoms form $B_{12}$ clusters. Near these clusters, the electronic density is high, while between clusters, the density localizes to much smaller values. An example of this behavior is shown in Fig.~\ref{fig:boron_density_viz}.

This spatial distribution of electronic density represents a notable difference from the metallic systems previously treated with \texttt{MALA} in Refs.~\citenum{malapaper}, \citenum{hyperparameterpaper}, \citenum{temperaturepaper}, \citenum{sizetransferpaper}, and \citenum{Lenz_Dissertation}. The DOS of boron, shown in Fig.~\ref{fig:B298_DOS}, also differs significantly, displaying a multitude of distinct peaks and features. This increased need for spatially and energetically accurate LDOS reconstruction leads to slightly reduced accuracy compared to results reported for metallic systems. Nonetheless, without further tuning of the model or bispectrum hyperparameters, a largely accurate model of $\alpha$-rhombohedral boron has been trained with \texttt{MALA}, using only a modest amount of \texttt{Python} code.

With the model’s accuracy and application boundaries established, it can now be employed in a modeling workflow. Currently, there are two main strategies for applying \texttt{MALA} models. First, \texttt{MALA} models can be queried directly using the standard \texttt{MALA} interface for LDOS and derived properties. This can be achieved with:

\begin{lstlisting}[language=Python]
from ase.io import read
import mala

# Load a trained network.
parameters, network, data_handler, predictor = mala.Predictor.load_run(
    run_name=..., path=...
)
parameters.targets.pseudopotential_path = ...

# Predict the LDOS from some atomic configuration.
atoms = read(...)
ldos = predictor.predict_for_atoms(atoms)

# Read the LDOS into a calculator.
ldos_calculator: mala.LDOS = predictor.target_calculator
ldos_calculator.read_from_array(ldos)

# Predict properties from the LDOS.
band_energy = ldos_calculator.band_energy
total_energy = ldos_calculator.total_energy
density = ldos_calculator.density
dos = ldos_calculator.density_of_states
\end{lstlisting}

The \lstinline|Predictor| class contains all computations for \texttt{MALA} model inference, yielding the LDOS. Derived properties, such as electronic density and electronic DOS, are accessible as properties of the \texttt{MALA} LDOS calculator object. Current research efforts are focused on extending the range of directly accessible observables, such as band structure and electrical conductivity. Direct access to the LDOS can also be valuable in the context of scanning tunnelling microscopy applications.

\texttt{MALA} models can also be used within an Atomic Simulation Environment (ASE)~\cite{ASE} workflow. ASE is a powerful library of \texttt{Python} routines and classes for general materials modeling, providing a unified interface for representing atomic structures and performing electronic structure calculations. These calculations are handled by \lstinline|Calculator| objects, which provide functions for calculating properties such as total free energy or ionic forces. A \lstinline|Calculator| can be implemented via file-based interfaces to popular DFT codes such as \texttt{Quantum ESPRESSO} or via a direct \texttt{Python} interface. \texttt{MALA} includes an ASE-compatible \lstinline|Calculator| interface, accessible with:

\begin{lstlisting}[language=Python]
from ase.io import read
import mala

# Load a trained network.
calculator = mala.load_run(run_name=..., path=...)
calculator.mala_parameters.targets.pseudopotential_path = ...

# Use the ASE calculator formalism to model a system.
atoms = read(...)
atoms.set_calculator(calculator)
mala.printout(atoms.get_potential_energy())
\end{lstlisting}

With the ASE interface, DFT-MD simulations are theoretically possible, although in the current version, ionic forces are not fully implemented within \texttt{MALA} (see Sec.~\ref{ss.observables}). There are, however, no conceptual barriers to performing MD simulations using \texttt{MALA} models through the ASE interface. Since \texttt{MALA} relies on \texttt{LAMMPS} for computationally intensive tasks, an interface to \texttt{LAMMPS} for MD simulations is also in development.

\section{Numerical benchmarks}
\label{s.benchmarks}

\subsection{Accuracy and transferability}
\label{ss.accuracy}

 We illustrate the capabilities of the \texttt{MALA} machine learning model with three specific examples that demonstrate transferability across length scales, temperature range and across the solid-liquid phase boundary. 

\subsubsection{Transferability across length scales}
We demonstrate the ability of the \texttt{MALA} machine learning model to make accurate predictions on larger length scales than those used in training. This transferability across length scales is achieved through the specific design of the \texttt{MALA} model, which uses local descriptors of the atomic configuration around a grid point in Cartesian space to predict the LDOS at that point. By using the principle of nearsightedness in electronic structure~\cite{Ko96,PrKo05}, we exploit the fact that a perturbation in the electronic structure at a given point has a finite spatial range that decays with distance, provided the system is weakly correlated. This property allows us to train the \texttt{MALA} model on local electronic environments and then apply it effectively to much larger systems.

To demonstrate this, we study beryllium atoms arranged in hcp supercells at room temperature and ambient mass density (1.896$,\mathrm{g\,cm^{-3}}$). The training and validation data set consists of 256 atoms and includes data from DFT simulations together with corresponding LDOS data. A \texttt{MALA} model is trained on this dataset and its predictive accuracy is evaluated by comparing its total free energy and electronic density predictions with DFT results for systems containing 512, 1024 and 2048 atoms. For each system size, the \texttt{MALA} model performs inference on ten different atomic configurations.
\begin{figure}[ht]
\centering
\includegraphics[width=1\linewidth]{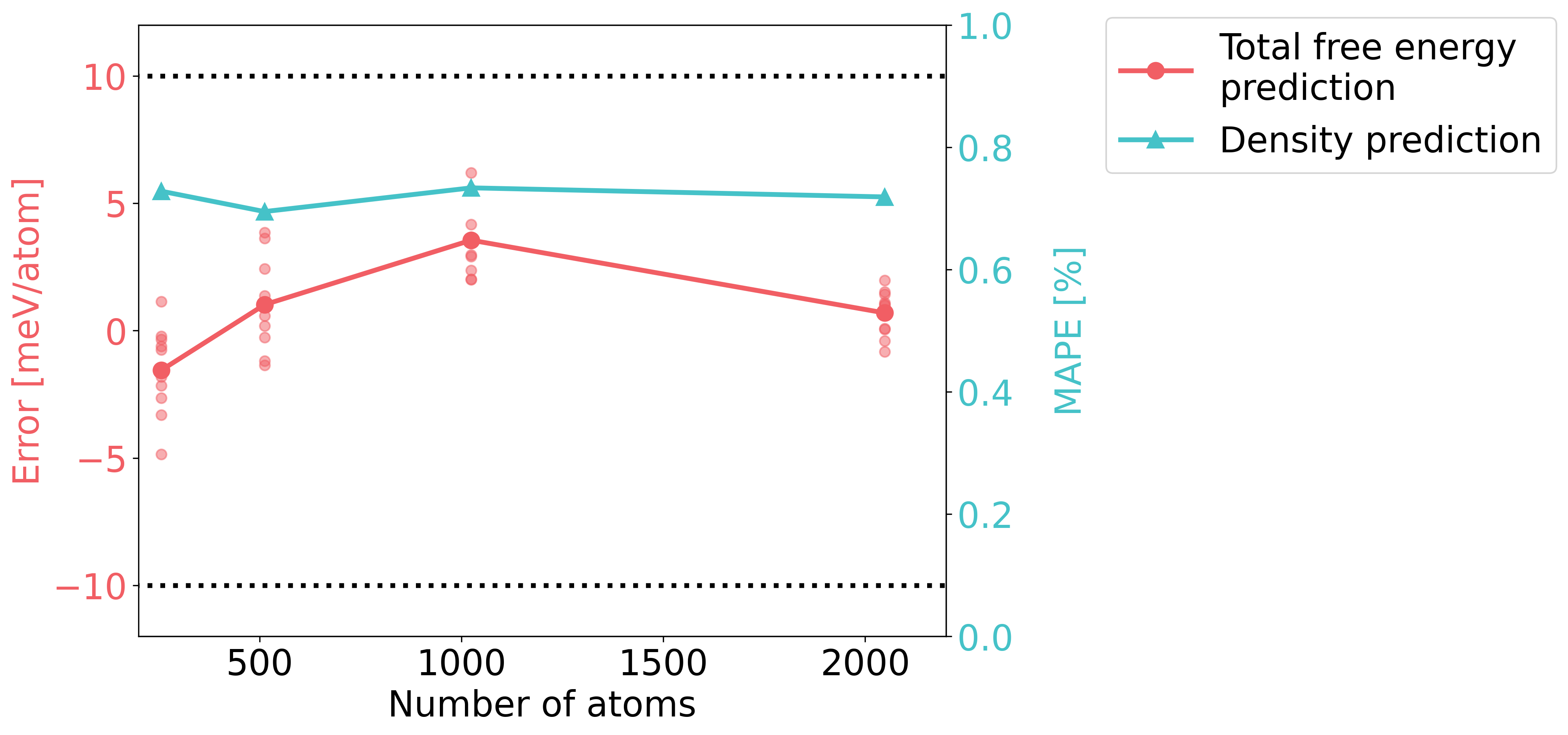}
\caption[Prediction errors for system size transfer.]{\texttt{MALA} model prediction errors are shown for a beryllium model trained on data from a 256-atom system and applied to larger systems. The red curve represents the total free energy error, and the blue curve represents the mean absolute percentage error (MAPE) of the electronic density prediction. For each system size, the total free energy error was evaluated over ten atomic configurations, with the solid red points indicating the mean error. Transparent dots represent individual configuration errors, illustrating the error distribution. The dotted black lines mark an absolute error threshold of $10 \,\mathrm{meV/atom}$, commonly considered acceptable for machine learning interatomic potentials.}	
\label{fig:sizetransfer_accuracy_1}
\end{figure}  

The results, presented in Fig.~\ref{fig:sizetransfer_accuracy_1}, show that the \texttt{MALA} model meets accuracy criteria for both total free energy and electronic density. 
Specifically, the total free energy error remains well below $10 \,\mathrm{meV/atom}$ across all $N\ion$, and the mean absolute percentage error (MAPE) of the electronic density does not exceed 1\%.

More significant than these absolute accuracy levels are the relative trends observed. For both metrics there is no clear increase or decrease in error as the number of atoms increases. The density MAPE remains almost constant with an almost entirely flat trend, while for the total free energy the error first increases slightly up to 1024 atoms and then decreases. This lack of a clear correlation between prediction error and system size underlines the feasibility of applying \texttt{MALA} models trained on smaller systems to significantly larger scales.

\begin{figure}[ht]
\centering
\includegraphics[width=1\linewidth]{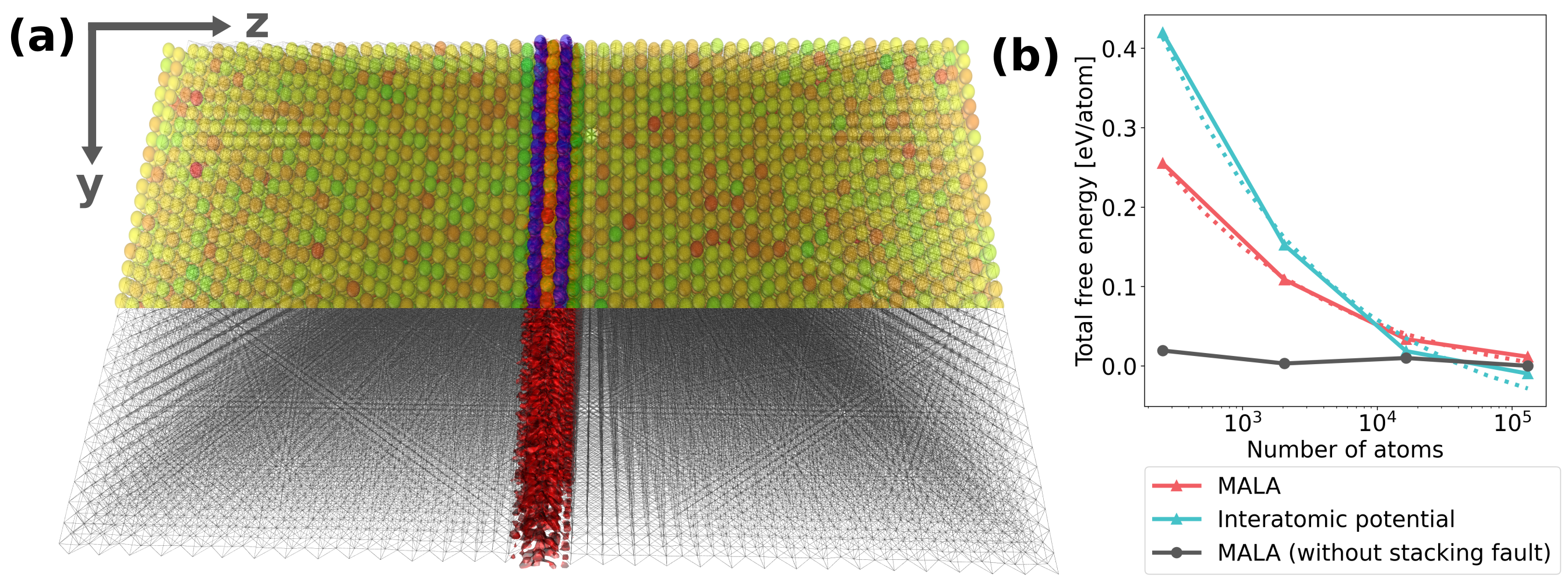}
\caption{Illustration of \texttt{MALA} predictions for a slab of beryllium with a stacking fault. Panel (a) displays a simulation cell containing 131,072 beryllium atoms, with a stacking fault localized in the center of the slab, and the corresponding electronic density (red); Panel (b) presents total free energies for simulation cells with a stacking fault, predicted using a \texttt{MALA} model (red), an interatomic potential (blue), and an ${N\ion}^{-\frac{1}{3}}$ fit as a dotted line. Additionally, the \texttt{MALA} predictions without a stacking fault are shown (gray). Figure adapted from original research published in Ref.~\citenum{sizetransferpaper}.}	
\label{fig:sizetransfer_stackingfault}
\end{figure}  

Next, we demonstrate the ability of \texttt{MALA} models to predict electronic structure on length scales beyond the reach of standard DFT calculations. We consider a slab of beryllium containing 131,072 atoms at room temperature. Fig.~\ref{fig:sizetransfer_stackingfault} (a) illustrates a beryllium slab in which a stacking fault has been introduced by locally changing the hcp crystal symmetry of the bulk (yellow, green and orange) to an fcc geometry (blue) in the center of the slab.

We apply the same \texttt{MALA} model, trained on atomic snapshots of 256 atoms, to predict the electronic structure of this large beryllium slab, including the stacking fault. The results show the difference in electronic density between a slab with a stacking fault and one without highlighted by the red point cloud in the middle bottom half of the figure, which is clearly localized around the stacking fault.

In addition to predicting the electronic density, we also assess whether the \texttt{MALA} model correctly captures the energetics. Although DFT reference data is not available for this large system, we can still assess the qualitative accuracy of the energetics by examining the energy trends in slabs with and without a stacking fault over different system sizes. Fig.~\ref{fig:sizetransfer_stackingfault}(b) shows the total energy per atom as a function of atomic number for beryllium slabs with and without a stacking fault. For slabs without a stacking fault, the total free energy per atom should remain approximately constant, as shown by the grey curve. For slabs with a stacking fault, dimensional analysis predicts a polynomial scaling behavior according to $N\ion^{-1/3}$. As shown in the figure, the \texttt{MALA} model (red) reproduces this expected behavior. In addition, we show that an interatomic potential \citenum{daw_embedded-atom_1984,BeEAM} used to generate the atomic configurations of the beryllium slab follows the same qualitative trend.

\subsubsection{Transferability across a phase boundary}

We investigate the transferability of \texttt{MALA} models across the solid-liquid phase boundary in a material. To this end, we consider 20 atomic snapshots of aluminum at ambient mass density 2.699$,\mathrm{g\,cm^{-3}}$ at the melting point, which is 933$\,\mathrm{K}$. At the melting point, we distinguish between atomic snapshots that are solid-like, retaining a high degree of crystal symmetry, and liquid-like snapshots, which exhibit significant structural disorder.

We train a \texttt{MALA} machine learning model using both solid-like and liquid-like snapshots, with ten atomic snapshots each for training, validation, and testing. In the following, we assess whether this model can make accurate predictions of the electronic structure and corresponding energetics for unseen snapshots. The following results are similar to those published in Ref.~\citenum{malapaper}, but using updated datasets and \texttt{MALA} models \citenum{Lenz_Dissertation}.

\begin{figure}[ht!]
\centering
\includegraphics[width=1\linewidth]{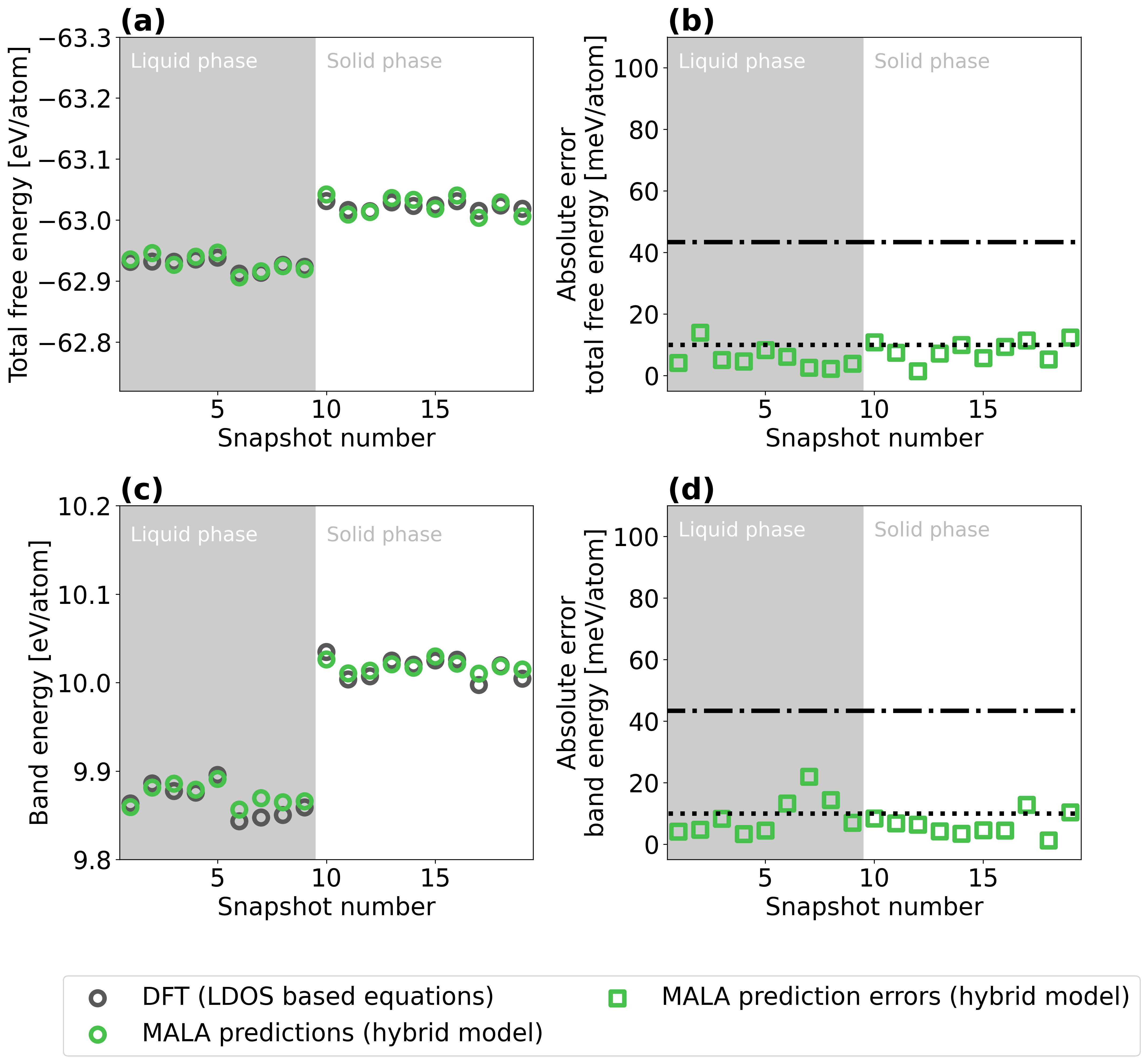}
\caption{Energy inference errors of the \texttt{MALA} model trained on multiple phases of aluminum at the melting point. The model was trained using six solid and six liquid configurations. Panels \textbf{(a)} and \textbf{(c)} display predicted versus actual values for the total free energy and band energy, while panels \textbf{(b)} and \textbf{(d)} show the corresponding prediction errors. The dotted line at $10 \,\mathrm{meV/atom}$ indicates the threshold commonly used to label a model as accurate, while the dashed-dotted line represents chemical accuracy, defined as $43.4,\mathrm{meV/atom}$. Figure adapted from original research published in Ref.~\citenum{malapaper}, with updated results given in Ref.~\citenum{Lenz_Dissertation}.}	
\label{fig:phaseboundary_multi_phase_results}
\end{figure} 

The results on the energetics are shown in Fig.~\ref{fig:phaseboundary_multi_phase_results}. The data illustrates that incorporating both phases yields a \texttt{MALA} model capable of accurate electronic structure prediction for either phase. In general, both the predicted total free energies and band energies in Fig.~\ref{fig:phaseboundary_multi_phase_results}\textbf{(a)} and Fig.~\ref{fig:phaseboundary_multi_phase_results}\textbf{(c)} agree well with the actual energies calculated using DFT. Although the $10 \,\mathrm{meV/atom}$ threshold is slightly exceeded for a few atomic configurations, the average prediction error lies below it. Another criterion for accuracy is the energy scale associated with the phase transition from the solid to the liquid phase. Here the gap is about $95.63,\mathrm{meV/atom}$ for the total free energy and $151.66,\mathrm{meV/atom}$ for the band energy, which are easily resolved given the inference errors of the \texttt{MALA} model.

\begin{figure}[ht!]
\centering
\includegraphics[width=1\linewidth]{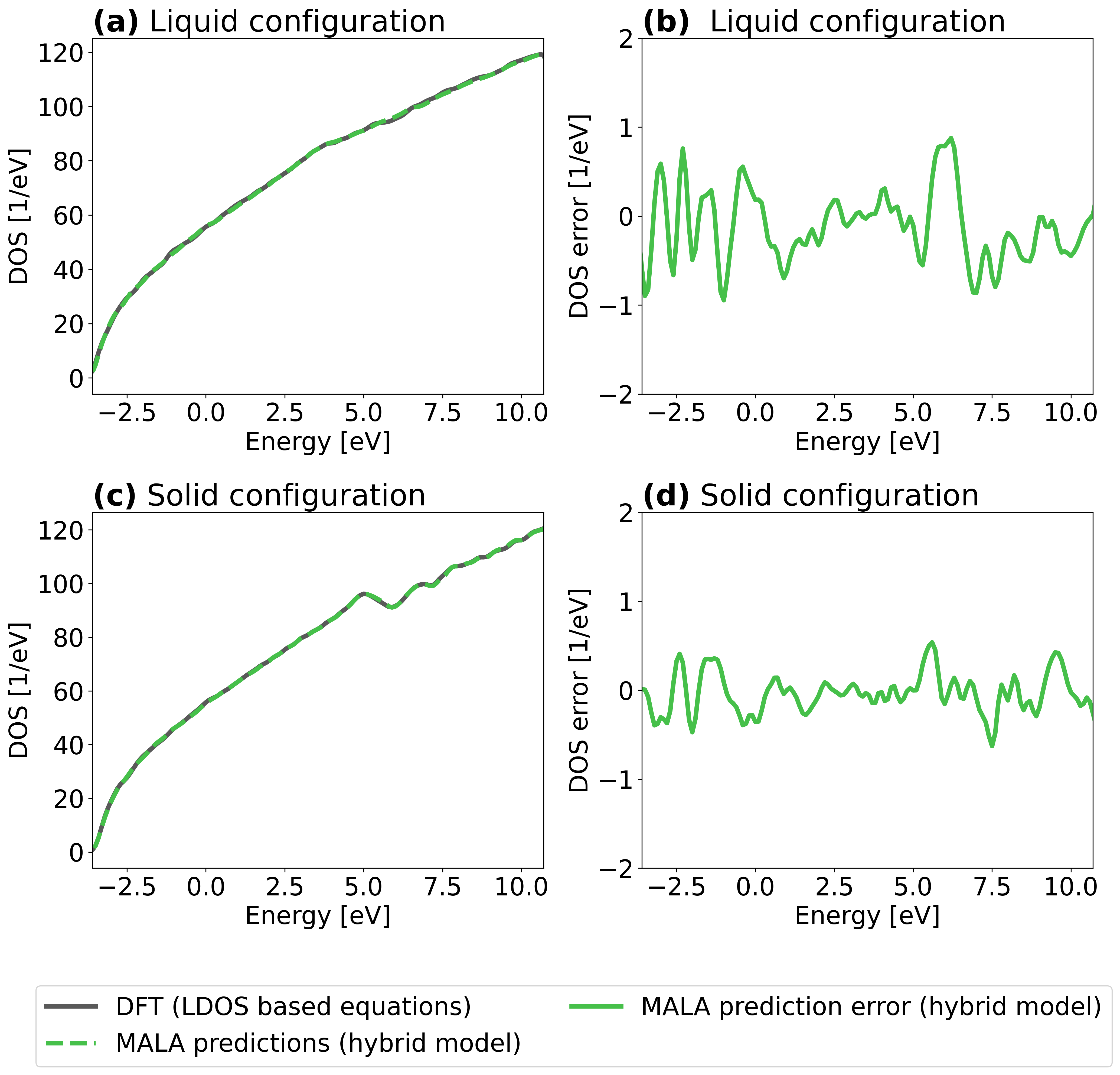}
\caption{DOS predictions and errors for a liquid-like atomic configuration (snapshot 7) and a solid-like atomic configuration (snapshot 17) of the same model shown in Fig.~\ref{fig:phaseboundary_multi_phase_results}. Panels \textbf{(a)} and \textbf{(c)} display the actual and predicted DOS, while panels \textbf{(b)} and \textbf{(d)} show the corresponding DOS prediction errors. Figure adapted from original research published in Ref.~\citenum{malapaper}, with updated results given in Ref.~\citenum{Lenz_Dissertation}.}	
\label{fig:phaseboundary_dos_results}
\end{figure}  

\begin{figure}[ht!]
\centering
\includegraphics[width=0.95\linewidth]{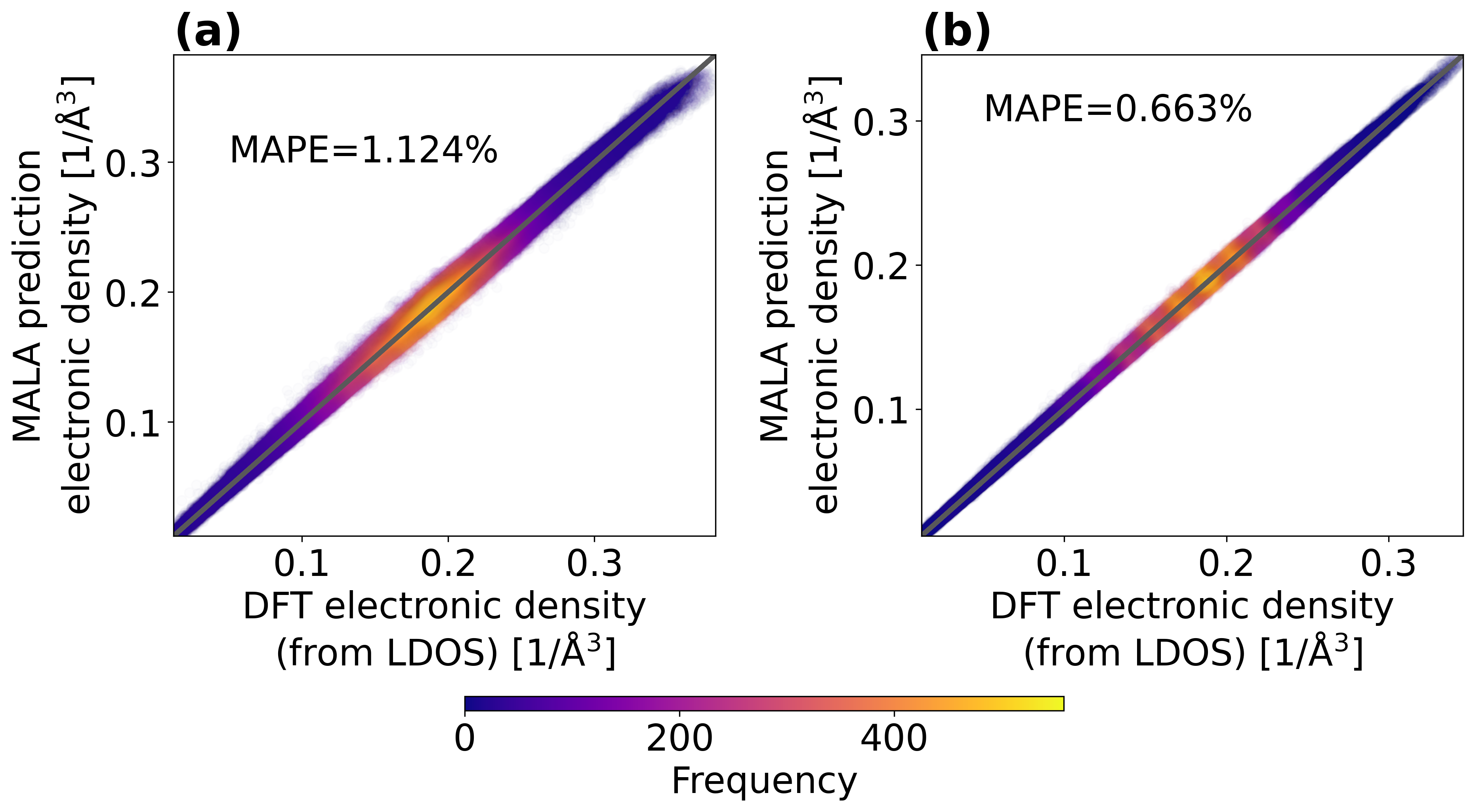}
\caption{Visualization of density predictions and corresponding MAPE for a liquid-like atomic configuration (snapshot \#7) and a solid-like atomic configuration (snapshot \#17) using the same \texttt{MALA} model as in Fig.~\ref{fig:phaseboundary_multi_phase_results}. Both panels show the electronic density predicted by \texttt{MALA} compared to the actual electronic density, represented as a colored point cloud. The color bar indicates the frequency of individual values within the point cloud, illustrating the distribution. An ideal prediction is indicated by the dark gray line. Figure adapted from original research published in Ref.~\citenum{malapaper}, with updated results given in Ref.~\citenum{Lenz_Dissertation}.}
\label{fig:phaseboundary_density_results}
\end{figure}

We further assess the accuracy of the \texttt{MALA} model by examining both the prediction of the DOS and the electronic density. To this end, Fig.~\ref{fig:phaseboundary_dos_results} and Fig.~\ref{fig:phaseboundary_density_results} illustrate DOS and density prediction errors for one atomic configuration corresponding to the liquid phase and one corresponding to the solid phase. Considering the DOS in Fig.~\ref{fig:phaseboundary_dos_results}\textbf{(a)} and \textbf{(c)}, we observe high accuracy in both phases, with only minor errors. Notably, the DOS prediction accuracy is slightly higher for the solid phase (see Fig.~\ref{fig:phaseboundary_dos_results}\textbf{(d)}) than for the liquid phase (see Fig.~\ref{fig:phaseboundary_dos_results}\textbf{(b}).

A similar trend is observed in the density predictions shown in Fig.~\ref{fig:phaseboundary_density_results}, which displays a correlation plot of the actual and the predicted electronic density values on each spatial grid point, yielding a point cloud. Fig.~\ref{fig:phaseboundary_density_results}\textbf{(a)} gives this correlation for a liquid and Fig.~\ref{fig:phaseboundary_density_results}\textbf{(b)} for a solid configuration. It is clearly evident that overall, the electronic density is predicted accurately for either phase, with slightly higher accuracy for the solid phase. This is likely because the liquid phase exhibits greater spatial variation in electronic density, corresponding to the larger spread along the $x$-axis of the point cloud in Fig.~\ref{fig:phaseboundary_density_results}\textbf{(a)}. This trend is further quantified via the MAPE, also provided in Fig.~\ref{fig:phaseboundary_density_results}. The MAPE only slightly exceeds 1\% for the liquid phase prediction and is even lower for the solid phase.

\subsubsection{Transferability across a temperature range}
As a final example, we demonstrate the transferability of \texttt{MALA} models across a temperature range. To this end, we construct a \texttt{MALA} model using training data from solid aluminum snapshots spanning temperatures from $100\,\mathrm{K}$ to $933\,\mathrm{K}$ in increments of $100\,\mathrm{K}$. An ensemble of five models with different initializations was trained to assess model robustness across this temperature range. Investigating this robustness is particularly important for temperature transferability, as greater data variability may arise from increasing temperatures and thermal fluctuations. In all DFT calculations used to generate the training data, the ionic and electronic temperatures were set to the same value, i.e., $\tau=\tau\ion$. Note that the following results build upon those published in Ref.~\citenum{temperaturepaper}, but with updated datasets and \texttt{MALA} models.

\begin{figure}[ht!]
\centering
\includegraphics[width=1\linewidth]{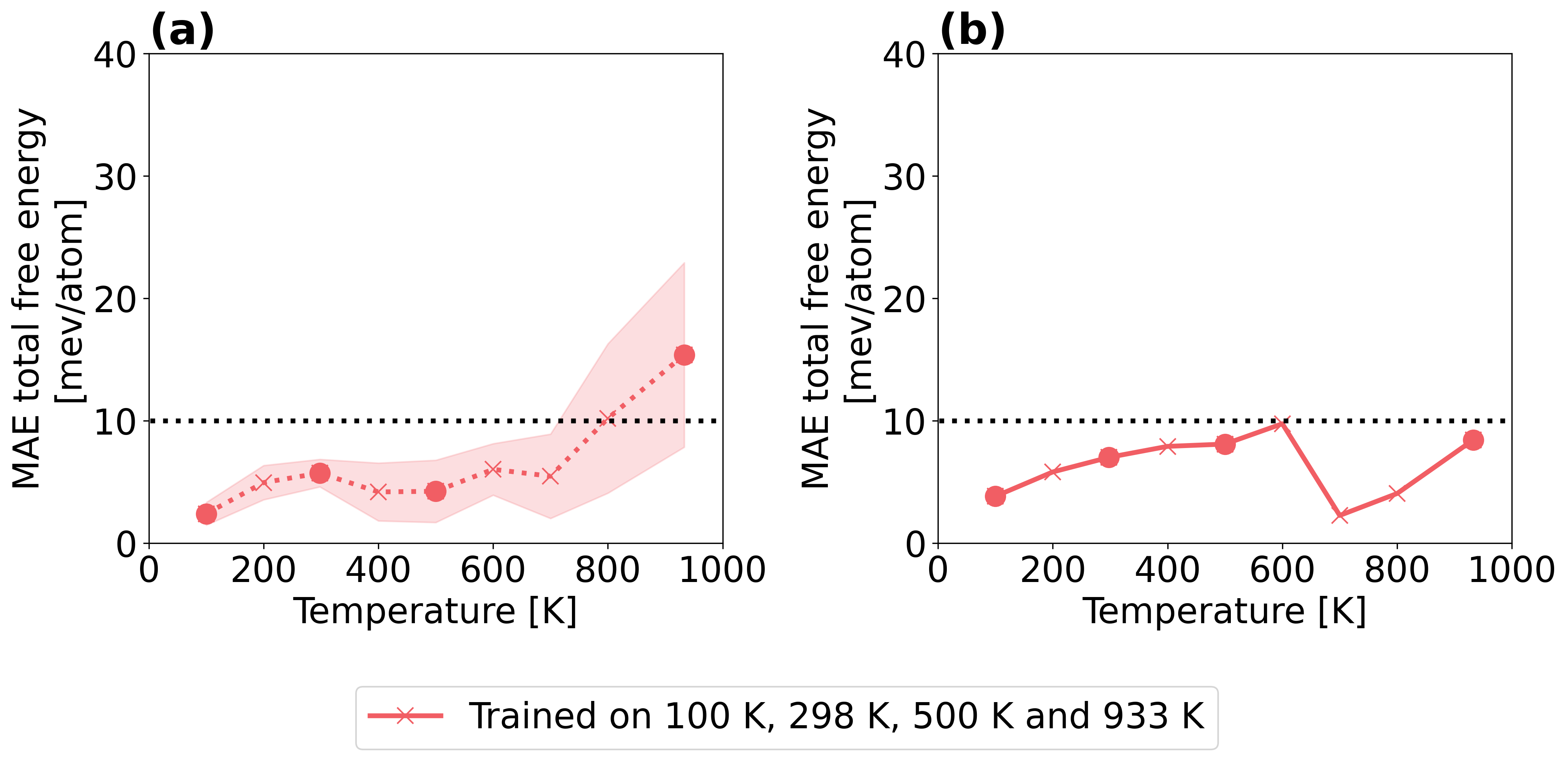}
\caption{Total free energy errors across a temperature range for a \texttt{MALA} model trained on data sampled at multiple temperatures. An ensemble of five models was trained, using four atomic configurations per training temperature. Panel \textbf{(a)} shows the ensemble average (dotted line) and standard deviation (shaded area), while panel \textbf{(b)} displays the performance of the best model, corresponding to the lowest overall inference error. The dotted horizontal lines indicate the common accuracy threshold for ML interatomic potentials of $10 \,\mathrm{meV/atom}$. Figure adapted from original research published in Ref.~\citenum{temperaturepaper}, with updated results \citenum{Lenz_Dissertation}.}	
\label{fig:temptransfer_superhybrid_results}
\end{figure}

The prediction errors for the \texttt{MALA} model trained on four temperatures ($100 \,\mathrm{K}$, $298 \,\mathrm{K}$, $500 \,\mathrm{K}$, and $933 \,\mathrm{K}$) are presented in Fig.~\ref{fig:temptransfer_superhybrid_results}. We observe competitive accuracy across the entire temperature range. In the best-performing model within the ensemble, shown in Fig.~\ref{fig:temptransfer_superhybrid_results}\textbf{(b)}, the $10 \,\mathrm{meV/atom}$ threshold holds consistently across all configurations. Notably, only a slight increase in average error with respect to temperature is observed, indicating that models trained in this manner sufficiently capture the electronic structure across the entire temperature range of interest. 

\subsection{Computational scaling}
\label{ss.scaling}

Having established the general transferability of \texttt{MALA} models, we can now examine their scalability properties. In the following, we discuss the scaling behavior of multi-GPU inferences within the \texttt{MALA} package. Note that scaling results for training routines are omitted here, as \texttt{MALA} uses a standard \texttt{PyTorch DDP} implementation for which the general observations about scaling performance apply~\cite{PytorchDDPTest1,PytorchDDPTest2}.

To provide context for the inference scalability results, an overview of the typical computational cost associated with model training is useful. Table~\ref{tab:model_costs} presents the training costs in GPU hours for published \texttt{MALA} models, as reported in Ref.~\citenum{Lenz_Dissertation}. These models utilize architectures originally introduced in Ref.~\citenum{malapaper}.

\begin{table}
\centering  
\caption{Average computational cost for training \texttt{MALA} models. The computational cost is averaged across multiple network initializations. All models were trained using a single Nvidia V100 GPU per model, except for the two-phase aluminum models at the melting point, which were trained on a single Nvidia A100 GPU per model.}\label{tab:model_costs}		
	\begin{tabularx}{\textwidth}{Xr}%cccccccc}
	\toprule
	\textbf{Model Type} & \textbf{Cost}  [GPUh]  \ \\\midrule
	Aluminum room temperature model~\cite{malapaper}  & $3.37$  \\
 	Beryllium room temperature model (128 atoms)~\cite{hyperparameterpaper}  & $1.08$  \\
    Beryllium room temperature model (256 atoms)~\cite{sizetransferpaper}  & $0.86$ 
     \\
     Beryllium room temperature model (256 atoms)~\cite{sizetransferpaper}  & $0.86$ 
     \\
     Aluminum melting temperature (933 K, liquid phase)~\cite{malapaper}  & $9.26$ 
     \\
     Aluminum melting temperature (933 K, solid phase)~\cite{malapaper}  & $1.90$ 
     \\
     Aluminum melting temperature (933 K, both phases)~\cite{malapaper}  & $9.59$ 
     \\
     Aluminum temperature transfer single-temperature model (100 K)~\cite{temperaturepaper}  & $3.24$ 
     \\
     Aluminum temperature transfer single-temperature model (500 K)~\cite{temperaturepaper}  & $2.19$ 
     \\
     Aluminum temperature transfer single-temperature model (933 K, solid phase)~\cite{temperaturepaper}  & $2.05$ 
     \\
     Aluminum temperature transfer two-temperature model (100 K, 933 K)~\cite{temperaturepaper}  & $7.51$ 
     \\
     Aluminum temperature transfer three-temperature model (100 K, 500 K, 933 K)~\cite{temperaturepaper}  & $15.19$
     \\
     Aluminum temperature transfer four-temperature model (100 K, 298 K, 500 K, 933 K)~\cite{temperaturepaper}  & $14.19$ 
     \\
     Aluminum temperature transfer optimized four-temperature model (100 K, 298 K, 500 K, 933 K)~\cite{temperaturepaper}  & $24.21$ \ \\\bottomrule
\end{tabularx}
\end{table}

In Tab.~\ref{tab:model_costs}, it is evident that the computational costs for neural network model training generally range from a few GPU hours up to approximately 24 GPU hours for the largest models. As these values represent the total training cost, models can typically be ready for use within a day of initiating the training procedure. This makes the time required for neural network training a relatively minor overhead in the \texttt{MALA} workflow. In contrast, data generation is often a more complex and time-intensive process. As described in Ref.~\citenum{sizetransferpaper}, DFT simulations for a single snapshot can take up to nearly a full day of wall time, with preceding DFT-MD simulations potentially requiring several days. However, with parallelization across multiple snapshots and trajectories, data generation can usually be completed within a few days of wall time. Furthermore, because the data generation process relies on standard DFT(-MD) workflows, existing approaches for automation and acceleration can be leveraged to streamline the process~\cite{aiida1,aiida2,aiida3}.

\begin{figure}[htp]
\centering
\includegraphics[width=0.95\columnwidth]{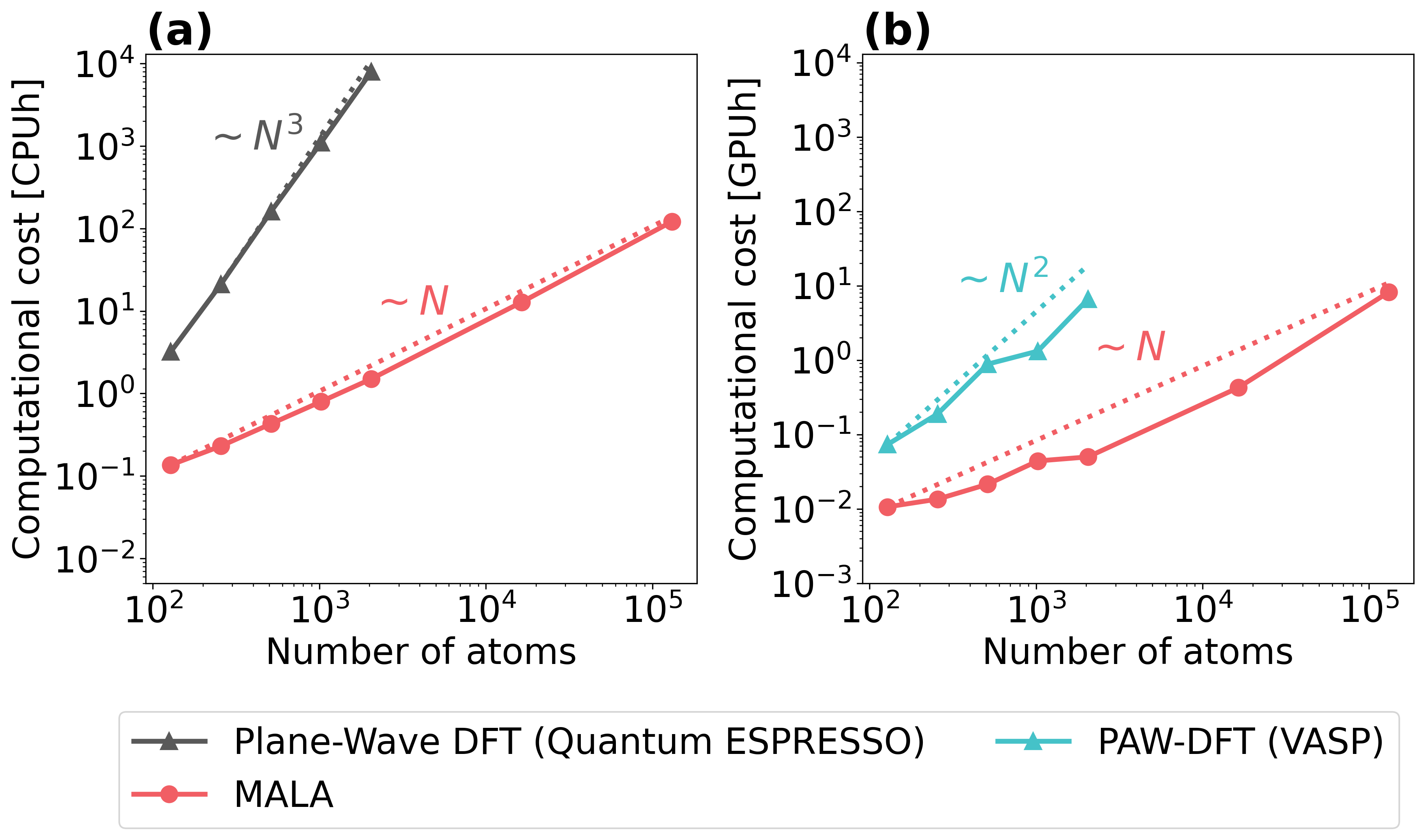}
\caption{Computational cost for DFT and \texttt{MALA} total energy predictions as a function of the number of beryllium atoms $N$. \textbf{(a)}: Comparison of \texttt{MALA} to standard CPU-based DFT simulations (plane-wave DFT calculations using \texttt{Quantum ESPRESSO}). \textbf{(b)}: Comparison of \texttt{MALA} to GPU-accelerated DFT simulations (projector augmented wave method DFT calculations using \texttt{VASP}). For a meaningful comparison, \texttt{MALA} inferences were conducted with multiple CPUs in \textbf{(a)} and in a GPU-accelerated manner in \textbf{(b)}.}
\label{fig:gpu_scaling_beryllium}
\end{figure}

As with any machine learning approach, the computational cost of model creation is justified by either a reduction in inference time for similar applications or the potential to enable new applications altogether. For \texttt{MALA} models, this benefit is particularly relevant given the scaling properties of DFT. Typically, DFT scales cubically with both system size and temperature~\cite{ofdftpaper}. However, the specific scaling behavior with respect to system size varies depending on the DFT implementation and calculation parameters, with sub-cubic scaling possible up to a certain atom count. For example, Fig.~\ref{fig:gpu_scaling_beryllium}\textbf{(b)} demonstrates that the efficient GPU implementation of the \texttt{VASP} code scales quadratically with the number of atoms. In some cases, linear scaling can be achieved in DFT simulations through approximations, as in linear-scaling DFT~\cite{LinearScaling1,LinearScaling2,LinearScaling3,LinearScaling4,LinearScaling5} or orbital-free DFT~\cite{ligneres_introduction_2005,chen_orbital-free_2008,wesolowski_recent_2012,karasiev_progress_2014}. However, since these methods rely on specific assumptions and approximations, they are not the focus of the comparisons presented here. Instead, this analysis centers on demonstrating the linear dependence of \texttt{MALA} calculations on system size in comparison to standard DFT simulations, which generally scale cubically or quadratically with atom count.

Additionally, \texttt{MALA} model inference times are unaffected by temperature, as electronic temperature effects are encoded within the (L)DOS, as discussed in Sec.~\ref{ss.accuracy}. Thus, they are not covered in this section. Results showing system size scaling are presented in Fig.~\ref{fig:gpu_scaling_beryllium}, with detailed computational parameters listed in Tab.~\ref{tab:details_cost_analysis_gpu} (GPU setup) and Tab.~\ref{tab:details_cost_analysis_cpu} (CPU setup).

\begin{table}
\centering  
\caption{Computational details for GPU scaling experiments on computational cost: All experiments were conducted on beryllium, with results presented in Fig.~\ref{fig:gpu_scaling_beryllium}\textbf{(a)}.}
\label{tab:details_cost_analysis_gpu}		
\begin{tabularx}{\textwidth}{XXXXXXX}%cccccccc}
\toprule
Number of Atoms &Number of gridpoints &Number of GPUs &Gridpoints per atom &Gridpoints per GPU \ \\\midrule
128 &    $6.22\times10^5$ & 4 & 4860 &  $1.56\times10^5$ \\
256 &    $1.24\times10^6$ & 4 & 4860 &  $3.11\times10^5$ \\
512 &    $2.49\times10^6$ & 4 & 4860 &  $6.22\times10^5$ \\
1024 &   $4.67\times10^6$ & 4 & 4556 &  $1.17\times10^6$ \\
2048 &   $9.33\times10^6$ & 9 & 4556 &  $1.04\times10^6$ \\
16384 &  $7.46\times10^7$ & 15 & 4556 & $4.98\times10^6$ \\
131072 & $3.01\times10^8$ & 30 & 2298 & $1.00\times10^7$ \\\bottomrule
\end{tabularx}
\end{table}

\begin{table}
\centering  
\caption{Computational details for CPU scaling experiments on computational cost: All experiments were conducted on beryllium, with results shown in Fig.~\ref{fig:gpu_scaling_beryllium}\textbf{(b)}.}
\label{tab:details_cost_analysis_cpu}		
\begin{tabularx}{\textwidth}{XXXXXXX}%cccccccc}
\toprule
Number of Atoms &Number of gridpoints &Number of CPUs &Gridpoints per atom &Gridpoints per CPU \ \\\midrule
128 &    $6.22\times10^5$& 24 & 4860 &  $2.59\times10^4$ \\
256 &    $1.24\times10^6$& 24 & 4860 &  $5.18\times10^4$ \\
512 &    $2.49\times10^6$& 24 & 4860 &  $1.04\times10^5$ \\
1024 &   $4.67\times10^6$& 45 & 4556 &  $1.04\times10^5$ \\
2048 &   $9.33\times10^6$& 45 & 4556 &  $2.07\times10^5$ \\
16384 &  $7.46\times10^7$& 150 & 4860 & $4.98\times10^5$ \\
131072 & $3.01\times10^8$& 150 & 4860 & $2.01\times10^6$  \\\bottomrule
\end{tabularx}
\end{table}

The \texttt{MALA} inference cost is compared with two popular DFT implementations: plane-wave DFT, as implemented in software such as \texttt{Quantum ESPRESSO}, and the projector augmented wave (PAW) method DFT, as implemented, for example, in \texttt{VASP}. For the latter, a GPU-accelerated version is used to establish an higher efficiency limit for standard DFT simulations. In this setup, the GPU implementation of \texttt{VASP} can achieve quadratic scaling under certain conditions.

In Fig.~\ref{fig:gpu_scaling_beryllium}, it is evident that \texttt{MALA} inferences not only scale linearly, contrasting with the cubic or quadratic scaling behaviors of DFT, but also that the baseline computational cost of \texttt{MALA} predictions is up to two orders of magnitude lower than that of DFT. Additionally, Fig.~\ref{fig:gpu_scaling_beryllium} shows that multi-GPU \texttt{MALA} inferences enable simulations involving hundreds of thousands of atoms, at a computational cost of only a few GPU hours. This cost is comparable to that of standard DFT simulations for systems of only a few thousand atoms. Even advanced linear-scaling DFT approaches~\cite{LinearScaling1,LinearScaling3} are an order of magnitude more expensive, as discussed in Ref.~\citenum{sizetransferpaper}.

\begin{figure}[htp]
\centering    
\includegraphics[width=0.95\columnwidth]{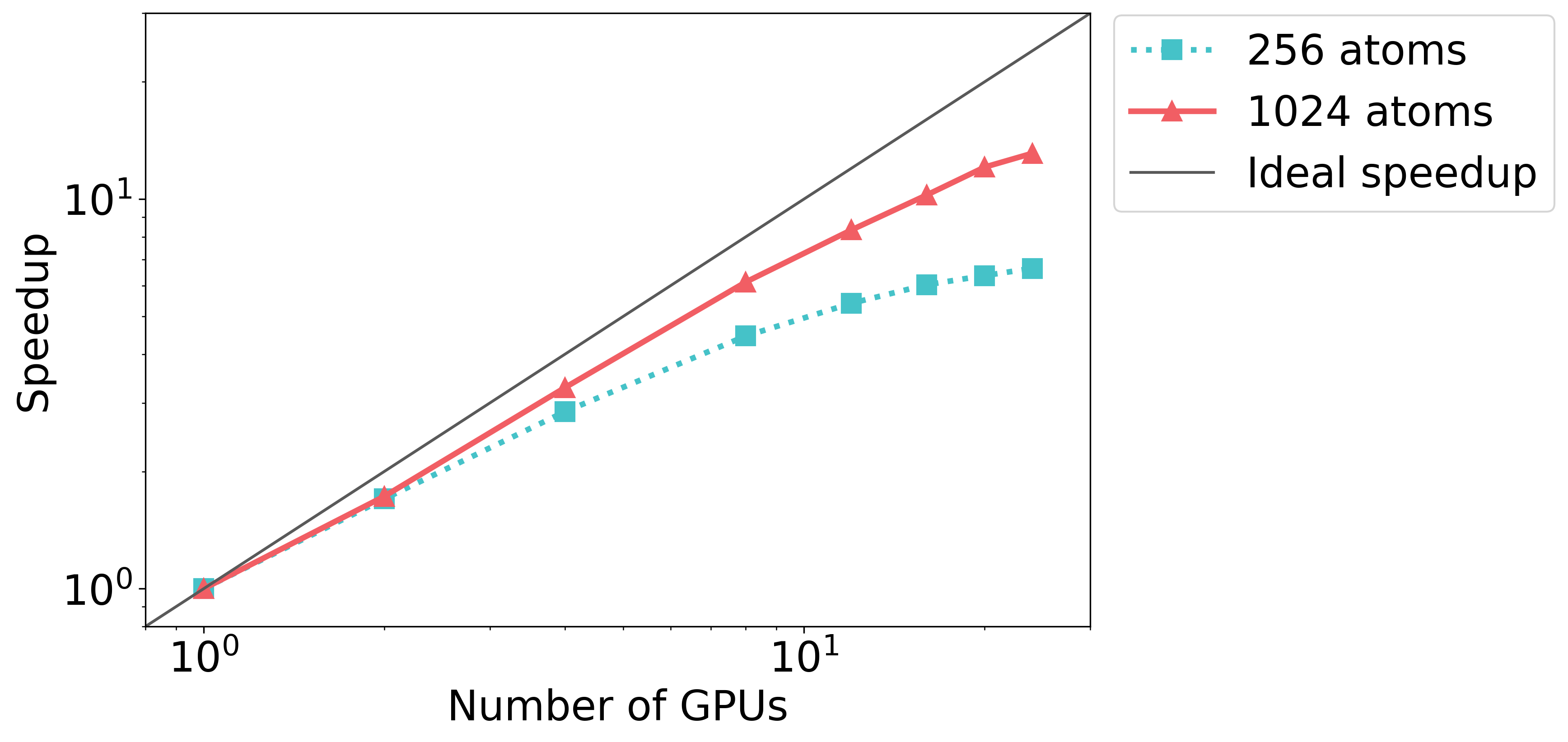}
\caption{Strong scaling results for \texttt{MALA} inference on an aluminum model at room temperature. Inferences were performed with either 256 or 1024 atoms, and speedup was calculated relative to inference times on a single GPU. The model used was trained as described in Ref.~\citenum{malapaper}.}
\label{fig:inference_strong_scaling}
\end{figure}

Of course, computational cost alone provides only a partial view of the feasibility of the \texttt{MALA} workflow. When \texttt{MALA} simulations are applied to large-scale systems, wall time becomes a more critical metric. This raises the question of how effectively the (linearly) growing computational cost can be distributed across multiple GPUs. To address this, we examined the strong scaling behavior of \texttt{MALA} inferences, with results presented in Fig.~\ref{fig:inference_strong_scaling} and computational parameters detailed in Tab.~\ref{tab:details_strong_scaling}.

\begin{table}
\centering  
\caption{Computational details for strong scaling experiments: All experiments were conducted on aluminum, with results shown in Figs.~\ref{fig:inference_strong_scaling}, \ref{fig:inference_strong_scaling_portions}, and \ref{fig:calculation_proportions}.}
\label{tab:details_strong_scaling}		
\begin{tabularx}{\textwidth}{XXXXXXX}%cccccccc}
\toprule
Number of Atoms &Number of gridpoints &Number of GPUs &Gridpoints per atom &Gridpoints per GPU \ \\\midrule
256 &$8.00\times10^{6}$ &1 &31250 &   $8.00\times10^{6}$ \\
256 &$8.00\times10^{6}$ &2 & 31250 &  $4.00\times10^{6}$ \\
256 &$8.00\times10^{6}$ &4 & 31250 &  $2.00\times10^{6}$ \\
256 &$8.00\times10^{6}$ &8 & 31250 &  $1.00\times10^{6}$ \\
256 &$8.00\times10^{6}$ &12 & 31250 & $6.67\times10^{5}$ \\
256 &$8.00\times10^{6}$ &16 & 31250 & $5.00\times10^{5}$ \\
256 &$8.00\times10^{6}$ &20 & 31250&  $4.00\times10^{5}$ \\
256 &$8.00\times10^{6}$ &24 & 31250 & $3.33\times10^{5}$ \\
1024 &$3.20\times10^7$ &1 & 31250 &   $3.20\times10^{7}$ \\
1024 &$3.20\times10^7$ &2 & 31250 &   $1.60\times10^{7}$ \\
1024 &$3.20\times10^7$ &4 & 31250 &   $8.00\times10^{6}$ \\
1024 &$3.20\times10^7$ &8 & 31250 &   $4.00\times10^{6}$ \\
1024 &$3.20\times10^7$ &12 & 31250 &  $2.67\times10^{6}$ \\
1024 &$3.20\times10^7$ &16 & 31250 &  $2.00\times10^{6}$ \\
1024 &$3.20\times10^7$ &20 & 31250 &  $1.60\times10^{6}$ \\
1024 &$3.20\times10^7$ &24 & 31250 &  $1.33\times10^{6}$ \ \\\bottomrule
\end{tabularx}
\end{table}

In a strong scaling test, the problem size remains constant while the number of computational resources increases~\cite{StrongScaling}. This approach allows for assessing the ability of a parallelized workflow to distribute workload effectively across computational nodes. In Fig.~\ref{fig:inference_strong_scaling}, we demonstrate this by selecting 256-atom and 1024-atom aluminum systems at room temperature and performing inferences using a \texttt{MALA} model trained following the methodology outlined in Ref.~\citenum{malapaper}. These inferences were conducted on up to 24 GPUs, with each GPU assigned to a separate computational node. For these inferences, the speedup is defined as
\begin{equation}
    \mathrm{Speedup} = \frac{t_0}{t_{N_\mathrm{G}}} \;, \label{eq:speedup}
\end{equation}
where $t_0$ the wall time for computation on a single node or GPU, and $t_{N_\mathrm{G}}$ is the wall time on $N_\mathrm{G}$ nodes or GPUs.

\begin{figure}[htp]
\centering
\includegraphics[width=0.95\columnwidth]{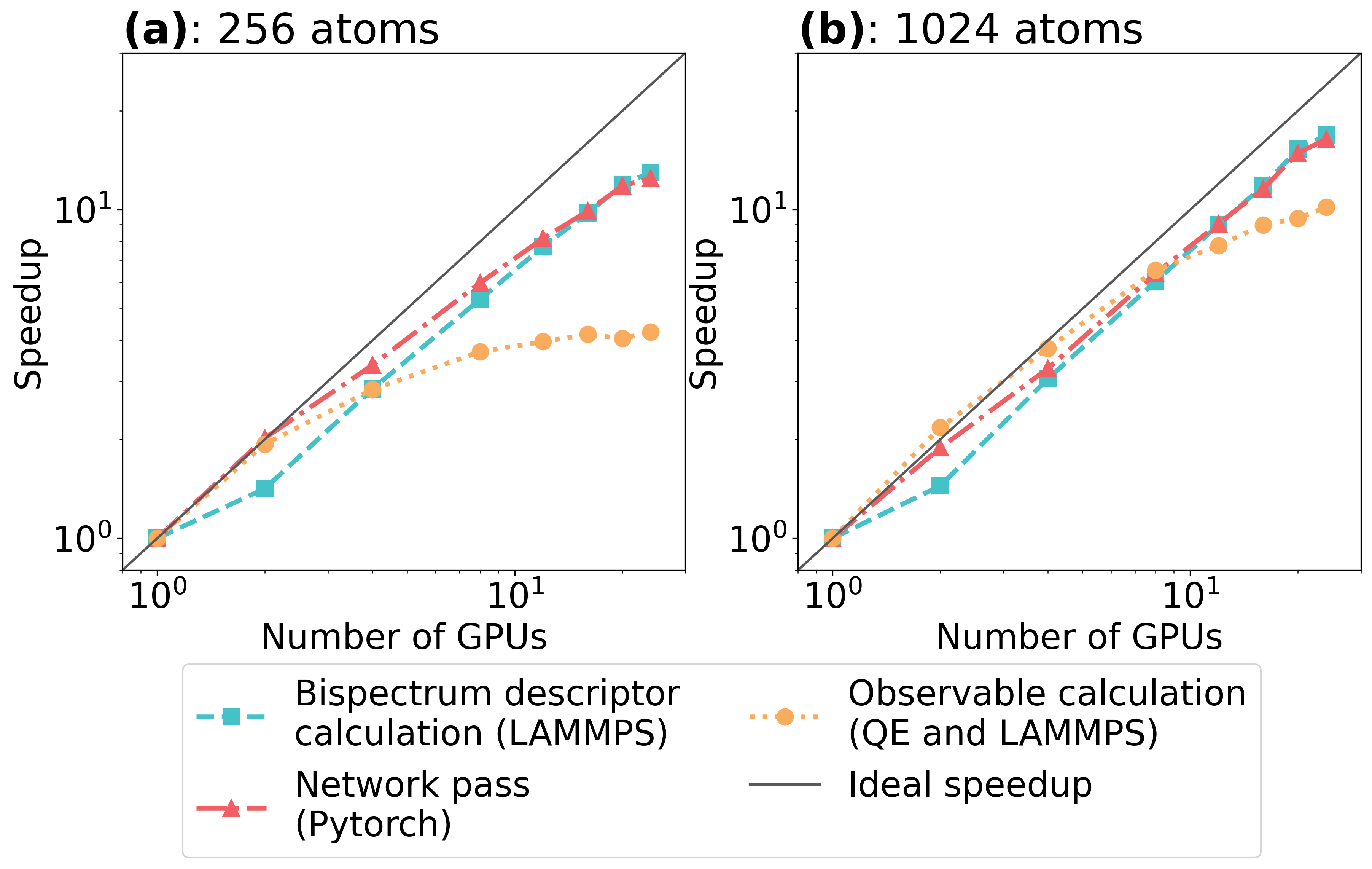}
\caption{Strong scaling results for individual components of \texttt{MALA} inferences on an aluminum model at room temperature. Inferences were performed on systems of either 256 atoms \textbf{(a)} or 1024 atoms \textbf{(b)}, with speedup calculated relative to single-GPU inference times. The model used was trained following the approach in Ref.~\citenum{malapaper}.}
\label{fig:inference_strong_scaling_portions}
\end{figure}

\begin{figure}[htp]
\centering
\includegraphics[width=0.95\columnwidth]{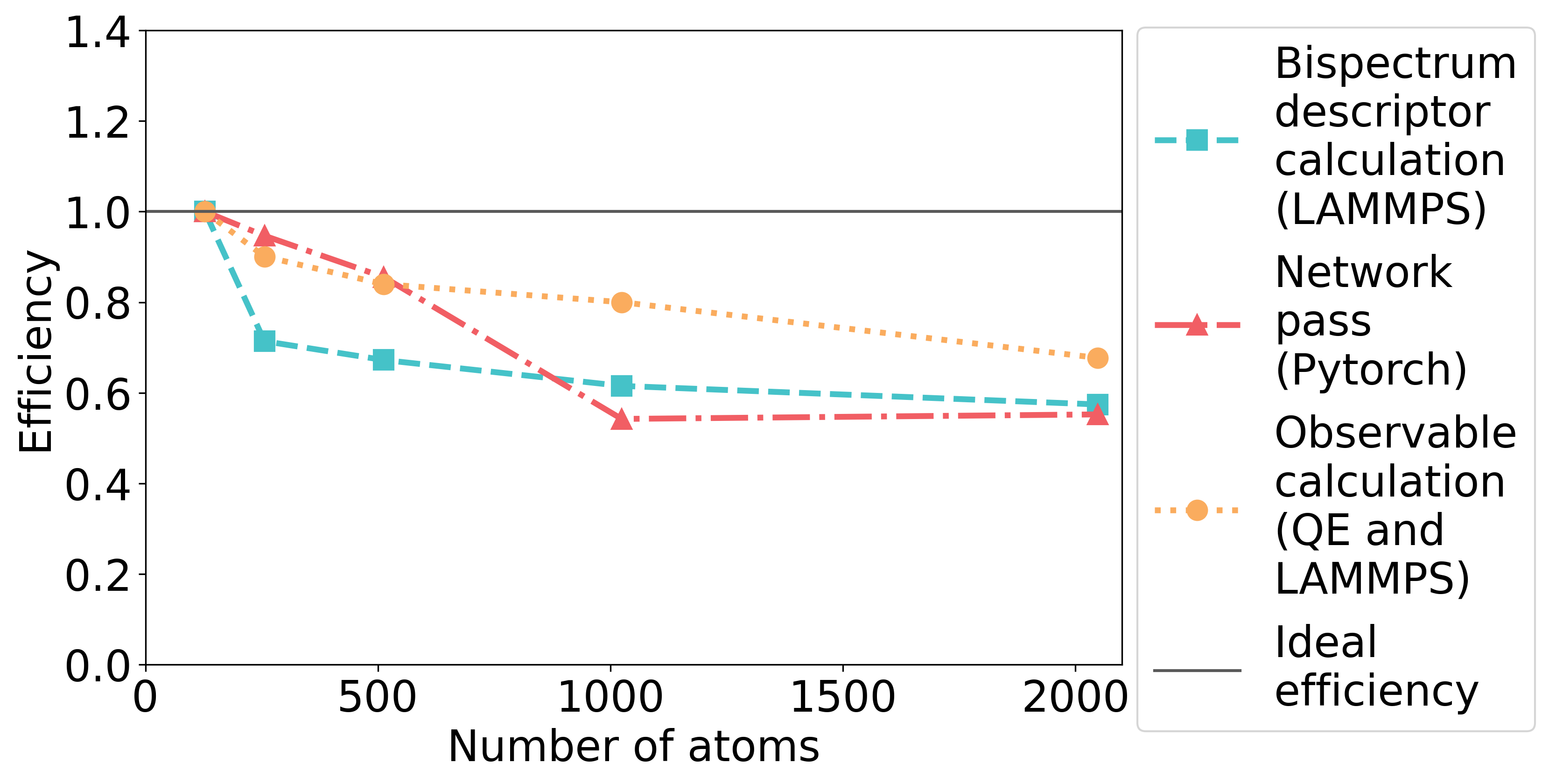}
\caption{Weak scaling results for individual components of \texttt{MALA} inferences based on a beryllium model at room temperature, as described in Ref.~\citenum{sizetransferpaper}. Each successively larger calculation was performed with double the number of GPUs compared to the previous one, beginning with a single GPU for a 256-atom system.}
\label{fig:inference_weak_scaling_portions}
\end{figure}

\begin{figure}[htp]
\centering
\includegraphics[width=0.95\columnwidth]{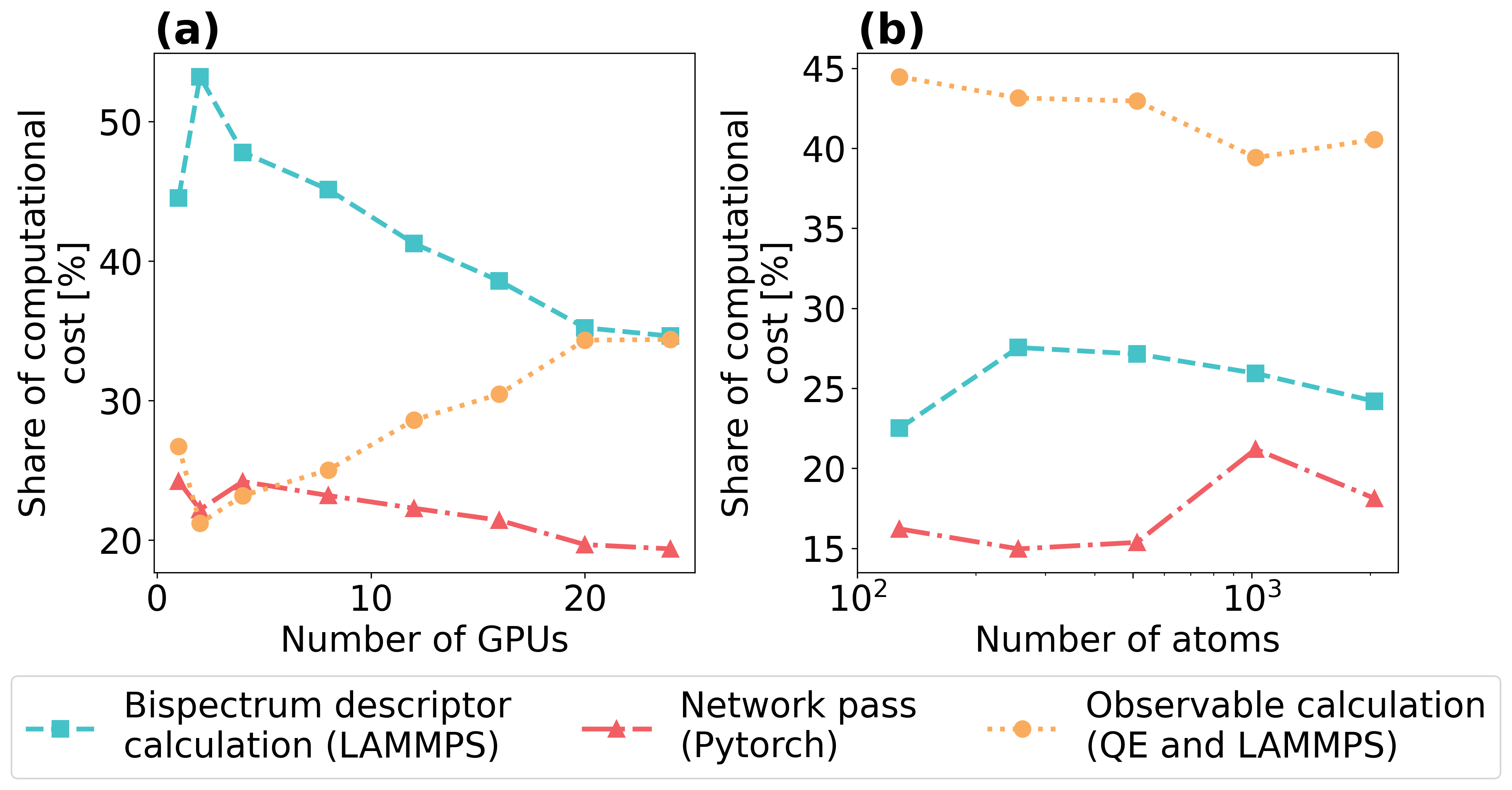}
\caption{Share of computation time for different components of \texttt{MALA} inference and their scaling behavior with respect to the number of GPUs \textbf{(a)} and the number of atoms \textbf{(b)}. For \textbf{(a)}, results are based on Fig.~\ref{fig:inference_strong_scaling_portions}\textbf{(b)}, representing 1024 aluminum atoms at room temperature. For \textbf{(b)}, results are derived from the weak scaling calculations shown in Fig.~\ref{fig:inference_weak_scaling_portions}, representing beryllium at room temperature.}
\label{fig:calculation_proportions}
\end{figure}

Comparing the reported inference speedups in Fig.~\ref{fig:inference_strong_scaling} to an ideal speedup reveals that beyond a certain number of GPUs, no additional speedup can be achieved. This limitation arises because, at high GPU counts, computational speedup is constrained by the serial portions of the code. To further investigate which parts of the code may be affected by serial computations, Fig.~\ref{fig:inference_strong_scaling_portions} presents the same results as Fig.~\ref{fig:inference_strong_scaling}, broken down into the three primary calculation components, namely
\begin{enumerate}
    \item Bispectrum descriptor calculation performed via \texttt{LAMMPS}, shown in blue in Figs.~\ref{fig:inference_strong_scaling_portions},~\ref{fig:inference_weak_scaling_portions}, ~\ref{fig:calculation_proportions},
    \item Neural network pass performed via \texttt{PyTorch}, shown in red in Figs.~\ref{fig:inference_strong_scaling_portions}, ~\ref{fig:inference_weak_scaling_portions}, ~\ref{fig:calculation_proportions},
    \item Observable calculation, which includes both the electronic density calculations performed by \texttt{Quantum ESPRESSO} and the calculation of Gaussian descriptors, shown in orange in Figs.~\ref{fig:inference_strong_scaling_portions},~\ref{fig:inference_weak_scaling_portions},~\ref{fig:calculation_proportions}.
\end{enumerate}

In this analysis, it is clear that the primary source of serial bottlenecks in the parallelized \texttt{MALA} inference lies within the observable calculation step. The first two parts of the \texttt{MALA} workflow maintain near-ideal speedup significantly longer than the final calculation step. For instance, in the case of 1024 atoms, Fig.~\ref{fig:inference_strong_scaling}\textbf{(b)} shows that nearly ideal speedup is achieved for the descriptor calculation and the neural network pass. This is because these initial steps of the \texttt{MALA} inference are highly parallelizable. Computations of grid-based descriptors and forward passes for individual grid points are mostly independent from each other.

In contrast, the observable calculation step (requiring Fourier transformations and grid-based integrations) hampers overall scalability. The Gaussian descriptor calculations, which are also performed in this step, are parallelized using the same strategies as those for the bispectrum descriptors. As a result, the observed scaling in the observable calculation step depends on the proportion of computational load attributed to the Gaussian descriptor calculations versus other tasks like Fourier transformations. For instance, in Fig.~\ref{fig:inference_strong_scaling}\textbf{(a)}, observable calculations show minimal speedup beyond a few nodes, whereas for the larger system in Fig.~\ref{fig:inference_strong_scaling}\textbf{(b)}, significant speedups are achieved with more than ten nodes. This difference occurs because in a larger system, the Gaussian descriptor calculation constitutes a more substantial portion of the observable calculation relative to Fourier transformations.

In addition to strong scaling, we also investigate weak scaling. Here, both the number of computational nodes and the problem size are increased, keeping the workload per node constant~\cite{WeakScaling}. This approach assesses how the efficiency of a parallel computation changes as computational resources increase. Mathematically, weak scaling efficiency is identical to the strong scaling speedup in Eq.~\eqref{eq:speedup}, with the important distinction that, since the system size varies, weak scaling efficiency is bounded between $\mathrm{Efficiency} = 0$ and $\mathrm{Efficiency} = 1$. Unlike strong scaling, where ideal behavior is eventually unachievable due to serial components, workflows that are fully data-parallel, allowing computational nodes to operate independently, can exhibit near-ideal efficiency in weak scaling scenarios.

\begin{figure}[htp]
\centering
\includegraphics[width=0.95\columnwidth]{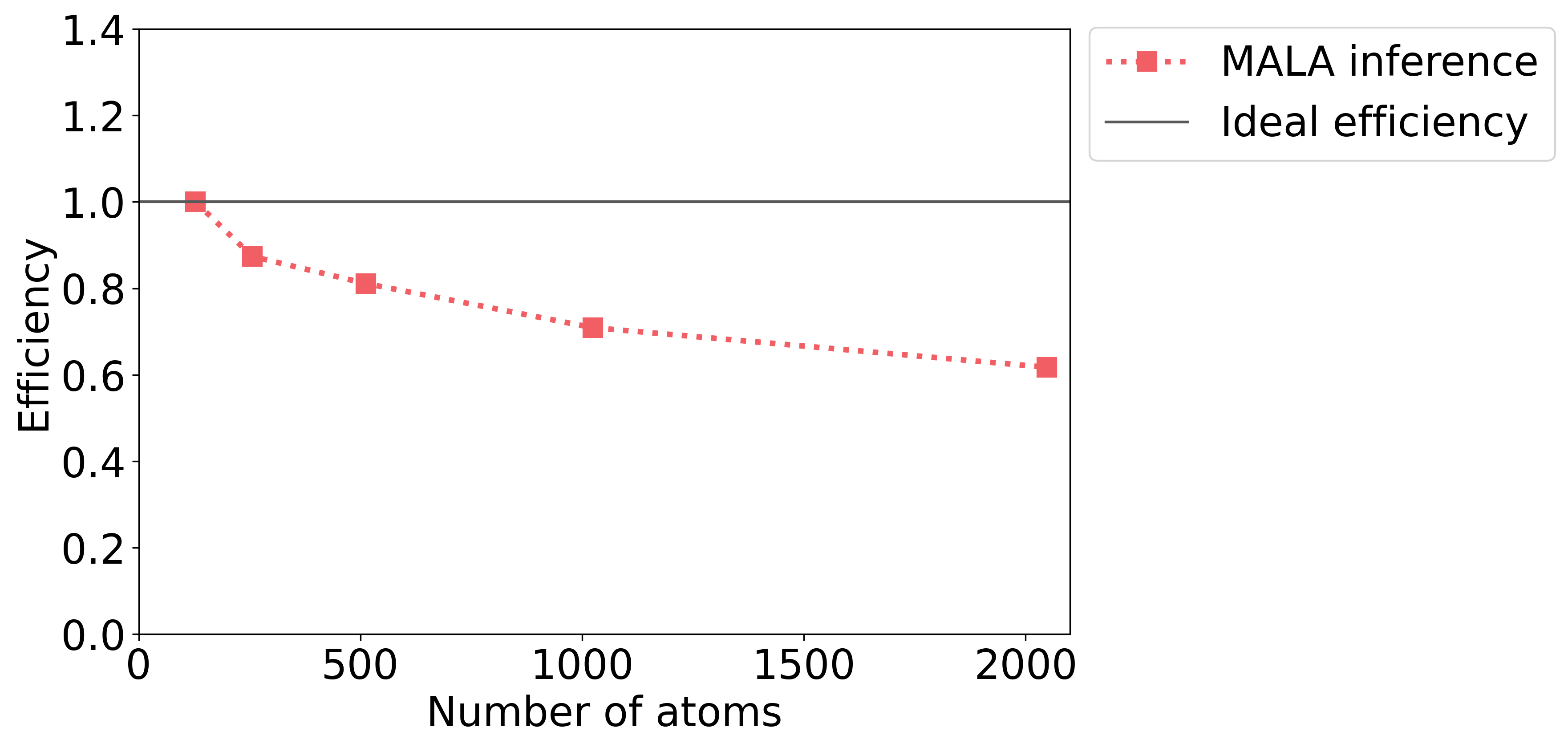}
\caption{Weak scaling for \texttt{MALA} inference based on a beryllium model at room temperature, as described in Ref.~\citenum{sizetransferpaper}. Each successively larger calculation was performed with double the number of GPUs compared to the previous one, starting with a single GPU for a 256-atom system.}
\label{fig:inference_weak_scaling}
\end{figure}

Weak scaling for \texttt{MALA} inferences was calculated using beryllium systems of varying sizes, as shown in Fig.~\ref{fig:inference_weak_scaling}, with computational parameters provided in Tab.~\ref{tab:details_weak_scaling}. The results indicate that although scaling efficiency decreases with increasing system size and computational resources, the rate of this decline lessens for larger systems. This behavior arises because substantial portions of the \texttt{MALA} inference process require minimal communication between nodes, allowing for highly data-parallel execution.

\begin{table}
\centering  
\caption{Computational details for weak scaling experiments: All experiments were conducted on beryllium. For systems with 1024 and 2048 atoms, a slightly smaller number of grid points per atom (and thus per GPU) was used. This adjustment reflects the fact that atom count is generally the more relevant parameter from a user perspective, so the number of atoms was varied directly rather than the grid point count. Additionally, the numerical grid size was chosen for consistency with \texttt{Quantum ESPRESSO} simulations, which use a slightly smaller grid at larger atom counts. The resulting difference in grid points per GPU is approximately 6\%, small enough to allow for a quantitative comparison of the results shown in Figs.~\ref{fig:inference_weak_scaling},~\ref{fig:inference_weak_scaling_portions} and ~\ref{fig:calculation_proportions}.}
\label{tab:details_weak_scaling}		
\begin{tabularx}{\textwidth}{XXXXXXX}%cccccccc}
\toprule
Number of Atoms &Number of gridpoints &Number of GPUs &Gridpoints per atom &Gridpoints per GPU \ \\\midrule
128 &  $6.22\times10^5$ &1 &4860 & $6.22\times10^{5}$ \\
256 &  $1.24\times10^6$ &2 &4860 & $6.22\times10^{5}$ \\
512 &  $2.49\times10^6$ &4 &4860 & $6.22\times10^{5}$ \\
1024 & $4.67\times10^6$ &8 &4556 & $5.83\times10^{5}$ \\
2048 & $9.33\times10^6$ &16 &4556 & $5.83\times10^{5}$ \\\bottomrule
\end{tabularx}
\end{table}

This behavior can be further analyzed by benchmarking the weak scaling efficiency of individual components of a \texttt{MALA} inference, as shown in Fig.~\ref{fig:inference_weak_scaling_portions}. The results indicate that the efficiency of all calculation components decreases relatively uniformly, with the efficiency of the observable calculation decreasing slightly more gradually. Overall, computational efficiency stabilizes at approximately 0.6 to 0.7, enabling effective utilization of large computational resources for extended systems.

Finally, the computational cost of individual inference components can be analyzed using weak and strong scaling results to gain insights into future parallelization potential. To this end, the relative computational cost of the three main calculation components is visualized in Fig.~\ref{fig:calculation_proportions} as a function of both the number of nodes used for inference and the system size.

A key takeaway from Fig.~\ref{fig:calculation_proportions}\textbf{(b)} is that the proportion of computation time for each task remains relatively stable as the number of atoms increases, provided the computational load per node is kept constant, as done here. Additionally, in these experiments, observable calculation is clearly the most computationally expensive task.

However, this trend is not universal, as shown in Fig.~\ref{fig:calculation_proportions}\textbf{(a)}. For small numbers of GPUs (i.e., when each node must process a large number of grid points), the bispectrum descriptor calculation dominates the computational cost. As more computational resources are added, the relative share of the bispectrum descriptor calculation decreases, while the observable calculation’s share rises by a similar amount. The computational cost of the neural network pass remains fairly constant and relatively low, due to \texttt{PyTorch}’s efficient neural network implementation.

This behavior highlights that, due to the current implementations of bispectrum descriptor and observable calculations, the former is more readily parallelized. Thus, with a large number of computational resources, the observable calculation emerges as the primary bottleneck in \texttt{MALA} inferences, particularly for extended systems where a constant workload per node is maintained. This analysis underscores the potential for future development in the \texttt{MALA} package, specifically enhancing parallel efficiency for observable calculations by addressing serial portions of the code.

\section{Conclusion}
\label{s.conclusion}

In this work, we have presented the technical details of the \texttt{MALA} package, a scalable and efficient machine learning framework designed to accelerate electronic structure calculations within the scope of density functional theory. \texttt{MALA} models are built using local descriptors of the ionic configuration around a point in space, enabling the package to deliver key electronic observables, including the local density of states (LDOS), electronic density, density of states (DOS), and total (free) energy.

We provided an in-depth overview of \texttt{MALA}’s software design and implementation. The \texttt{MALA} package offers a unified software library that integrates data sampling, model training, and scalable inference, while ensuring seamless compatibility with state-of-the-art simulation tools such as \texttt{Quantum ESPRESSO} and \texttt{LAMMPS}.

To illustrate the capabilities of the \texttt{MALA} package, we included a tutorial-style example for training a model on boron clusters. Additionally, we demonstrated \texttt{MALA}’s versatility in accurately modeling systems across the solid-liquid phase boundary and over a range of temperatures. More importantly, we showed that \texttt{MALA} models can predict electronic structures at scales far exceeding those accessible to standard DFT calculations. Specifically, we demonstrated that a \texttt{MALA} model can predict the electronic structure of a stacking fault in a 131,072-atom beryllium slab, reproducing the system’s energetics at a qualitative level.

Our analysis of the \texttt{MALA} package’s computational scaling, under both strong and weak scaling conditions, highlights its efficiency and scalability, and provides insights into computational bottlenecks that will be addressed in future development.

We anticipate that the \texttt{MALA} package will be a valuable resource for materials modeling, particularly in applications where the quantum mechanical properties of the electronic structure are essential. \texttt{MALA}’s potential to resolve the electronic structure at large scales efficiently makes it especially suitable for systems beyond the reach of standard DFT. We envision applying \texttt{MALA} to model materials with complex electronic structures at large scales, such as twisted bilayer and multilayer systems. Additionally, \texttt{MALA}’s ability to compute the LDOS at large scales can be used to generate scanning tunneling microscopy (STM) images, which are otherwise challenging to compute for extended systems.

Future developments for \texttt{MALA} include extending its functionality to compute band structures and transport properties, such as electrical conductivity. We also plan to implement force calculations based on \texttt{MALA}’s predicted total energy, enabling machine learning-driven DFT-MD simulations. This approach would complement machine-learning interatomic potentials with the added capability of resolving the electronic structure at each time step at large scales. Additionally, we envision hybrid models that pair \texttt{MALA} with atomistic interatomic potentials.

In summary, we have provided a comprehensive overview of the \texttt{MALA} package and demonstrated its capabilities as a versatile solution for diverse computational materials research needs.

\section*{Acknowledgments}

Sandia National Laboratories is a multi-mission laboratory managed and operated by National Technology \& Engineering Solutions of Sandia, LLC (NTESS), a wholly owned subsidiary of Honeywell International Inc., for the U.S. Department of Energy’s National Nuclear Security Administration (DOE/NNSA) under contract DE-NA0003525. This written work is authored by an employee of NTESS. The employee, not NTESS, owns the right, title and interest in and to the written work and is responsible for its contents. Any subjective views or opinions that might be expressed in the written work do not necessarily represent the views of the U.S. Government. The publisher acknowledges that the U.S. Government retains a non-exclusive, paid-up, irrevocable, world-wide license to publish or reproduce the published form of this written work or allow others to do so, for U.S. Government purposes. The DOE will provide public access to results of federally sponsored research in accordance with the DOE Public Access Plan.

This work was partially supported by the Center for Advanced Systems Understanding (CASUS) which is financed by Germany's Federal Ministry of Education and Research (BMBF) and by the Saxon state government out of the State budget approved by the Saxon State Parliament. Computations were performed on a Bull Cluster at the Center for Information Services and High-Performance Computing (ZIH) at Technische Universit\"at Dresden and on the cluster Hemera at Helmholtz-Zentrum Dresden-Rossendorf (HZDR).

\appendix

\section{Thermal electronic ensembles}
\label{sec:appendix.thermal.electronic.ensembles}
At finite electronic temperature, we work within the quantum statistical mechanical description of the electronic structure problem, where properties at thermodynamic equilibrium are provided by the free energy operator
\begin{equation}
\label{eq:freeenergyoperator}
\hat{A} = \hat{H} - \tau \hat{S} \; 
\end{equation}
with $\hat{H}$ denoting the Born-Oppenheimer Hamiltonian given in Eq.~\eqref{eq:hamiltonian} and $\hat{S}$ the entropy operator. The latter is defined in terms of the density matrix $\hat{\Gamma}$ which is defined in terms of projection operators 
\begin{equation}
\label{eq:densitymatrix}
\hat{\Gamma} = \sum_{j} w_{j} \ket{\Psi_{j}}\bra{\Psi_{j}} \; , \end{equation}
where $w_{j}$ are statistical weights. 

In Eq.~\eqref{eq:freeenergyoperator}, the entropy operator is connected to the temperature $\tau$ of the system. Due to working within the Born-Oppenheimer approximation we have separated the coupled interacting electron-ion system into the subsystem of ions and electrons, where we explicitly distinguish between the temperature of ionic subsystem $\tau\ion$ and that of the electronic subsystem $\tau$.

In the context of quantum statistical mechanics~\cite{ToKuKuSa83}, the calculation of any arbitrary observable $O$ is given by the density matrix as 
\begin{equation}
O[\hat{\Gamma}] = \Tr{\left(\hat{\Gamma}\hat{O}\right)} \; ,
\end{equation}
and thus, the free energy of the system is expressed as 
\begin{equation}
A[\hat{\Gamma}] = \Tr{\left(\hat{\Gamma}\hat{A}\right)} \; . \label{eq:grandcanonicalpotential_qsm}
\end{equation}

Thermodynamic equilibrium for such an ensemble is the density matrix $\hat{\Gamma}$ yielding the minimal free energy $A$. This corresponds to the set of statistical weights for Eq.~\eqref{eq:densitymatrix} given by 
\begin{equation}
w^0_{\nu} = \frac{\exp{\left(-\frac{E^0_{\nu}}{\kB \tau}\right)}}{\sum_{\nu} \exp{\left(-\frac{E^0_{\nu}}{\kB \tau}\right)}} \; ,
\end{equation}
with the ground state eigenvalues $E^0_{\nu}$ of the Born-Oppenheimer Hamiltonian. 

Solving the quantum statistical electronic problem suffers in essence from the same computational infeasibility as solving the non-relativistic Schrödinger equation for the electronic ground state defined in Eq.~\eqref{eq:se}. 

A computational feasible solution of the electronic structure problem for a thermal ensemble is given by the thermal generalization of DFT~\cite{mermin_thermal_1965,PPFS11,PPGB14}. 
Here, the ground-state energy functional given by Eq.~\eqref{eq:energy.ks} is replaced by a free energy functional
\begin{equation}
A[n] = T\s[n] - \tau S\s[n] + E\Hartree[n] + E\eleion[n] + A\xc[n] + E\ionion\; , \label{eq:KSfreeenergy}
\end{equation}
where the XC energy $E\xc$ becomes the XC free energy $A\xc$ and an electronic entropy term $S\s$ appears. The electronic entropy is defined in terms of the Kohn-Sham eigenvalues as 
\begin{equation}
S\s = - \kB \sum_{j=1}^{N\ele'} \Big[ f_\beta(\epsilon_j) \ln\big(f_\beta(\epsilon_j)\big) + \big(1-f_\beta(\epsilon_j)\big)\ln\big(1-f_\beta(\epsilon_j)\big) \Big] \; , \label{eq:electronicentropy}
\end{equation}
or likewise
\begin{equation}
S\s = - \kB \int d\epsilon\; \Big[ f_\beta(\epsilon) \ln\big(f_\beta(\epsilon)\big) + \big(1-f_\beta(\epsilon)\big)\ln\big(1-f_\beta(\epsilon)\big) \Big] D(\epsilon) \; , \label{eq:electronicentropyDOS}
\end{equation}
when expressed in terms of the DOS, where $N\ele' \neq N\ele$ is the number of wave functions included in the Kohn-Sham system. For practical calculations, $N\ele'$ has to be set higher than $N\ele$ to account for thermal excitations of electrons. It further scales with temperature, giving rise to the unfavorable scaling properties of DFT with temperature. The Fermi-Dirac distribution in Eq.~\eqref{eq:electronicentropy} is given by
\begin{equation}
f_\beta(\epsilon) = \frac{1}{1+ \exp{\left[\beta(\epsilon-\epsilon_\mathrm{F})\right]}} \; , \label{eq:fermidistribution}
\end{equation}
with $\beta = 1/(k_{B}\tau)$ denoting the inverse electronic temperature, $\epsilon_\mathrm{F}$ the Fermi energy, and $\kB$ the Boltzmann constant. The Fermi energy takes the place of the chemical potential $\mu$, which is, as mentioned above, formally not constant in the context of the canonical ensemble. Rather, during the Kohn-Sham self-consistency cycle, $\epsilon_\mathrm{F}$ is determined numerically based on the number of electrons. Observables are thus calculated with a $\epsilon_\mathrm{F}$ that reproduces the correct number of electrons with the self-consistent Kohn-Sham density. 

While the overall structure of the Kohn-Sham approach is retained in the finite-temperature case, one notable difference is in the calculation of the density. The definition of the electronic density is extended by the inclusion of a thermal excitations
\begin{equation}
n(\br) = \sum_{j=1}^{N\ele'} f_\beta(\epsilon_j) \psi^*_j(\br)\psi_j(\br) \; . \label{eq:kohnshamdensity_T}
\end{equation}
in terms of the Fermi-Dirac distribution.

A second notable difference to ground-state Kohn-Sham DFT is the temperature dependence of several terms involved. The Kohn-Sham eigenvalues $\epsilon_j$ become temperature dependent in the finite-temperature case, as does the XC free energy. In case of the latter, temperature-dependent XC effects are a topic of ongoing research~\cite{PPFS11,PPGB14,TempXC1,TempXC2,TempXC3,TempXC4}, and an accurate computational treatment of high temperature systems requires the development of explicitly temperature-depdendent XC functionals~\cite{KoPeBu23}.

%% Statement on the use of generative AI
\vspace{1cm}
During the preparation of this work the authors used \texttt{DeepL} in order to check the text for grammar mistakes. After using this service, the authors reviewed and edited the content as needed and take full responsibility for the content of the publication.

\bibliographystyle{elsarticle-num} 
\bibliography{references.bib}


%% \begin{thebibliography}{00}


%% \bibitem{}

%% \end{thebibliography}
\end{document}